\documentclass[useAMS,usenatbib]{mnras} 
\usepackage{graphicx}
\usepackage{epsf}
\usepackage{epsfig}
\usepackage{times}
\usepackage{amssymb, amsmath}
\usepackage{url} 
\usepackage{aas_macros}
\usepackage{color}

\title[Inhomogeneous subgrid clumping]
  {The impact of inhomogeneous subgrid clumping on cosmic reionization}

\author[Y.~Mao et al.]
  {Yi Mao,$^{1}$\thanks{Email: ymao@tsinghua.edu.cn}
  Jun Koda,$^{2}$  
  Paul R. Shapiro,$^3$ 
  Ilian T. Iliev,$^4$ 
  Garrelt Mellema,$^5$  
  \newauthor 
  Hyunbae Park,$^{6}$ 
  Kyungjin Ahn,$^{7}$ 
  and Michele Bianco $^4$ \\   
  $^1$ Department of Astronomy and Tsinghua Center for Astrophysics, Tsinghua University, Beijing 100084, China \\
  $^2$ Dipartimento di Matematica e Fisica, Universit\`a degli Studi Roma Tre, Via della Vasca Navale 84, 00146 Rome, Italy \\
  $^3$ Department of Astronomy and Texas Cosmology Center, University of Texas, Austin, TX 78712, USA \\
  $^4$ Astronomy Centre, Department of Physics \& Astronomy, Pevensey II Building, University of Sussex, Falmer, Brighton BN1 9QH, UK \\
  $^5$ Department of Astronomy and Oskar Klein Centre, AlbaNova, Stockholm University, SE-106 91 Stockholm, Sweden \\
  $^6$ Kavli Institute for the Physics and Mathematics of the Universe (WPI), The University of Tokyo Institutes for Advanced Study, The University of Tokyo, \\  Kashiwa, Chiba 277-8583, Japan \\ 
  $^7$ Department of Earth Science, Chosun University, Gwangju 501-759, Korea 
  } 
\date{Accepted 2019 October 20. Received 2019 September 17; in original form 2019 June 6.}

\pubyear{2019}

\begin{document}
\label{firstpage}
\pagerange{\pageref{firstpage}--\pageref{lastpage}} 
\maketitle

\begin{abstract}

Cosmic reionization was driven by the imbalance between early sources and sinks of ionizing radiation, 
both of which were dominated by small-scale structure and are thus usually treated in cosmological reionization 
simulations by subgrid modelling. The recombination rate of intergalactic hydrogen is customarily boosted by
a subgrid {\it clumping factor}, ${\left<n^2\right>/\left<n\right>^2}$, 
which corrects for unresolved fluctuations in gas density ${n}$ 
on scales below the grid-spacing of coarse-grained simulations. 
We investigate in detail the impact of this 
{\it inhomogeneous} subgrid clumping on reionization and its observables, as follows: 
(1) Previous attempts generally underestimated the clumping factor because of
insufficient mass resolution.  We perform a high-resolution $N$-body simulation that resolves haloes 
down to the pre-reionization Jeans mass to derive the time-dependent, spatially-varying {\it local} 
clumping factor and a fitting formula for its correlation with local overdensity. 
(2) We then perform a large-scale $N$-body and radiative transfer simulation that accounts for
this inhomogeneous subgrid clumping by applying this clumping factor-overdensity correlation. 
Boosting recombination significantly slows the expansion of ionized regions, which delays
completion of reionization and suppresses 21~cm power spectra on large scales in the later stages of reionization. 
(3) We also consider a simplified prescription in which the globally-averaged, time-evolving
clumping factor from the same high-resolution $N$-body simulation 
is applied uniformly to all cells in the reionization simulation,
instead. Observables computed with this model 
agree fairly well with those from the inhomogeneous clumping model,  e.g. predicting 21~cm power spectra to 
within 20\% error, suggesting it may be a useful approximation.

\end{abstract}

\begin{keywords}
  Cosmology: theory--reionization-- 
  methods: numerical--
  galaxies: intergalactic medium
\end{keywords}


\section{Introduction}

Observational astronomy has recently made important progress in advancing our knowledge frontier of the epoch of reionization (EOR) (e.g.\ \citealt{2010Natur.468..796B,2013MNRAS.433..639P,2013A&amp;A...550A.136Y,2014ApJ...788..106P,2014A&amp;A...568A.101J,2015ApJ...801...51J,2015PhRvD..91b3002D,2015ApJ...802L..19R,2015ApJ...809...61A,2015ApJ...809...62P,2017ApJ...838...65P,2019MNRAS.488.4271G}), and will in the foreseeable future answer key open questions such as: When did the EOR begin and end? Over what time period the abundance of neutral hydrogen drops significantly? What is the characteristic size distribution of H~II regions, and its evolution? Does the reionization on average proceed inside-out, with higher density regions first, or outside-in?

A bottleneck in the theoretical quest for answers to those questions is the estimate of hydrogen recombination rate during reionization. Hydrogen recombination is a process wherein ionized hydrogen and free electrons recombine into neutral hydrogen atoms. To reionize these recombined atoms, a fraction of the ionizing photons is used. This means that less ionizing photons are available for increasing the sizes of H~II regions, slowing down the overall reionization process. The quantitative calculation of the recombination rate, nevertheless, is intrinsically difficult, because the recombination rate is affected by the clumpiness of local ionized gas in the intergalactic medium (IGM). Specifically, the rate is proportional to $\left<n_{\rm HII,IGM}^2\right>_{\rm cell}$ (the cellwise local average of the {\it square} of H~II density in the IGM) which, in principle, can be calculated only if the gas density fluctuations at {\it all} scales are resolved.  
In practice, it is customary to define the subgrid {\it clumping factor}, 
\begin{equation}
C_{\rm HII,IGM,cell} \equiv \left<n_{\rm HII,IGM}^2\right>_{\rm cell}/\left<n_{\rm HII,IGM}\right>_{\rm cell}^2\,,
\end{equation}
which relates the physical but nonlinear quantity $\left<n_{\rm HII,IGM}^2\right>_{\rm cell}$ to the linear quantity $\left<n_{\rm HII,IGM}\right>_{\rm cell}$ (the cellwise local average of the H~II density in the IGM). The latter can be calculated from theory or simulations by smoothing density fluctuations over the coarse-grained resolution. As such, the gas clumping factor corrects for the unresolved density fluctuations. 

Early analytical and semianalytic models of reionization either assumed a constant (in space and time) clumping factor (e.g.\ \citealt{2003ApJ...591...12C}; \citealt*{2007MNRAS.375..324Z}), a clumping factor based on linear theory (e.g.\ \citealt*{2000ApJ...530....1M}), or ignore clumping altogether ($C=1$; \citealt{2004ApJ...610....1O}).  Attempts have been made to improve upon this in numerical radiative transfer (RT) simulations (e.g., \citealt{2006MNRAS.372..679M,2007MNRAS.376..534I}; \citealt*{2007ApJ...657...15K}) and some semianalytic models (e.g.\ \citealt{2013MNRAS.433.2900D}), by using a single, globally uniform but time varying, clumping factor, derived from high resolution small box simulations. However, all these simplified treatments generally fail to take into account the inhomogeneous nature of the clumping factor. In fact, simulations in other contexts, e.g.\ cosmological simulations of galaxy formation \citep{2015MNRAS.446.3330T}, also often accounted for unresolved structures with oversimplified treatments of subgrid clumping factor. 

The small-scale inhomogeneities can be divided into two major types, depending on whether they are caused by self-shielded virialized haloes or unshielded filamentary IGM regions. For haloes, two distinct populations can be defined by the virial temperature before reionization, $T_{\rm vir}=10^4\,{\rm K}$. 
Here we assume that the gas contains only the metal-free, primordial composition. The dominant source of photons for reionization is formed by the group of the {\it atomically cooling haloes} (``ACHs'') in the mass range $M \gtrsim 10^8 M_\odot$ (with $T_{\rm vir}\gtrsim 10^4\,{\rm K}$), in which gas radiatively cools through collisionally excited lines of atomic hydrogen. Minihaloes --- haloes in the mass range $10^4 M_\odot \lesssim M \lesssim 10^8 M_\odot$ (with $T_{\rm vir}\lesssim 10^4\,{\rm K}$) --- are the other type of populations which are only able to form stars by using ${\rm H}_2$ molecules as the gas coolant through rotational-vibrational line excitations. However, ${\rm H}_2$ molecules are vulnerable to dissociation by UV photons in the Lyman-Werner bands, which are produced by the
first stars, long before a significant fraction of neutral hydrogen is reionized by the ionizing radiation from such stars (e.g.\ \citealt*{1997ApJ...476..458H}; \citealt{2012ApJ...756L..16A}). While minihaloes generically cannot make a significant contribution to the ionizing background, a minihalo can trap the intergalactic ionization front (I-front) by photoevaporating all of its baryonic gas when the I-front sweeps through a neutral patch containing both filamentary IGM and haloes \citep*{2004MNRAS.348..753S,2005MNRAS.361..405I}. However, minihaloes are biased relative to the matter density field in such a way that they are highly clustered around the more massive haloes, which are themselves clustered around density peaks in the matter distribution, where reionization starts and from which the intergalactic I-fronts propagate outward (\citealt*{2005ApJ...624..491I}; \citealt{2006ApJ...648..922S,2006MNRAS.366..689C}). As a result, in large-scale RT simulations of reionization, which are too coarse-grained to resolve the minihalo scale, the impact of the minihalo photoevaporation as a sink of ionizing photons emitted by ACHs is approximately accounted for in the escape fraction parameter assigned to those ACHs.  In effect, the minihaloes partially ``shield'' the ACHs, so their contribution to the absorption of ionizing starlight from the ACHs is degenerate with the uncertain value of the escape fraction assigned to each halo.

However, the filamentary IGM outside the evaporating minihaloes is better represented by explicitly accounting for it in the clumping factor of the ionized gas overtaken by the global I-fronts in the IGM. \cite{2011MNRAS.412L..16R} argued that since clumping is a measure of inhomogeneity in the density field, the aforementioned simplified treatment in which the clumping factor is modelled as spatially uniform does not account for variations of local unresolved density gradient, and may overestimate the importance of recombinations. 
This was confirmed by \cite{2014ApJ...787..146K}, using an analytical approach as a variant of the excursion set model of reionization \citep*{2004ApJ...613....1F}, and by \cite{2014MNRAS.440.1662S}, \cite{2016MNRAS.457.1550H} and \cite{2019MNRAS.484..933P}, using semi-numerical simulations of reionization based on the same analytical approximations.  

In this paper, we attempt to incorporate in a self-consistent manner the spatial variations of the local subgrid clumping factor in the {\it full} numerical N-body and RT simulations of reionization. The local IGM clumping factor varies in space in a way which is correlated with the variation of the locally-averaged mean matter density sampled with coarse-grained resolution by the N-body+RT simulations. This correlation was considered by \cite{2015ApJ...810..154K}, however qualitatively, and only at a single time ($z=5.7$). To make this correlation utilizable in numerical reionization simulations, we shall quantify this correlation with a fitting formula over a wide range of redshifts, using data of high-resolution N-body simulations. 
It is worth noting that many previous simulations (e.g., \citealt{2009MNRAS.394.1812P,2011MNRAS.412L..16R}; \citealt*{2011ApJ...743...82M}; \citealt{2012MNRAS.427.2464F,2014ApJ...789..149S}) adopted insufficient mass resolutions, with minimal halo masses comparable to the Jeans mass after reionization $\sim 10^9\,M_\odot$, but orders of magnitude larger than the Jeans mass before reionization $\sim 10^4\,M_\odot$  \citep{1994ApJ...427...25S,2008cosm.book.....W}.\footnote{Jeans-smoothing of the pre-reionization
baryons results in a gradual filtering of the baryonic mass fraction $\xi$ of dark-matter-dominated haloes as a 
function of the total (dark and baryonic) halo mass $M$.  For $z > 150$, for example, when the
baryon and the cosmic microwave background temperatures are still the same, coupled by Compton scattering, the baryon Jeans mass is independent of
redshift, corresponding to a total halo mass $M = 6\times 10^5 M_{\odot}$, and ${\xi = 1/[1 + (M_J/M)^{2/3}]}$ 
\citep{1994ApJ...427...25S,2008cosm.book.....W}.  
At $z < 150$, however, the IGM temperature drops 
adiabatically like $(1+z)^2$, so $M_J$ drops like $(1+z)^{3/2}$, 
but during this phase, the linear evolution of $\xi$ 
is more complicated because $M_J$ is no longer independent of redshift, so we do not quote it here.  
Eventually, if some reheating
of the IGM begins to halt the decline of its temperature without significantly reionizing it, 
such as the recoil heating associated with the Lyman $\alpha$ pumping of its 21~cm level population by the
Wouthuysen-Field mechanism \citep{2007ApJ...655..843C}
or heating by early X-ray sources, the Jeans mass will, thereafter, halt its decline and begin to 
increase with time.  As a result of preheating, we might then expect the prereionization filter scale to increase from 
$M = 10^4 M_{\sun}$, the value at $z = 10$ with no reheating, to a value as high as $\sim 10^5 M_{\sun}$.} 
The small-scale inhomogeneities in their simulations, and, hence, the clumping factor, were likely significantly underestimated. In our paper, we use a small-box high resolution N-body simulation with dark matter particle mass of $5\times 10^3\,M_\odot$, minimal resolved halo mass of $10^5\,M_\odot$, and a spatial resolution of less than $0.2\,h^{-1}$ comoving ${\rm kpc}$, in such a way as to allow gas (assuming the gas follows the dark matter distributions) to be Jeans smoothed on small scales, in the spirit of the small-box simulations in \cite{2005ApJ...624..491I,2006MNRAS.372..679M,2007MNRAS.376..534I,2007ApJ...657...15K}.

Since we only run $N$-body (i.e.\ no hydrodynamics) simulations for clumping factor, we will neglect RT and hydrodynamical effects in the clumping factor calculations. \cite{2012MNRAS.427.2464F} used cosmological hydrodynamic simulations which incorporate a treatment for self-shielding within Lyman limit systems, and showed that the clumping factor in the H~II regions can be suppressed, because the gas in the most overdense regions that is likely to be ionized earlier is self-shielded. Note, however, that we partially bypass this problem by excising N-body particles in the haloes from the clumping factor calculation, because haloes are generally self-shielded, as first done by \cite{2001AIPC..586..219S,2003MNRAS.341...81I,2005ApJ...624..491I,2006MNRAS.369.1625I}. Also, photoionization heating may further reduce the clumping factor because the increased pressure support may smooth out density fluctuations on small scales, as shown by cosmological hydrodynamic simulations in \cite{2009MNRAS.394.1812P,2012MNRAS.427.2464F,2016ApJ...831...86P}. Our estimate of the clumping factor, therefore, serves to represent the effect of the maximum IGM clumping.

\cite{2013ApJ...763..146E} demonstrated that it is necessary to resolve
the prereionization Jeans scale in order to take proper account of 
small structure in computing the clumping factor.  They did
this by post-processing with radiative transfer a time-slice of a high-resolution hydrodynamical simulation (with no radiation)
in a very small, sub-Mpc-sized box only as large as a single 
cell in a large-scale reionization simulation like our {\tt C$^2$-Ray}
simulations. The density field of the gas was taken as fixed
and non-evolving. This made it possible, however,
to distinguish the ionized gas from the
neutral and self-shielded regions in their tiny box, 
at the initial time of its exposure to ionizing radiation.
\cite{2016ApJ...831...86P} performed 
fully-coupled radiation-hydrodynamics of a similarly small-box with the 
same high-resolution necessary to resolve this scale, to follow its 
subsequent response to the arrival of ionization radiation during reionization. This followed the hydrodynamical back-reaction of the 
gas to its photoheating,
from the time of its first exposure to reionization, including
the time-dependent evolution of the clumping factor in this 
small box as the self-shielded regions photoevaporated.
Here, we will exclude such evaporating self-shielded regions from our 
treatment of the IGM, by excising the volumes inside haloes
before we compute the clumping factor.  

The purpose of this paper is to explore the impact of inhomogeneous IGM clumpiness on cosmic reionization. In order to demonstrate
the importance of resolving this small-scale structure
to account fully for the enhancement of the recombination rate
of the IGM that results, we have chosen here to maximize the 
effect by neglecting the time-dependence caused by hydrodynamical back-reaction.  We will base our clumping factor on the density
field in our high-resolution N-body simulation which resolves the
Jeans scale in the prereionization IGM, from which we excise the
regions inside haloes. Although the N-body simulation is a much
smaller volume than the reionization simulations to which we
will apply the results, it is much larger than those simulations
mentioned above that also resolved the prereionization Jeans
scale.   This makes it possible for us to consider the full range
of variation of the clumping factor with respect to the local
overdensity of the coarse-grained cells we will encounter in
such a large-scale reionization simulation that does not 
resolve this subgrid structure.   We will therefore be able
to derive a fitting formula for the correlation between this 
clumping factor and the local overdensity of the coarse-grained 
cells over which we compute it. By exploiting this 
clumping-overdensity correlation fitting formula, we shall perform a series of full numerical RT simulations of reionization, including one simulation that takes into account the inhomogeneous subgrid clumping factor,\footnote{\cite{2007MNRAS.377.1043M} performed a large-scale RT simulation with inhomogeneous subgrid clumping factor. However, their clumping models, in which either the subgrid clumping factor decreases when local density increases (their {\it C4} case), or the global average of clumping factor is constant in time  (their {\it C5} case), are less physical.}
and investigate how inhomogeneous clumping has an impact on the observables of reionization, including the reionization history, the cosmic microwave background (CMB) Thomson optical depth, the redshifted 21~cm signal, the kinetic Sunyaev-Zel'dovich effect, and the post-reionization Lyman-limit opacity. In addition, we will numerically implement the prescription of homogeneous clumping criticized by \cite{2011MNRAS.412L..16R}, but explore its ``comfort zone'', i.e.\ the condition under which the observational predictions by this simplified model virtually agree with the inhomogeneous clumping model. 

The rest of this paper is organized as follows. In Section~\ref{sec:smallbox}, we describe our small-box high resolution N-body simulation, and how we compute the subgrid clumping factor by smoothing N-body particle data using an adaptive kernel. From these results we derive a fitting formula for the correlation between local clumping factor and local overdensity. In Section~\ref{sec:RTsim}, we run a series of large-scale N-body+RT simulations, including one with inhomogeneous clumping, one with homogeneous clumping, and two with no clumping but with different photon production efficiencies. We explore the observational signatures of inhomogeneous clumping and investigate the comfort zone of the homogeneous clumping model in Section~\ref{sec:RTresult}. We end with concluding remarks in Section~\ref{sec:conclusion}.

\begin{table*}
\begin{center}
\begin{minipage}{1\linewidth}
\caption{N-body simulation parameters.}
\label{tab:summary_N-body_table}
\begin{tabular}{@{}lllllllll}
\hline
box size & $N_{\rm particle}$   & mesh   & spatial resolution\footnote{The force smoothing length is fixed to $1/20$ of the mean inter-particle spacing.} & $m_{\rm particle}$ & $M_{\rm halo,min}$ & coarse-grained mesh & coarse-grained cell size \\ 
\hline
$114\,h^{-1} {\rm Mpc}$ & $3072^3$ & $6144^3$ & $1.86\, {\rm kpc}/h$ & $5.47\times10^6\,M_\odot$ & $1.09\times 10^8\,M_\odot$ & $256^3$ & $0.45\,h^{-1} {\rm Mpc}$ \\
6.3 $\,h^{-1}$Mpc & $1728^3$ & $3456^3$ & $0.182\, {\rm kpc}/h$ & $5.12\times10^3\,M_\odot$ & $1.02\times 10^5\,M_\odot$ & $14^3$ & $0.45\,h^{-1} {\rm Mpc}$ \\
\hline
\end{tabular}
\end{minipage}
\end{center}
\end{table*}

\section{Computing clumping factor from small-scale high-resolution N-body simulations}
\label{sec:smallbox}

\subsection{N-body simulations}

We start by performing N-body simulations of the high-redshift structure formation in the $\Lambda$CDM cosmology (see Table~\ref{tab:summary_N-body_table}), using the {\tt CUBEP$^3$M} code. We briefly describe the simulation below, but refer readers to \cite{2012MNRAS.423.2222I} and \cite{2013MNRAS.436..540H} for details of the N-body simulations and the {\tt CUBEP$^3$M} code. We first run a large-box N-body simulation in a comoving volume of $114\,h^{-1} {\rm Mpc}$ on each side, using $3072^3$=29 billion particles. To find haloes, we use a spherical overdensity halo finder with overdensity parameter fixed to 178 of mean density, and require haloes to consist of at least 20 N-body particles, so we can resolve all ACHs ($\gtrsim 10^8 M_\odot$). 
We grid the density and velocity fields both for the total mass and for mass in the IGM (i.e.\ excluding N-body particles inside haloes) on a $256^3$ grid by smoothing N-body particle data with an adaptive kernel. Halo lists and density fields on the coarse-grained mesh are used by the RT code {\tt C$^2$-Ray} (see \S\ref{sec:RTsim}).

This mass resolution in the large box N-body simulation, however, is not enough to capture the density fluctuations at the Jeans mass scale before reionization ($\lesssim 10^5 M_\odot$). In order to calculate the subgrid clumping factor, we perform a small box N-body simulation in a comoving volume of $6.3\,h^{-1} {\rm Mpc}$ on each side, using $1728^3$=5.2 billion particles, which can resolve haloes at that Jeans scale. 
In principle, the subgrid clumping factor depends on a number of factors such as the redshift, the local overdensity, the grid resolution (or the scale over which the density field is smoothed), and the ionization level of the gas.\footnote{\citet*{2014MNRAS.443.2722J} employs numerical RT simulations of reionization with high resolution but in a rather small simulation volume, with focus on the redshift evolution of the {\it global mean} clumping factor and its dependence on these various factors. } 
In observing the dependence on the grid resolution, the density and velocity for the small box simulation is gridded on a $14^3$ coarse-grained mesh, the cell size of which is designed to match that of the large box N-body simulation ($\sim 0.45\,h^{-1} {\rm Mpc}$). We then compute the clumping factor on the grid using the small box high resolution N-body simulation data (see \S\ref{subsec:SPH-like-smoothing} below), and find empirical fitting formula for the correlation between the cellwise clumping factor and cellwise overdensity at each redshift (see \S\ref{subsec:fitting} below). This fitting formula from the small box simulation will be applied to the RT simulations in the large box because of matching coarse-grained cell size. Also, we assume in this paper that the subgrid clumping factor is independent of the cellwise ionized fraction. Similar approach was used elsewhere \citep{2012ApJ...756L..16A,2015MNRAS.450.1486A} to model the abundance of minihaloes, unresolved in a box of $114\,h^{-1} {\rm Mpc}$ on each side, by using the empirical relation between the local overdensity and the number of minihaloes found from smaller box ($6.3\,h^{-1} {\rm Mpc}$ and $20\,h^{-1} {\rm Mpc}$ on each side) simulations. 

In what follows, we use a $\Lambda$CDM cosmology with parameters $\Omega_\Lambda=0.73$, $\Omega_{\rm M}=0.27$, $\Omega_{\rm b}=0.044$, $H_0 = 100h$\,km\,s$^{-1}$\,Mpc$^{-1}$ with $h=0.7$, $\sigma_8=0.8$, $n_\mathrm{s}=0.96$, and $\eta_{\rm He} = 0.074$ (cosmic Helium abundance by number), consistent with the {\it WMAP} seven-year result \citep{2011ApJS..192...18K} and the {\it Planck} 1-year result \citep{2014A&amp;A...571A..16P}\footnote{We note that, while the values of cosmological parameters we use are not the most preferred, they are compatible with the {\it Planck} 2015 and 2018 results \citep{2016A&A...594A..13P,2018arXiv180706209P}.}.

\subsection{SPH-like smoothing with adaptive kernel} 
\label{subsec:SPH-like-smoothing}

In this section, we briefly describe how we smooth N-body particle data adaptively onto a grid, 
by a technique that resembles Smoothed-Particle-Hydrodynamics (SPH), with a focus on the calculation of the 
density field and the clumping factor. We refer readers to \cite{1996ApJS..103..269S} for a comprehensive 
discussion of SPH with an adaptive kernel. A brief description of this technique was also given in \cite{2012MNRAS.422..926M} which focused on velocity and velocity gradient fields. 

The reason that we adopt the SPH-like smoothing method, as opposed to a fixed smoothing kernel approach like the Cloud-in-Cell technique, is as follows. The N-body simulation is Lagrangian, by definition, so only an adaptive form of smoothing, which adjusts the length resolution locally to match the mean separation of particles, can retain the full dynamic range of density variations contained in the particle data.  Also, in regions that are underdense, assigning particle mass to a uniformly-spaced grid when individual cells are empty or contain too few particles gives incorrect results due to shot noise. The adaptive smoothing kernel approach avoids this.

Suppose we know the location ${\bf r}_i$ ($i=1,\ldots,N_{\rm particle}$) of $N_{\rm particle}$ N-body particles (each with the same mass $m_{\rm particle}$). To smooth the particle data, we define a particle's kernel $h_i$ to be the distance between the particle $i$ and its 32$^{\rm nd}$ nearest neighbor particle. We employ the triangular shaped cloud (TSC) kernel function but with {\it adaptive} kernel size $h$, $W({\bf r};h) = f_h(x) f_h(y) f_h(z)$, centered at the particle location. The 1D kernel function $f_h(x)$ is triangular-shaped with width $2h$, i.e.  
\begin{equation}
f_h(x) = \biggl\{   
  \begin{tabular}{lcl} 
    $(1-|x|/h)/h $ & , & $|x| \le h$ \\
    0 & , & otherwise 
  \end{tabular}  
\end{equation}
We smooth the particle data with the ``scatter'' approach \citep{1996ApJS..103..269S}, i.e.\ a field point ${\bf r}=(x,y,z)$ is influenced by a particle $i$ if this particle's {\it own} zone of influence covers this field point (e.g., for TSC, the condition is that $|x - x_i |\le h_i$, $|y - y_i |\le h_i$, and $|z - z_i |\le h_i$, simultaneously satisfied for the particle $i$'s own kernel $h_i$). 

The smoothed number density field of N-body particles is defined as  
\begin{equation}\label{eqn:n}
n_{\rm N,total}({\bf r}) = \sum_{{\rm all}\,i} W({\bf r}-{\bf r}_i;h_i)\,.
\end{equation} 
(Throughout this paper, the symbol $n$ always denotes the comoving density, i.e. $n_{\rm proper}/(1+z)^3$.)   
We assume that baryons follow the dark matter distribution, so each N-body particle contains a fixed number of hydrogen atoms, $m_{\rm particle} (\Omega_b/\Omega_m)/(\mu_{\rm H} m_{\rm H})$. Here  $\mu_{\rm H} = 1+4\eta_{\rm He}/(1-\eta_{\rm He})= 1.32$ is the mean molecular weight for gas in primordial composition, $m_{\rm H}$ is the mass of a hydrogen atom,
$\eta_{\rm He}$ is the fraction of the baryons in helium nuclei,
and the mass of an N-body particle $m_{\rm particle} = \bar{\rho}_{m,0} V_{\rm box}/N_{\rm particle}$, where $\bar{\rho}_{m,0}$ is the present mean matter density, $V_{\rm box}$ is the total comoving volume of the simulation box. 
The number density of hydrogen atoms $n_{\rm H,total}$ is related to that of N-body particles $n_{\rm N,total}$ by $n_{\rm H,total} = n_{\rm N,total} m_{\rm particle} (\Omega_b/\Omega_m)/(\mu_{\rm H} m_{\rm H})$. Henceforth the subscripts ``H'' and ``N'' refer to ``hydrogen atoms'' and ``N-body particles'', respectively, and the subscript ``total'' indicates that the density includes all (i.e.\ IGM and halo) N-body particles. 

In this paper we are interested in the subgrid clumping of the IGM and not in the contribution of recombinations inside haloes. This is because the number of ionizing photons per unit time released from haloes into the IGM (see \S\ref{subsec:RTsim}) already includes the effect of recombinations inside haloes whose internal structure is anyway not well resolved in the simulations. Therefore, it is necessary to restrict the clumping factor for use in reionization simulations to that of the IGM, by ``excising the halo regions'' from the general density field in determining the IGM density field, as first done by \cite{2001AIPC..586..219S,2003MNRAS.341...81I,2005ApJ...624..491I,2006MNRAS.369.1625I}.
We define an IGM density in a way that is similar to that for the total density, but 
this time the summation excludes N-body particles that reside inside haloes, 
\begin{equation}\label{eqn:nIGM}
n_{\rm N, IGM}({\bf r}) = \sum_{i\in {\rm IGM}} W({\bf r}-{\bf r}_i;h_i)\,,
\end{equation} 
In this case, we use the subscript ``IGM''. In the IGM, the number density of hydrogen is related to the N-body particle number density by $n_{\rm H,IGM} = n_{\rm N,IGM} m_{\rm particle} (\Omega_b/\Omega_m)/(\mu_{\rm H} m_{\rm H})$.

We smooth particle data onto a regular coarse-grained mesh and compute the cellwise number density, according to 
\begin{eqnarray}
\left< n_{\rm N,total} \right>_{\rm cell} & \equiv & \frac{1}{V_{\rm cell}} \int_{\rm cell} n_{\rm N,total}({\bf r}) d^3r \nonumber \\
& =& \frac{1}{V_{\rm cell}} \sum_{{\rm all}\,i} \int_{\rm cell} W({\bf r}-{\bf r}_i;h_i)d^3r\,, \\ \label{eqn:nint}
\left< n_{\rm N,IGM} \right>_{\rm cell} & \equiv & \frac{1}{V_{\rm cell}} \int_{\rm cell} n_{\rm N,IGM}({\bf r}) d^3r \nonumber \\
& = & \frac{1}{V_{\rm cell}} \sum_{i\in {\rm IGM}} \int_{\rm cell} W({\bf r}-{\bf r}_i;h_i)d^3r\,,\label{eqn:nNIGM}
\end{eqnarray}
where $V_{\rm cell}$ is the comoving volume of a cell. Throughout this paper $\left< A \right>_{\rm cell}$ denotes the smoothed value of field $A$ for a given cell, and $\int_{\rm cell} \ldots\,d^3r$ denotes an integration over the volume of a given cell. The integral $\int_{\rm cell} W({\bf r}-{\bf r}_i;h_i)d^3r$ can be evaluated analytically, and is only a function of $h_i$ and the relative location between the particle $i$ and the cell boundaries.

Smoothing the quadratic density ($n^2$) field onto a grid, however, is computationally cumbersome. Strictly speaking, it involves double summation over particles. To avoid this, we adopt a standard method in SPH to calculate any smoothed field $A({\bf r})$ from the contribution $A_i$ of particle $i$, by 
\begin{equation}
A({\bf r}) = \sum_i \frac{A_i}{n_{i,{\rm total}}} W({\bf r}-{\bf r}_i;h_i)\,,
\end{equation}
where $n_{i,{\rm total}} = n_{\rm N,total}({\bf r}_i)$ (the smoothed density field at particle $i$'s location). Hence, the smoothed $n^2$ field in the IGM can be approximately written as  
\begin{equation}\label{eqn:n2}
n^2_{\rm N,IGM}({\bf r}) = \sum_{i\in {\rm IGM}} n_{i,{\rm IGM}} W({\bf r}-{\bf r}_i;h_i)\,,
\end{equation}
where $n_{i,{\rm IGM}} = n_{\rm N,IGM}({\bf r}_i)$, i.e.\ we only sum over the contributions from the IGM particles. 
The virtue of equation~(\ref{eqn:n2}) is that smoothing the quadratic density field onto a grid is now a two-step process, each step involving a single summation over particles: the first step is to compute $n_{i,{\rm IGM}}$ as per equation~(\ref{eqn:nIGM}), and the second step is to smooth $n_{i,{\rm IGM}}$ onto a grid according to   
\begin{eqnarray}
\left< n^2_{\rm N,IGM} \right>_{\rm cell} & \equiv & \frac{1}{V_{\rm cell}} \int_{\rm cell} n^2_{\rm N,IGM}({\bf r}) d^3r \label{eqn:n2igmdef}\\
& =& \frac{1}{V_{\rm cell}} \sum_{i\in {\rm IGM}} n_{i,{\rm IGM}} \int_{\rm cell} W({\bf r}-{\bf r}_i;h_i)d^3r\,.\nonumber \\
\label{eqn:n2int}
\end{eqnarray}

Once the cellwise smoothed $n^2$ field is available, we can compute the cellwise clumping factor in the IGM which is defined as  
\begin{equation}
C_{\rm IGM,cell} \equiv \left<n_{\rm N,IGM}^2\right>_{\rm cell}/\left<n_{\rm N,IGM}\right>_{\rm cell}^2\,.
\end{equation}
(Again, the subscript ``cell'' indicates the cellwise value for a given cell.) 
The IGM clumping factor relates the local IGM quadratic density field and the local IGM density field. 

\begin{figure}
\begin{center}
  \includegraphics[height=0.35\textheight]{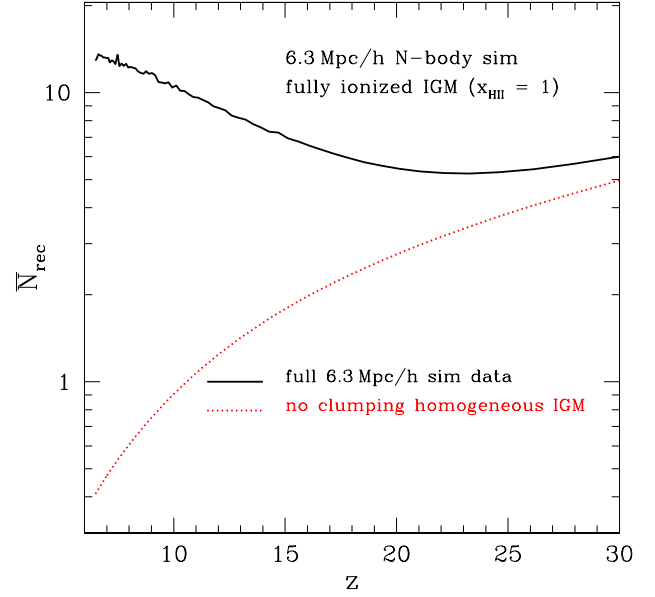} 
\end{center}
\caption{The spatially averaged number of recombinations per mean hydrogen atom per Hubble time at that redshift, $\bar{N}_{\rm rec}$, as a function of redshift. Here we assume the fully ionized IGM ($x_{\rm HII}=1$). We apply density fluctuation data from the $6.3\,h^{-1}\,{\rm Mpc}$ N-body simulation. The results are obtained  from two approaches: (1)
(solid/black) SPH-smoothing the full N-body particle data according to equation~(\ref{eqn:n2int}); (2) (dotted/red) simply assuming $\left<n_{\rm N,IGM}^2\right>_{\rm cell} = \bar{n}_{\rm N,IGM}^2$ for all cells. 
}
\label{fig:nbody_recomb}
\end{figure}

\begin{figure*}
\begin{center}
  \includegraphics[height=0.5\textheight]{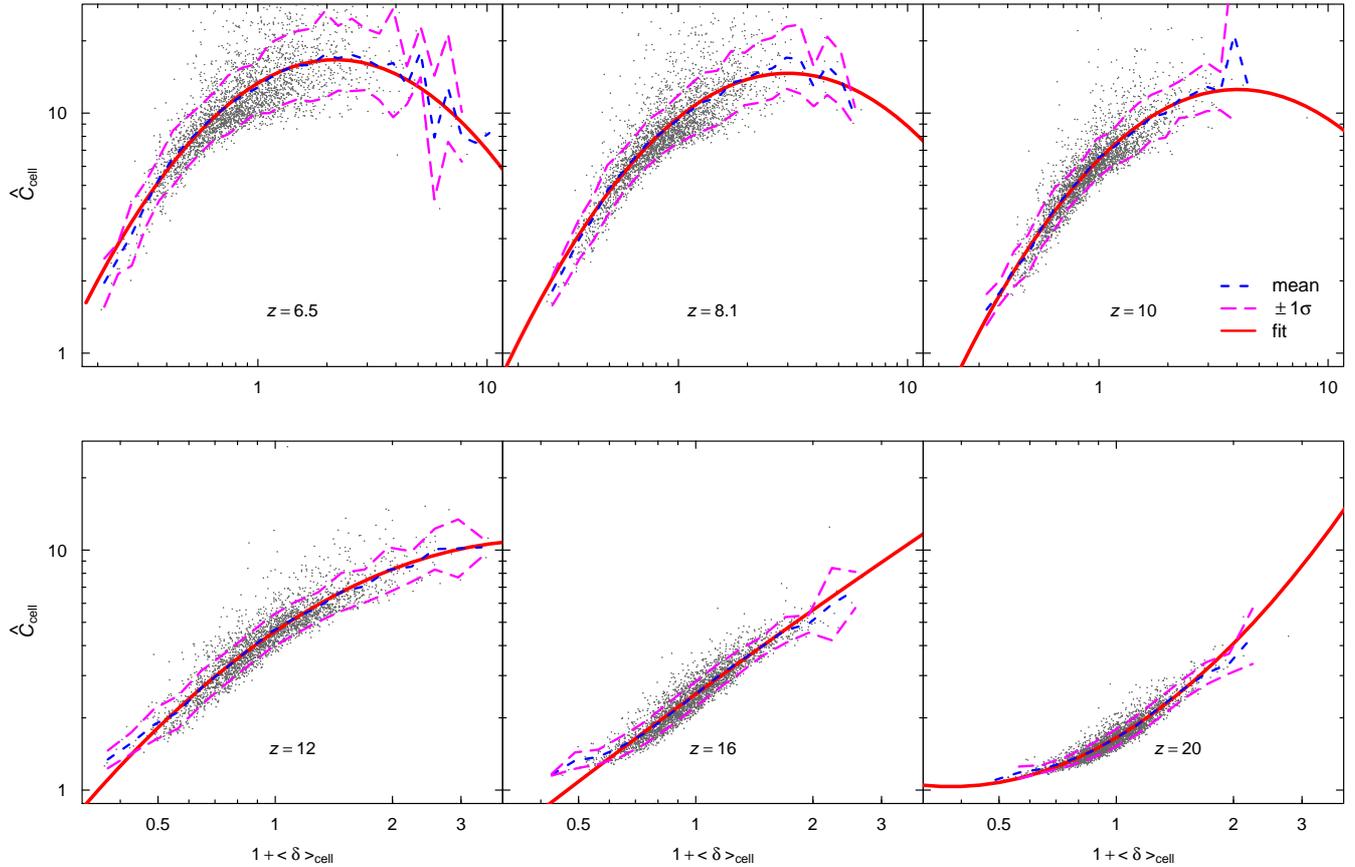} 
\end{center}
\caption{The correlation of local pseudo-clumping factor $\hat{C}_{\rm cell}$ and local density $1+\left<\delta\right>_{\rm cell}$, calculated directly from the $6.3\,h^{-1}$\,Mpc simulation with the coarse-grained cell size of $0.45\,h^{-1}$\,Mpc, at a number of representative redshift slices. Shown are the scattering of all cellwise data (dots), the bin-wise average of $\hat{C}_{\rm cell}$ for each overdensity bin (short dashed/blue), the minimum and maximum lines enveloping the $1\sigma$ variations (long dashed/magenta). We also show the fitting formula in equation~(\ref{eq:fit}) (solid/red) with the best-fit coefficients listed in Table~\ref{tab:fitting}. 
}
\label{fig:scatterplot}
\end{figure*}

\subsection{Recombination in a clumpy universe}

The cellwise hydrogen recombination rate (i.e.\ the number of recombinations for IGM hydrogen atoms in a cell per unit physical time) is 
\begin{equation}\label{eqn:recdef}
\left< \mathcal{R} \right>_{\rm cell} \equiv (1+z)^3 \int_{\rm cell} \alpha_{\rm B}({\bf r})\, n_e({\bf r}) n_{\rm HII}({\bf r}) d^3 r \,.
\end{equation}
We assume that helium is always singly ionized with the same fraction as hydrogen, i.e.\ $n_{\rm HeII}/n_{\rm He} = n_{\rm HII}/n_{\rm H}=x_{\rm HII}$, and is never doubly ionized, $n_{\rm HeIII}=0$ (cf.\ \citealt{2014MNRAS.443.2722J}). Therefore, $n_e = n_{\rm HII}+n_{\rm HeII}=n_{\rm HII}/(1-\eta_{\rm He})=1.08\,n_{\rm HII}$. This assumption is reasonable due to the soft stellar spectra considered here. We also assume that the gas temperature is homogeneous so that the Case B recombination coefficient for hydrogen is uniform, $\alpha_{\rm B} = 2.59\times 10^{-13} {\rm cm}^3\,{\rm s}^{-1}$ at $T=10^4\,{\rm K}$.  
In equation~(\ref{eqn:recdef}), the number densities in the RHS are comoving quantities, with the prefactor $(1+z)^3$ correcting for the conversion from the comoving to proper frame. 
If the coarse-grained mesh considered here is the RT mesh, then we make an approximation that the ionized fraction is uniform within a cell and equal to the cellwise value $\left< x_{\rm HII} \right>_{\rm cell}$, so that 
$\left< n_{\rm HII,IGM}^2\right>_{\rm cell} \approx \left< x_{\rm HII} \right>_{\rm cell}^2\,\left<n_{\rm H,IGM}^2\right>_{\rm cell}$, and therefore, 
\begin{equation}
\left< \mathcal{R} \right>_{\rm cell} = 1.08\,\alpha_{\rm B}\,(1+z)^3 V_{\rm cell} \left< x_{\rm HII} \right>_{\rm cell}^2\left<n_{\rm H,IGM}^2\right>_{\rm cell}.\label{eqn:recomb2} 
\end{equation}

We define the (dimensionless) number of recombinations per mean hydrogen atom per Hubble time, 
\begin{equation}
\left< N_{\rm rec} \right>_{\rm cell} \equiv \frac{\left< \mathcal{R} \right>_{\rm cell} \cdot \left(2/3\right) H^{-1}(z)}{ V_{\rm cell} \bar{\rho}_{m,0} (\Omega_b/\Omega_m) /(\mu_{\rm H} m_{\rm H})}. \label{eqn:Nrec}
\end{equation}

To illustrate the impact of subgrid clumping on recombination, we employ the $6.3~h^{-1}{\rm Mpc}$ N-body simulation data to compute the cellwise $\left< n^2_{\rm N,IGM} \right>_{\rm cell}$ and $\left< N_{\rm rec} \right>_{\rm cell}$, and show in Figure~\ref{fig:nbody_recomb} the mean\footnote{Throughout this paper, unless otherwise noted, the bar as in  $\bar{A}$ indicates the spatially averaged value over the entire simulation volume, i.e.\ $\bar{A} \equiv \sum_{\rm cell} \left<A\right>_{\rm cell} /N_{\rm mesh}$ where $N_{\rm mesh}$ is the total number of coarse-grained mesh cells.} recombination $\bar{N}_{\rm rec}$, i.e.\ global average of $\left< N_{\rm rec} \right>_{\rm cell}$. We compare the results from two methods: (1) as the benchmark, we follow the SPH-smoothing method in equation~(\ref{eqn:n2int}) to compute $\left< n^2_{\rm N,IGM} \right>_{\rm cell}$ and then average it over the simulation box; (2) as an overly simplistic prescription, we set $\left<n_{\rm N,IGM}^2\right>_{\rm cell} = \bar{n}_{\rm N,IGM}^2$ where $\bar{n}_{\rm N,IGM}$ is the global mean N-body particle number density of the IGM. For illustrative purpose, we assume in this subsection the fully ionized scenario ($x_{\rm HII}=1$) for all positions and time. As such, the first method is equivalent to directly SPH-smoothing all IGM N-body particle data over the simulation volume. The result, therefore, is independent of the coarse-grained mesh resolution we used here, but it does depend on the N-body particle mass resolution. Since our small box simulation resolves haloes down to the Jeans mass before reionization, the first method represents the maximum inhomogeneous clumping. In contrast, the second method gives the least clumping, because the density field in the IGM is assumed to be homogeneous on {\it all} scales. 
Figure~\ref{fig:nbody_recomb} shows that the second, overly simplistic, prescription always underestimates the recombination rate by a factor ranging from $0.8$ (at high redshift $z\sim 30$) to $25$ (at low redshift $z\sim 6.5$). This comparison demonstrates the importance of modelling clumping carefully.

\subsection{Correlation between local clumping and local overdensity}
\label{subsec:fitting}

The subgrid clumping is determined by gravitational dynamics, because gravitational instability results in the density fluctuations of {\it all} (i.e.\ IGM and halo) matter. The local IGM clumpiness $\left<n_{\rm N,IGM}^2\right>_{\rm cell}$, therefore, should be correlated with local {\it total} density $\left<n_{\rm N,total}\right>_{\rm cell}$. For this reason, we define a cellwise ``pseudo-clumping factor''
\begin{equation}
\hat{C}_{\rm cell} \equiv \left<n_{\rm N,IGM}^2\right>_{\rm cell}/\left<n_{\rm N,total}\right>_{\rm cell}^2\,.\label{eqn:Ccelldef}
\end{equation}
(Note that the denominator is the cellwise {\it total} matter density squared.)  The pseudo-clumping factor relates the local IGM quadratic density field and the local {\it total} density field, and, hence, should be correlated with local total density. 

The subgrid pseudo-clumping factor is related to the IGM clumping factor by 
\begin{equation}
\hat{C}_{\rm cell} = C_{\rm IGM,cell} \, f_{\rm IGM,cell}^2
\end{equation}
where $f_{\rm IGM,cell}\equiv \left<n_{\rm N,IGM}\right>_{\rm cell}/\left<n_{\rm N,total}\right>_{\rm cell}=1-f_{\rm coll,cell}$ and $f_{\rm coll,cell}$ is the fraction of the mass in a cell which is collapsed into haloes.
On {\it average}, $f_{\rm coll}$ is about a few per cent at the high redshifts of 
the EOR, but in few cells that contain large haloes, the collapsed fraction may be much higher. 
(Since we assume that baryons trace the dark matter distribution, there is no distinction herein between the collapsed fraction of baryons and that of dark matter.)

We smooth the $6.3\,h^{-1}$\,Mpc N-body particle data onto the $14^3$ coarse-grained mesh using the SPH-like approach, and compute the cellwise pseudo-clumping factor $\hat{C}_{\rm cell}$ and cellwise density $1+\left<\delta\right>_{\rm cell} \equiv \left<n_{\rm N,total}\right>_{\rm cell}/\bar{n}_{\rm N,total}$ for each cell, where $\bar{n}_{\rm N,total}$ is the mean N-body particle number density (global average of $\left<n_{\rm N,total}\right>_{\rm cell}$). In Figure~\ref{fig:scatterplot}, we plot the scattered distribution of $\hat{C}_{\rm cell}$ vs $1+\left<\delta\right>_{\rm cell}$, and find a strong, redshift-dependent, correlation between them. 

From the curve of the bin-wise mean $\hat{C}_{\rm cell}$ (averaged over data in the same overdensity bin) vs overdensity, we find that, at high redshift ($z\gtrsim 10$), $\hat{C}_{\rm cell}$ increases monotonically as local density gets larger, because gravity pulls matter toward the center of overdense regions, enhancing the overdensity and the gradient of matter density (therefore larger clumping) at the same time. At low redshift ($6\lesssim z \lesssim 10$), however, we find a concave correlation curve, and the peak shifts slightly to the smaller overdensity at lower redshift, e.g.\ the clumping peak appears at $1+\left<\delta\right>_{\rm cell}\approx 3$ (2) when $z=10$ (6.5). The clumping peak might be due to the fact that higher density regions are likely to form more haloes, and particles inside haloes are excised from our IGM clumping calculation. This also means that the clumping peak might depend on the definition of haloes in N-body simulations, e.g., if the minimum number of N-body particles that is required to resolve haloes is increased, then small overdense regions that would otherwise be identified as haloes could increase the IGM clumping factor, and shift the clumping peak to the larger overdensity. Further investigation is necessary to fully understand the clumping peak. 

To quantify this correlation, we consider a polynomial fit
\begin{equation}\label{eq:fit}
y = a_0 + a_1\,x + a_2\,x^2 \,,
\end{equation}
where $y=\log_{10} \hat{C}_{\rm cell}$ and $x = \log_{10}(1+\left<\delta\right>_{\rm cell})^2$, to fit to the scattered data at each redshift. Not only does the quadratic term make a second-order correction to the linear term, but it can characterize the concave nature in the correlation at low redshift. The coefficients $a_0$, $a_1$ and $a_2$ are redshift-dependent. Their best-fit values, using the least square method, are tabulated in Table~\ref{tab:fitting}. We find that the fitting curve tracks the bin-wise mean pseudo-clumping factor most of the time, but smoothes over numerical fluctuations when they are caused by the rareness of events. As shown in Fig.~\ref{fig:scatterplot}, this formula works well at both high (convex curve) and low (concave curve) redshifts. 

We can apply this fitting formula for calculating local pseudo-clumping factor from local overdensity in large box simulations with insufficient mass resolution, if both large and small box simulations have matching cell size in the coarse-grained mesh. For example, coefficients in Table~\ref{tab:fitting} are best-fit for comoving cell size of $0.45\,h^{-1} {\rm Mpc}$. The subgrid clumping factor calculated using this fitting formula is inhomogeneous because the overdensity of the coarse-grained cells varies from cell to cell. 
However, we note that this method does not account for the stochasticity of clumping for a given overdensity. 
To see this, we mark the $1\sigma$ variations enveloping the bin-wise mean $\hat{C}_{\rm cell}$ in Figure~\ref{fig:scatterplot}. Obviously, the stochasticity becomes larger as more nonlinear 
structures form at low redshift, and the variations can be quite significant. 
It is technically unfeasible to run a cosmological ($\gtrsim 100\,{\rm Mpc}$) 
N-body simulation with minimum halo mass resolved down to the Jeans mass scale before reionization 
($\lesssim 10^5\,M_\odot$). Therefore, it is beyond the scope of this paper to investigate how 
the stochasticity of clumping affects the recombination during the reionization. 
However, in \S\ref{subsec:global_clumping} we approach this problem partially by making a consistency check. 

\begin{figure}
\begin{center}
  \includegraphics[height=0.35\textheight]{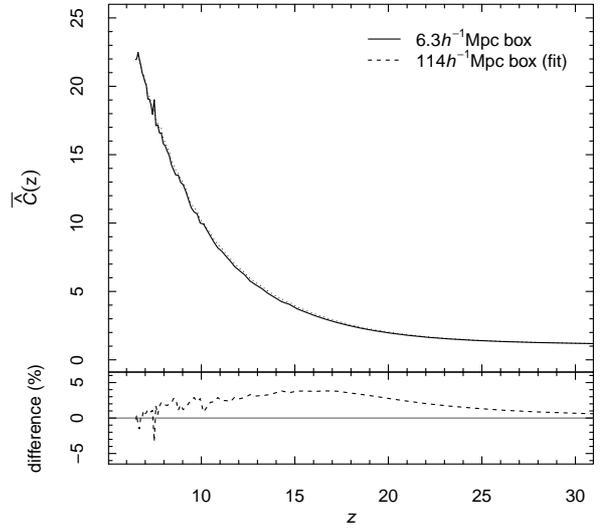} 
\end{center}
\caption{Top: The global mean pseudo-clumping factor $\overline{\hat{C}}(z)$, calculated (i) (solid) directly using the $6.3\,h^{-1}$\,Mpc simulation particle data, and (ii) (dashed) by using the $\hat{C}_{\rm cell}$-$\left<\delta\right>_{\rm cell}$ fitting formula with the local overdensity from the $114\,h^{-1}$\,Mpc simulation coarse-grained mesh data. Bottom: the relative error of the latter with respect to the former in per cent.  
}
\label{fig:meanC}
\end{figure}

\subsection{Global mean pseudo-clumping factor}
\label{subsec:global_clumping}

We define the global mean pseudo-clumping factor $\overline{\hat{C}}$ as 
\begin{equation}
\overline{\hat{C}}      \equiv \overline{n^2}_{\rm N,IGM}/\bar{n}_{\rm N,total}^2\,,\label{eqn:barC}
\end{equation}
where $\overline{n^2}_{\rm N,IGM}$ is the global average of the IGM quadratic density field $\left<n^2_{\rm N,IGM}\right>_{\rm cell}$
(note that $\overline{\hat{C}}$ is not the volume-weighted global average of $\hat{C}_{\rm cell}$). 

For the small box ($6.3~h^{-1}{\rm Mpc}$) N-body simulation, we smooth the full N-body particle data with SPH-like adaptive kernel to compute the cellwise $\left< n^2_{\rm N,IGM} \right>_{\rm cell}$ using equation~(\ref{eqn:n2int}), average it over the whole simulation volume to get $\overline{n^2}_{\rm N,IGM}$, and compute $\overline{\hat{C}}$. Note that this approach is equivalent to directly smoothing over all N-body particle data in the simulation volume, so the values of $\overline{\hat{C}}(z)$ depend only on the N-body fine cell resolution (e.g. $2^3\,N_{\rm particle} = 3456^3$ fine cells for the $6.3~h^{-1}{\rm Mpc}$ N-body simulation), not on the coarse-grained mesh resolution. We tabulate the results in Table~\ref{tab:meanC}. 

On the other hand, for the large box ($114~h^{-1}{\rm Mpc}$) simulation where the gas clumping is poorly resolved, we use the small box ($6.3~h^{-1}{\rm Mpc}$) result as the subgrid recipe as follows. We obtain the local clumping factor $\hat{C}_{\rm cell}$, for each cell of the large box coarse-grained mesh, from the coarse-grained mesh data of overdensity using the fitting formula (equation~\ref{eq:fit} and Table~\ref{tab:fitting}), compute the cellwise $\left< n^2_{\rm N,IGM} \right>_{\rm cell}= \hat{C}_{\rm cell} \left<n_{\rm N,total}\right>_{\rm cell}^2$, and then average it to compute $\overline{n^2}_{\rm N,IGM}$ and  $\overline{\hat{C}}$. 

While the first approach accounts for the {\it full} stochasticity of inhomogeneous clumping in the {\it small} simulation volume, the second approach only takes into account the inhomogeneity of clumping due to density variations across coarse-grained mesh cells, but neglects the stochasticity of clumping for a given local density. As a consistency check, we compare both results of $\overline{\hat{C}}$ in Figure~\ref{fig:meanC}, and find a good agreement ($< 4\%$ relative error) between them. This implies that the stochasticity of clumping for a given local density might affect the {\it mean} clumping only insignificantly. However, this stochasticity should play an important role on the {\it fluctuations} of local recombination and reionization at small scales (e.g.\ on the H~II region size distribution). We leave it to future work to investigate this effect more carefully.

\begin{table*}
\begin{center}
\begin{minipage}{1\linewidth}
\caption{Reionization simulation parameters and global reionization history results. All RT simulations in this paper are in a comoving volume of $114\,h^{-1}\,{\rm Mpc}$ on each side, which is coarse-grained onto a $256^3$ mesh. The minimum mass source is haloes with $10^8 M_\odot$, but haloes with $10^8 M_\odot \le M \le 10^9 M_\odot$ are vulnerable to suppression if they formed inside an already ionized region. $z_{x\%}$ refer to the redshift when the ionized fraction reaches $x\%$. $z_{\rm ov}$ is the overlap redshift, which we define by $\bar{x}_{\rm HII,m} = 0.99$.}
\label{tab:summary_RT_table}
\begin{tabular}{@{}llllllllllll}
\hline
label \footnote{``Clumping'' refers to {\it subgrid} clumping throughout this paper.}  & acronym &  $f_{\gamma}$ & $f_{\gamma}$ & subgrid &  $\tau_{\rm es}$ & $z_{10\%}$ & $z_{20\%}$ &$z_{50\%}$& $z_{75\%}$ & $z_{90\%}$&$z_{\rm ov}$ \\
       &  &  HMACH & LMACH & clumping factor \footnote{In all cases, the density that is multiplied by the clumping factor to compute the recombination rate is the inhomogeneous coarse-grained cellwise density in the RT mesh.}   &  &   &&& \\
\hline
``no clumping high efficiency'' & NCHE & 10 & 150 & $C_{\rm IGM,cell} = 1$ & 0.082&13.3& 11.5 & 9.5& 8.9 & 8.6  &8.4 \\
``no clumping low efficiency''  & NCLE & 2  & 10  & $C_{\rm IGM,cell} = 1$ & 0.058&9.9 & 8.9 & 7.6& 7.1 & 6.9  &6.7 \\
``inhomogeneous clumping''      & IC   & 10 & 150 & $\hat{C}_{\rm cell}$ via fitting \footnote{The cellwise subgrid pseudo-clumping factor is interpolated from the local overdensity using equation~(\ref{eq:fit}) with best-fit coefficients in Table~\ref{tab:fitting}.} 
& 0.069 & 12.6 & 10.3  & 8.2 & 7.7&  7.4 & 7.3 \\
``biased homogeneous clumping'' & BHC  & 10 & 150 & $C_{\rm IGM,cell} = \overline{\hat{C}}$ \footnote{The IGM subgrid clumping factor is everywhere equal to the global mean pseudo-clumping factor tabulated in Table~\ref{tab:meanC}.} & 0.067 & 12.6 & 9.9 & 8.0 & 7.3 &  7.1 & 6.9 \\
\hline
\end{tabular}
\end{minipage}
\end{center}
\end{table*}

\section{Applying clumping factor to large-scale reionization simulations} 
\label{sec:RTsim}

\subsection{Reionization simulations}
\label{subsec:RTsim}

To simulate cosmic reionization with statistically meaningful results, we employ the large box (comoving $114\,h^{-1} {\rm Mpc}$ on each side) N-body simulation data, which provides the spatial distribution of cosmological structures and their evolution in time. Assuming that the gas traces exactly the distribution of cold dark matter (CDM) particles, the halo lists and the IGM density fields on the coarse-grained mesh are employed as input to a full 3D RT simulation of cosmic reionization, using the code {\tt C$^2$-Ray} \citep{2006NewA...11..374M}, as described in \cite{2012MNRAS.423.2222I}. 
For the source of ionizing photons, we consider ACHs only. These haloes have their masses above the mass range of minihaloes ($> 10^8\,M_\odot$), corresponding to the virial temperature of gas above $10^4\,{\rm K}$. 
ACHs are assigned ionizing luminosities in proportion to their mass. The ionizing photon production efficiency $f_\gamma$, which is defined as the number of ionizing photons released by a halo per baryon per $\Delta t = 11.53$\,Myr, also depends on whether the halo mass is above or below $10^9\,M_\odot$. We call haloes with $M \ge 10^9\,M_\odot$ the high-mass ACHs (``HMACHs''), and those with $10^8\,M_\odot \le M \le 10^9\,M_\odot$ the low-mass ACHs (``LMACHs''). To incorporate feedback from reionization, LMACHs located in ionized regions (for ionized fraction higher than 10\%) do not produce any photons, due to Jeans-mass filtering \citep{2007MNRAS.376..534I}, which corresponds to the aggressive suppression case
in \citet{2016MNRAS.456.3011D}. We refer readers to \cite{2012MNRAS.423.2222I,2016MNRAS.456.3011D} for more details of the simulation code and the feedback processes. 

We have performed a series of RT simulations with varying assumptions on the source efficiencies and subgrid clumping factor, as summarized in Table~\ref{tab:summary_RT_table}. Specifically, we consider three prescriptions to approximate the quadratic density field $\left<n_{\rm N,IGM}^2\right>_{\rm cell}$ when calculating the recombination rate using equation~(\ref{eqn:recomb2}), as follows. (``Clumping'' in this paper always refers to subgrid clumping; sometimes we drop the word ``subgrid'' for brevity.)

(i) {\it no subgrid clumping} (``NC''): 
the simplest approximation is to set $C_{\rm IGM,cell} = 1 $ for all RT cells, as if the IGM density is uniform inside each RT cell. However, we still account for the fact that the recombination rate varies from cell to cell, since the resolved, coarse-grained IGM density field of the reionization simulation fluctuates {\it amongst the cells} (i.e. the mean density inside each cell varies from cell to cell), by using
\begin{equation}
\label{eqn:no-clumping}
\left<n_{\rm N,IGM}^2\right>_{\rm cell}  =   \left<n_{\rm N,IGM}\right>_{\rm cell}^2\,.
\end{equation}
For this NC prescription, we consider two scenarios with high and low source efficiencies, ``NCHE'' and ``NCLE'', respectively, which represent the early and late completion of reionization.

(ii) {\it inhomogeneous subgrid clumping} (``IC''): 
The recombination rate depends on the density fluctuations at two distinctive levels --- the variation from RT cell
to RT cell of the mean density inside each RT cell, and the density fluctuations within each RT cell on scales
which are below the RT grid-spacing (and, hence, are unresolved by the RT grid). 
While the former is accounted for by using the {\it cellwise} density as in equation~(\ref{eqn:no-clumping}), 
the latter is encoded in the subgrid clumping factor. 
In what we call the {\it inhomogeneous subgrid clumping} model, we compute the recombination rate by using 
\begin{equation}
\left<n_{\rm N,IGM}^2\right>_{\rm cell}  =   \hat{C}_{\rm cell} \left<n_{\rm N,total}\right>_{\rm cell}^2 \,, 
\end{equation}
where the cellwise pseudo-clumping factor $\hat{C}_{\rm cell}$ is obtained from the cellwise {\it total} overdensity for that cell in the RT mesh, exploiting the fitting formula in \S\ref{subsec:fitting}. 

The subgrid clumping factor in the IC model has two important features. 
First, it increases in time on average. Specifically, as Figure~\ref{fig:meanC} shows, 
$\overline{\hat{C}} \gtrsim 1$ at the early time $z\gtrsim 20$, 
but is rapidly boosted to $\gtrsim 10$ later at $z\lesssim 10$, 
and is as large as $\sim 22$ at the end of reionization $z \sim 6$. 
Secondly, the subgrid clumping factor at a given redshift differs from one RT cell to another. 
Except in highly overdense regions at low redshift, the subgrid clumping factor 
in an overdense (underdense) cell is generally larger (smaller) than the average. 
Because density varies amongst RT cells at a given redshift, 
so does the subgrid clumping factor, as reflected in the clumping-overdensity correlation. 

(iii) {\it biased homogeneous subgrid clumping} (``BHC''): 
As a more approximate treatment of clumping that neglects the {\it inhomogeneity}
of the subgrid clumping factor,
we improve upon the ``no subgrid clumping'' model, by introducing a global, 
homogeneous, redshift-dependent subgrid clumping factor
with which to multiply the square of the spatially-varying cellwise densities
of the reionization simulation, which we refer to as the 
{\it biased homogeneous subgrid clumping} model. In this case, 
we assume $C_{\rm IGM,cell}= \overline{\hat{C}}$, i.e.\ the cellwise IGM subgrid
clumping factor is equal to the same precomputed function of redshift everywhere in the RT mesh. 
The inhomogeneous recombination rate is then calculated using 
\begin{equation}
\left<n_{\rm N,IGM}^2\right>_{\rm cell}  =  \overline{\hat{C}} \left<n_{\rm N,IGM}\right>_{\rm cell}^2\,.
\label{eqn:BHC_recomb}
\end{equation}
Previous RT simulations (e.g.\ \citealt{2006MNRAS.372..679M,2007MNRAS.376..534I,2007ApJ...657...15K}) have employed a similar approximation (with different formulae for the mean clumping factor because of different cosmologies
and different N-body resolutions). Here we use the global mean pseudo-clumping factor from our small-box,
high-resolution N-body simulation (tabulated in Table~\ref{tab:meanC}).  We refer to this model as ``biased''
homogeneous subgrid clumping to distinguish it from the ``unbiased'' homogeneous case in which the cellwise
IGM density in Eq.~(\ref{eqn:BHC_recomb}) is replaced by the globally-averaged (cosmic-mean) IGM density at that
redshift (i.e., the same for all cells), a model we shall discuss again in \S\ref{subsec:UHC}.

Note that the source efficiencies for the IC and BHC model are assumed to be the same as that for the NCHE model, to isolate the effect of clumping factor from that of source efficiency. 
Also, the inclusion of the NCLE model is to test possible degeneracy between low source efficiency and subgrid clumping, both of which may delay the completion of reionization. In fact, as shown in Table \ref{tab:summary_RT_table}, we choose the source efficiencies for the NCLE model in such a way that the NCLE model ``overlaps'' (which we define by $\bar{x}_{\rm HII,m} = 0.99$) roughly at the same time as the BHC model.

\subsection{Observational signatures}

We briefly describe our methodology for predicting the observational signatures from the reionization simulation data. These include: 

(i) The mass-weighted global mean ionized fraction $\bar{x}_{\rm HII,m}(z)$ as a function of redshift. 

(ii) {\it Thomson optical depth}: 
\begin{equation}
\tau_{\rm es}(z) = c\,\sigma_T \int_{z}^{0} dz' (1+z')^3 n_e (z') (dt'/dz')
\end{equation}
is the mean optical depth along a line-of-sight (LOS) between an observer at $z = 0$ and a redshift $z$ due to Thomson scattering by free electrons in the post-recombination universe, where $\sigma_T = 6.65\times 10^{-25}\,{\rm cm}^2$ is the Thomson scattering cross section, 
$n_e(z) = n_{{\rm H}}(z) \bar{x}_{\rm HII,m}(z) \chi_{\rm eff}$ is the mean comoving number density of free electrons at redshift $z$, and $n_{{\rm H}}$ is the mean comoving number density of hydrogen. In this paper, we assume that helium is singly ionized to He~II at the same rate that hydrogen is ionized to H~II, i.e.\ $n_{\rm HeII}/n_{\rm He} = n_{\rm HII}/n_{\rm H}=x_{\rm HII}$, and helium is never doubly ionized to He~III directly. After $z\le 3$, He~II is assumed to be fully ionized to He~III. Therefore, $\chi_{\rm eff} = 1+p \,\eta_{\rm He}/(1-\eta_{\rm He})$, where $p=2(1)$ for $z\le 3$ ($z>3$), so $\chi_{\rm eff} = 1.16 (1.08) $ for $z\le 3$ ($z>3$).

(iii) {\it The 21~cm brightness temperature: mean, root-mean-square (RMS) fluctuation, and power spectrum.} 
In the optically thin approximation, the 21~cm differential brightness temperature (i.e.\ the 21~cm brightness temperature relative to the CMB temperature) at the observed frequency $\nu_{\rm obs}$, which corresponds to redshift $z$ of the emitter and its real location ${\bf r}$, is 
\begin{equation}
\delta T_b(z,{\bf r}) = \widehat{\delta T}_b (z) \, \frac{1+\delta_{\rho_{\rm HI}}(z,{\bf r})}{\left|1+\delta_{\partial_r v}(z,{\bf r})\right|}\,\left[1 -\frac{ T_{\rm CMB}(z)}{T_s(z,{\bf r})}\right]\,. 
\label{eqn:dtb1}
\end{equation}
In this paper, we focus on the limit where the spin temperature $T_s \gg T_{\rm CMB}$, which is equivalent to assuming efficient heating by X-ray sources before substantial reionization ($>10\%$) is achieved.
As such, the dependence of 21~cm brightness temperature on spin temperature can be neglected. In this limit, the mean of the 21~cm differential brightness temperature is equal to the normalization $\widehat{\delta T}_b$, 
\begin{equation}
\widehat{\delta T}_b (z) = (23.88\,{\rm mK}) \left(\frac{\Omega_{\rm b}h^2}{0.02}\right)\sqrt{\frac{0.15}{\Omega_{\rm M} h^2}\frac{1+z}{10}}\bar{x}_{\rm HI}(z)\,.
\end{equation}

The fluctuations of the 21~cm brightness temperature depends on the {\it neutral} hydrogen density fluctuations in real space $\delta_{\rho_{\rm HI}}({\bf r}) = n_{\rm HI}({\bf r})/\bar{n}_{\rm HI}-1$ , and the velocity gradient   $\delta_{\partial_r v}({\bf r}) \equiv \frac{1+z}{H(z)} \frac{dv_\parallel}{dr_\parallel }({\bf r})$ [more precisely, the gradient of the proper radial peculiar velocity along the LOS, normalized by the conformal Hubble constant $H/(1+z)$]. 

Since the observed wavelength is redshifted both cosmologically and by the Doppler shift associated with peculiar velocity ${\bf v}({\bf r})$, the observed 21~cm brightness temperature is in redshift space, in which the coordinate ${\bf s}$ is shifted from the {\it real} comoving coordinate $r$ along the LOS due to the Doppler shift, ${\bf s} = {\bf r} + \frac{(1+z)}{H(z)} v_\parallel(t,{\bf r}) \, {\bf n}$, where ${\bf n}$ is the LOS unit vector. To compute the 3D power spectrum of the 21~cm brightness temperature fluctuations (hereafter, ``21~cm power spectrum'') in redshift space from our RT simulation data, we employ the {\it Mesh-to-Mesh Real-to-Redshift-Space-Mapping (MM-RRM)} scheme (see \citealt{2012MNRAS.422..926M}). 

The variance is the integral of the 21~cm power spectrum. (For the detectability of 21~cm variance, see, e.g.\ \citealt{2014MNRAS.443.1113P}).

(iv) {\it The kinetic Sunyaev-Zel'dovich (kSZ) effect}:
During the Thomson scattering, the bulk peculiar velocity of free electrons induces Doppler shifts in the energy of CMB photons. Inhomogeneity in the density and velocity of free electrons will induce temperature fluctuations in the CMB given by 
\begin{equation}
\Delta T({\bf n})/T = - \int d\tau'_{\rm es} e^{-\tau'_{\rm es}}\, {\bf n} \cdot \mathbf{v}/c \,,
\end{equation}
where ${\bf n}$ is the LOS unit vector and $\mathbf{v}$ is the peculiar velocity field. We shall report the magnitude of the fluctuations in terms of the angular power spectrum, $D_l \equiv l(l+1)C_l/(2 \pi)$, where $C_l \equiv \frac{1}{2l+1}\Sigma_m |a_{lm}|^2$. Here, $a_{lm}\equiv \int d^2\mathbf{n} \Delta T (\mathbf{n})Y^*_{lm}(\mathbf{n})$ is the coefficient of spherical-harmonics mode, $Y_{lm}$, of $\Delta T$. Detailed description of how we compute $D_l$ of the kSZ signal from the reionization models of this work is in Section 2 of \cite{2013ApJ...769...93P}.

(v) {\it Post-reionization Lyman-limit opacity}: we can estimate the Lyman-limit optical depth over the LOS across the entire box along the $x$, $y$, and $z$-axis, respectively, by $\tau_{\rm LL} = (1+z)^2\,\int  \sigma_H\, x_{\rm HI}\, n_{\rm H}\, dr$, where $\sigma_H = 6.3\times 10^{-18}\,{\rm cm}^2$ is the H~I cross section at the Lyman limit. Here we neglect the redshifting of Lyman-limit photons during the travel. The exact expression when redshifting is accounted for depends on the emission redshift of the source in question (see \citealt{1980ApJ...241....1S}). 
The mean free path of the IGM to H~I ionizing radiation is $\lambda^{912}_{\rm mfp} = L_{\rm box,proper}/\tau_{\rm LL}$. 
We will discuss how the late-time Lyman-limit opacity is affected by reionization in Section \ref{subsubsec:LLO}.

\begin{figure}
\begin{center}
  \includegraphics[height=0.3\textheight]{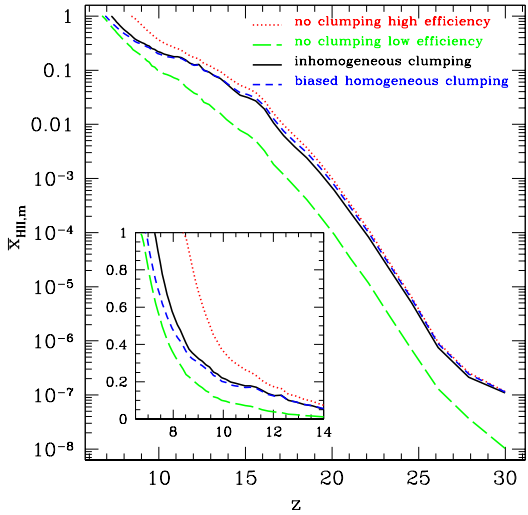} 
\end{center}
\caption{The history of reionization: the mass-weighted mean ionized fraction $\bar{x}_{\rm HII,m}$ (in the logarithmic scale) vs $z$ . Inset: history of $\bar{x}_{\rm HII,m}$ (in the linear scale) at $6.5<z\le 14$, corresponding to $1 \ge \bar{x}_{\rm HII,m} \ge 0.01$.  
}
\label{fig:histories}
\end{figure}

\begin{figure}
\begin{center}
  \includegraphics[height=0.3\textheight]{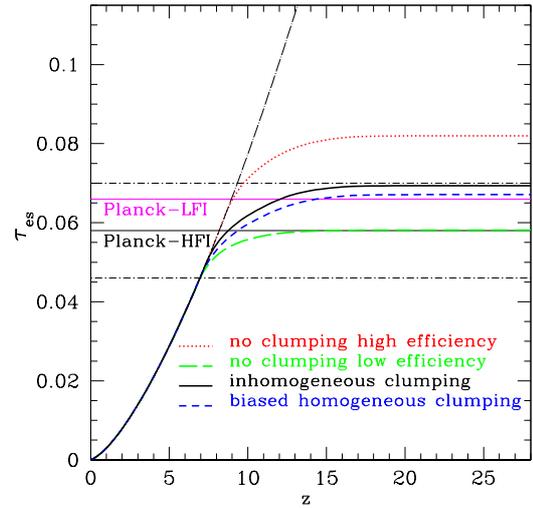} 
\end{center}
\caption{Thomson optical depth $\tau_{\rm es}(z)$, integrated from redshift $0$ to redshift $z$, for various reionization models. Horizontal lines are the best fit value (thin black solid) and 68\% confidence limits (thin black dot-short dashed), $\tau_{\rm es} = 0.058\pm 0.012$, constrained by the {\it Planck}-HFI 2016 CMB temperature and polarization data (``lollipop+Planck TT'') \citep{2016A&A...596A.108P}, and the best fit value (thin magenta solid), $\tau_{\rm es} = 0.066$, from the {\it Planck}-LFI 2015 result (``Planck TT+lowP+lensing+BAO'') \citep{2016A&A...594A..13P}. The dot-long dashed line shows the value of $\tau_{\rm es}$ for a fully-ionized universe. 
}
\label{fig:tau}
\end{figure}

\section{Results and Discussions}
\label{sec:RTresult}

In this section, we first investigate the effect of inhomogeneous clumping on cosmic reionization, by comparing the result of the IC model with that of the NCHE and NCLE models. 
To understand the effect of inhomogeneity in subgrid clumping, we also compare the results of the BHC and IC model.

\subsection{Reionization history}
\label{subsec:reion-history}

We plot the reionization history in Figure~\ref{fig:histories} and the corresponding $\tau_{\rm es}$ in Figure~\ref{fig:tau}, and list the redshifts for some key stages of reionization in Table~\ref{tab:summary_RT_table}. The NCHE model both starts and completes reionization at early times, e.g.\ $\bar{x}_{\rm HII,m}=0.10$ at $z=13.3$, and $z_{\rm ov}=8.4$. Here we define $z_{\rm ov}$ as the redshift at the stage $\bar{x}_{\rm HII,m}=0.99$. However, this early completion of reionization may be disfavored by the observation of high-redshift quasar absorption spectra (see, e.g.\ \citealt*{2006ARA&A..44..415F}; \citealt{2010ApJ...723..869O,2011ApJ...734..119K,2011Natur.474..616M,2019MNRAS.484.5094G,2019MNRAS.484.5142P}). The NCLE model, on the other hand, both starts and completes reionization at late times, e.g.\ $\bar{x}_{\rm HII,m}=0.10$ at $z=9.9$, and $z_{\rm ov}=6.7$. This yields a small $\tau_{\rm es}=0.058$.  

In comparison, both IC and BHC models yield more extended reionization histories than the NCHE and NCLE model,
\footnote{
There can be other scenarios which make reionization extended. 
For example, if a model with no subgrid clumping assumes high LMACH efficiency and low HMACH efficiency, 
then reionization can start at early times (due to high LMACH efficiency) but finish 
at late times (due to low HMACH efficiency, after self-regulation suppresses the LMACHS and
reionization is left to the HMACHs). 
The observational signatures of this model should be similar to those of the 
NCHE model at the early stages of reionization, and to those of the NCLE model at the late stages. 
Reionization can also be extended when there exist extra sources, 
such as stars inside minihaloes (``MHs'') \citep{2012ApJ...756L..16A}. With MH sources, 
reionization starts earlier but MH sources are suppressed by the rising UV 
background of ${\rm H}_{\rm 2}$-dissociating radiation in the Lyman-Werner bands and by photoevaporation, 
long before they can finish reionization, delaying its completion until the later rise of the
ACHs.} 
because the clumping factor increases significantly with time, thereby enhancing the recombination rate and slowing down the reionization. Specifically, the IC model yields a later end of reionization $z_{\rm ov}=7.3$ than the NCHE model, but a higher optical depth $\tau_{\rm es}=0.069$ than the NCLE model. The $\tau_{\rm es}$ from the IC model is close to the best fit value ($\tau_{\rm es} = 0.066\pm 0.013$) constrained by the {\it Planck}-LFI 2015 result (``Planck TT+lowP+lensing+BAO'') \citep{2016A&A...594A..13P}, 
and consistent with the latest constraint ($\tau_{\rm es} = 0.058\pm 0.012$) from the {\it Planck}-HFI 2016 result (``lollipop+Planck TT'') \citep{2016A&A...596A.108P}.

Recall that we set the values of the lower-efficiency case NCLE source efficiencies
so as to make reionization end approximately at the same redshift as the IC case with its higher efficiencies.
According to Figure \ref{fig:histories}, in fact, the ionized volume in case IC starts its
early rise at the same time and at the same rate as the higher-efficiency case NCHE, which
has the same efficiencies as IC but no subgrid clumping, until $\bar{x}_{\rm HII,m}\cong 0.1$. Thereafter, 
the rising subgrid clumping in IC slows the rate of further increase of its ionized fraction relative
to that of NCHE by an ever-increasing amount until, after $\bar{x}_{\rm HII,m}\cong 0.5$, it matches
that of the lower-efficiency case NCLE, instead.  
Not only does subgrid clumping make the duration of
reionization more extended, therefore, but a more complete characterization is that, with 
inhomogeneous subgrid clumping, the rate of reionization initially follows the case with
the same efficiencies but no clumping, while at later times it shifts to follow that of a case with
no clumping but much lower efficiencies, with the latter set so as to end reionization at the same 
redshift.  

The reionization histories of the BHC and IC models are close. 
Their difference is much smaller than that between the IC and ``no clumping'' models. 
For the BHC model, reionization proceeds slightly faster in the early phase but slower in the late phase than the IC model, and thus its reionization history is more extended. The histories of these two models cross at $z\simeq 9 - 10$, corresponding to $\bar{x}_{\rm HII,m} \simeq 0.2 - 0.3$. The BHC model yields a 
slightly later $z_{\rm ov}$($=6.9$), and a smaller $\tau_{es}$($=0.067$), than the IC model. 
These can be explained as follows. The BHC model assumes the global mean pseudo-clumping factor, and thus, roughly speaking, it underestimates (overestimates) the recombination rate in overdense (underdense) regions, compared to the IC model. When reionization just starts, overdense regions are reionized earlier on average, so the reionization proceeds faster due to the underestimate of recombination in the BHC model. On the other hand, at the late stage of reionization, the I-fronts reach underdense regions, so the reionization finishes more slowly 
due to the overestimate of recombination in the BHC model. 

\begin{figure}
\begin{center}
  \includegraphics[height=0.33\textheight]{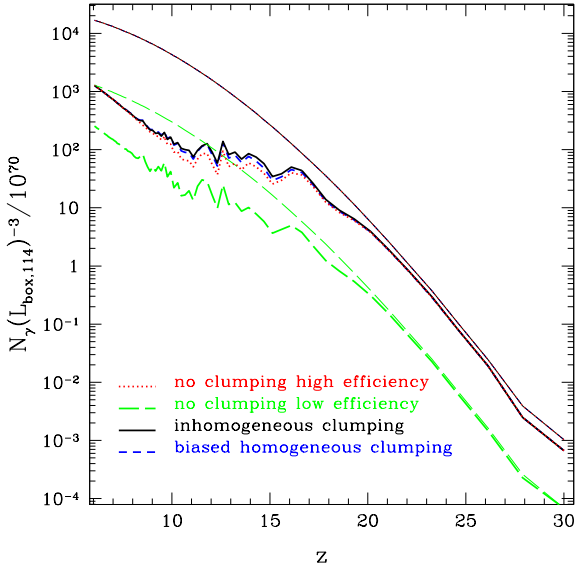} 
\end{center}
\caption{Number of ionizing photons emitted by all sources if there were no suppression (thin lines), and by all active sources (thick lines), in the simulation volume per time step. 
}
\label{fig:lum}
\end{figure}

\begin{figure}
\begin{center}
  \includegraphics[height=0.33\textheight]{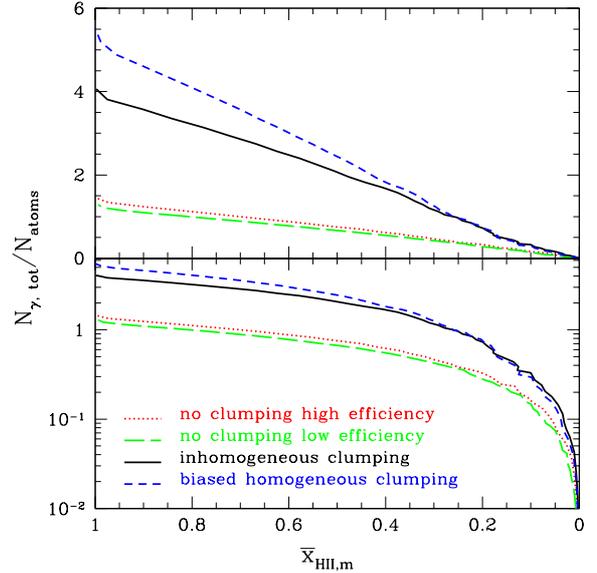} 
\end{center}
\caption{Cumulative number of photons per total gas atom released into the IGM, in the linear (top) and logarithmic (bottom) scales, for each reionization model in their respective history of reionization. 
}
\label{fig:lum2}
\end{figure}

\subsection{Ionizing radiation, clumping factor, and recombination}
\label{subsec:recomb}

\begin{figure*}
\begin{center}
  \includegraphics[height=0.3\textheight]{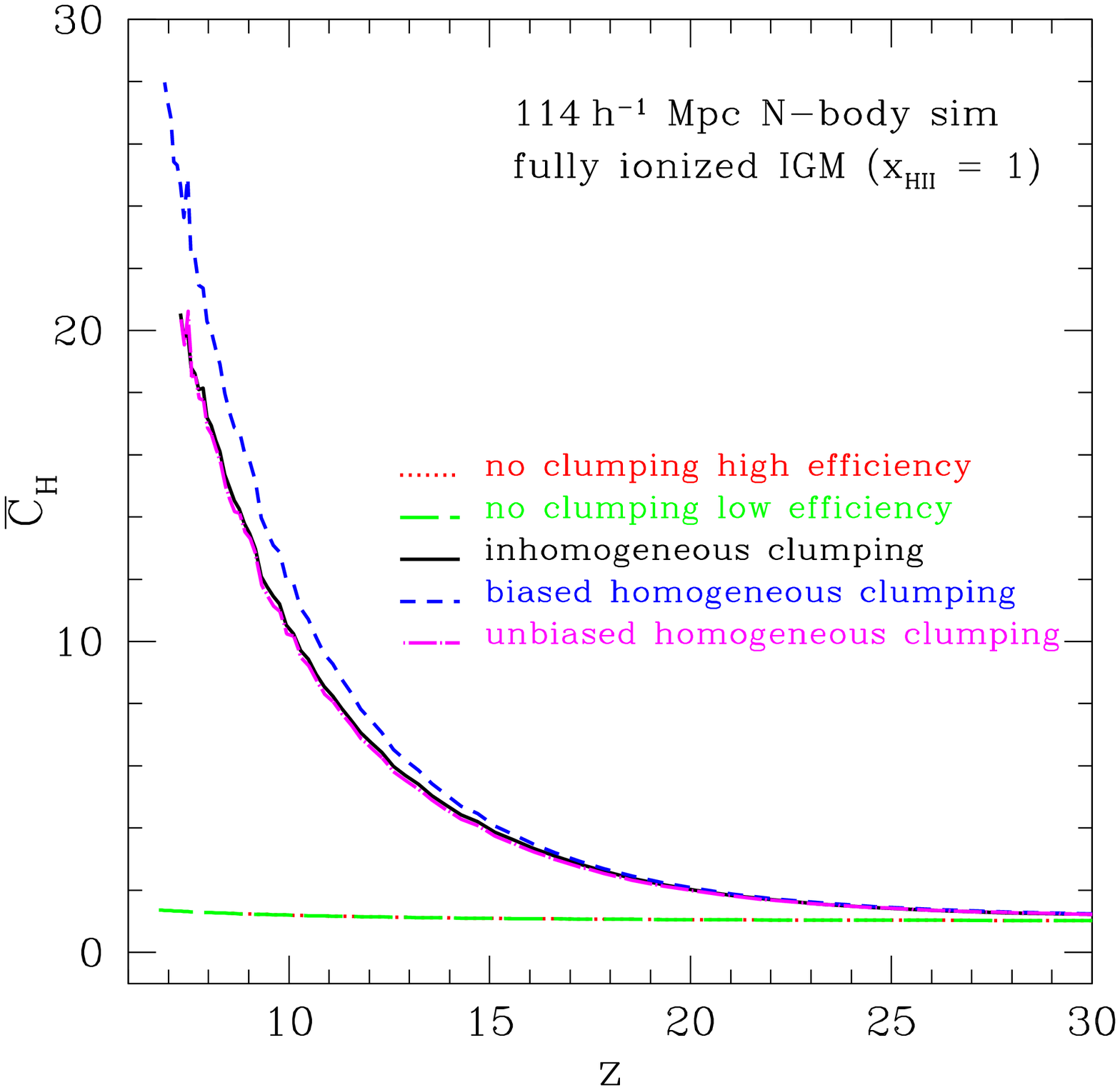} 
  \includegraphics[height=0.3\textheight]{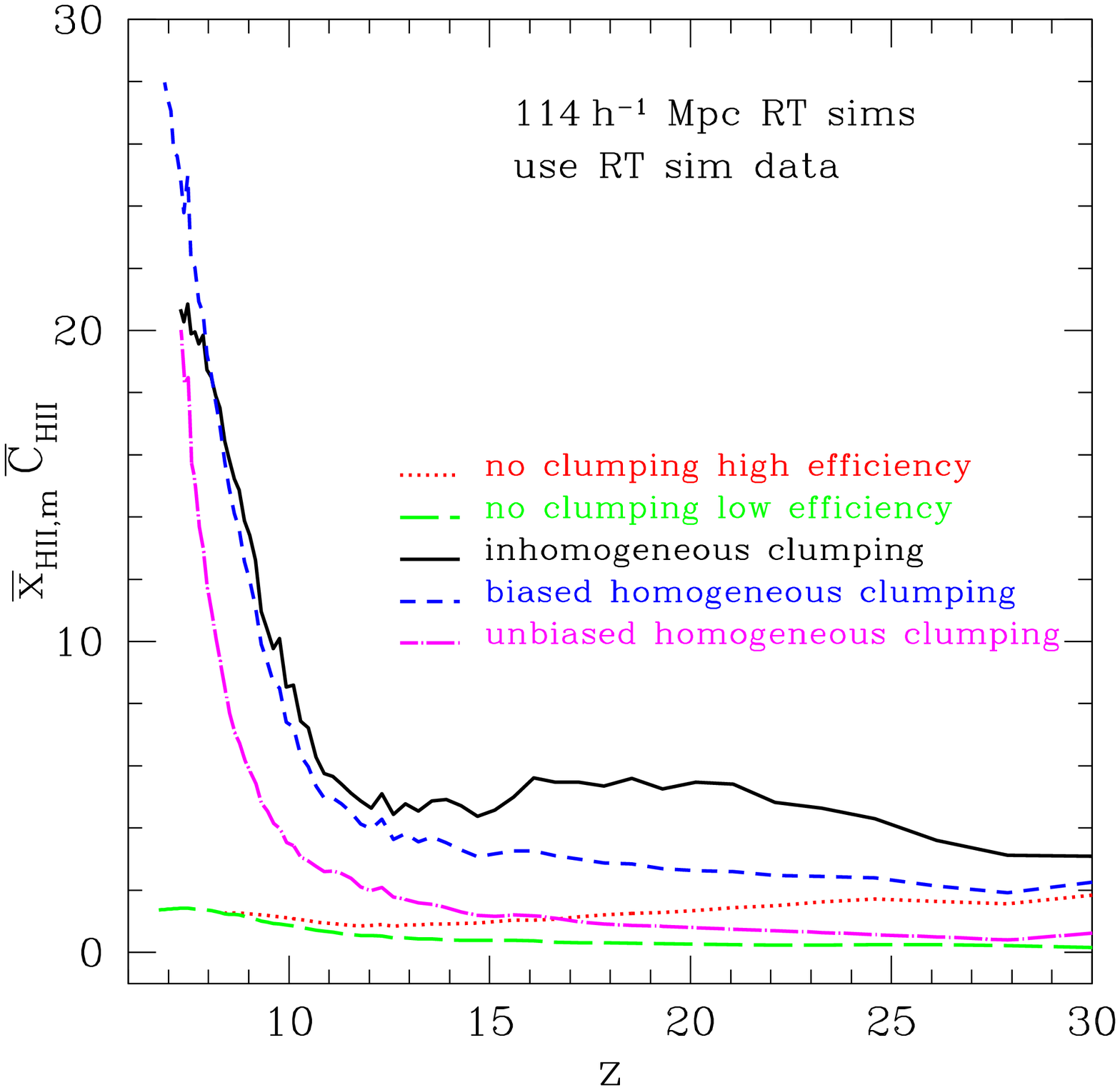}
\end{center}
\caption{The mean IGM H~II clumping factor, $\bar{C}_{\rm HII}\equiv\overline{n^2}_{\rm HII,IGM}/\bar{n}^2_{\rm HII,IGM}$ as a function of redshift, for various models with different assumptions on source efficiency and/or subgrid clumping factor. Left panel: for illustrative purpose, we assume the fully ionized IGM ($x_{\rm HII}=1$ everywhere). Right panel: we plot $\bar{x}_{\rm HII,m}\,\bar{C}_{\rm HII}$, applying for RT simulation data. In all cases, the density that is multiplied by the subgrid clumping factor to compute the cellwise recombination rate is the inhomogeneous cellwise density in the coarse-grained RT mesh.  }
\label{fig:CHII_sim}
\end{figure*}

\subsubsection{Ionizing radiation}
\label{subsec:ion-rad}

Figure~\ref{fig:lum} shows the number of ionizing photons,
for all (i.e. \ hydrogen and helium) atoms, 
released per simulation time step from all sources in the simulation volume. 
If there were no suppression of haloes, the photon number from {\it all} sources (thin lines) 
is exactly the same for the NCHE, IC, and BHC models because they assume the same source efficiency. 
(Note that these RT simulations are postprocessed from the same N-body simulation, 
and therefore have the same halo abundances and distributions at each redshift.) 
LMACHs, however, are subject to Jeans mass 
filtering if they are located inside an ionized region, which causes them to be
suppressed as {\it active} sources. As a result, only those LMACH haloes which 
are not suppressed at a given time are {\it active} as sources.
This is the essence of their ``self-regulation'': as their abundance rises,
the more ionized volume they create, within which they and other LMACHs that 
form are subsequently suppressed.  So the number of ionizing photons from active 
sources (thick lines) is slightly different amongst the NCHE, IC, and BHC models, 
but the difference between the former two models is larger than that between the latter two. 
Basically, most of the ionizing sources in the early stage of reionization are LMACHs. 
At that time, in the order of the IC, BHC, and NCHE model, reionization proceeds from more 
slowly to more rapidly, so the LMACH suppressed fractions of these cases proceed
from less suppressed to more suppressed, and their global rates of releasing ionizing photons 
proceed from more released to less released. Nevertheless, 
the difference in the early stage is small: 
the recombination rate is not important at that time 
because the ionized regions are small and the time available for recombination is short. 
At late times, HMACHs become more numerous and dominate the sources of ionizing photons. 
Since HMACHs are not self-regulated, the actual photon number in these three models converges. 
The upshot is that the actual number of ionizing photons released 
per time step for different clumping models is similar if they assume the same source efficiency. 

However, reionization is governed by the competition between ionizations and recombinations. 
Figure~\ref{fig:lum2} shows the cumulative number of photons per total gas atom in the IGM 
as a function of the mass-weighted mean ionized fraction. We find that even though the NCHE and 
NCLE models assume rather different source efficiencies, it takes them almost the same 
integrated number of photons to ionize an atom and keep it ionized, to achieve the same global ionized fraction. 
Also, the NCHE model always needs less photons to keep the same number of atoms (corresponding to a fixed $\bar{x}_{\rm HII,m}$) ionized than the IC and BHC models, just because the recombination rate in the NCHE model is always smaller. The comparison between the BHC and IC model depends on the phase of reionization: the number of the consumed photons in the BHC model is slightly smaller at the beginning, but significantly larger towards the end of reionization, than in the IC model. At the outset of reionization, the BHC model underestimates the clumping factor against the IC model, so the former needs less ionizing photons. Starting from the middle stage ($x_{\rm HII}\gtrsim 0.2 - 0.3$), however, the former overestimates the clumping factor, so it requires more photons. Specifically, it takes $\sim 5.4$ /$4$ /$1.4$ photons per atom on average to complete the reionization in the BHC/IC/NCHE model, respectively, i.e.\ wasting 4.4/3/0.4 photons on average in reionizing a previously recombined atom (through the repeated process of ionizations and recombinations). 

\begin{figure*}
\begin{center}
  \includegraphics[height=0.3\textheight]{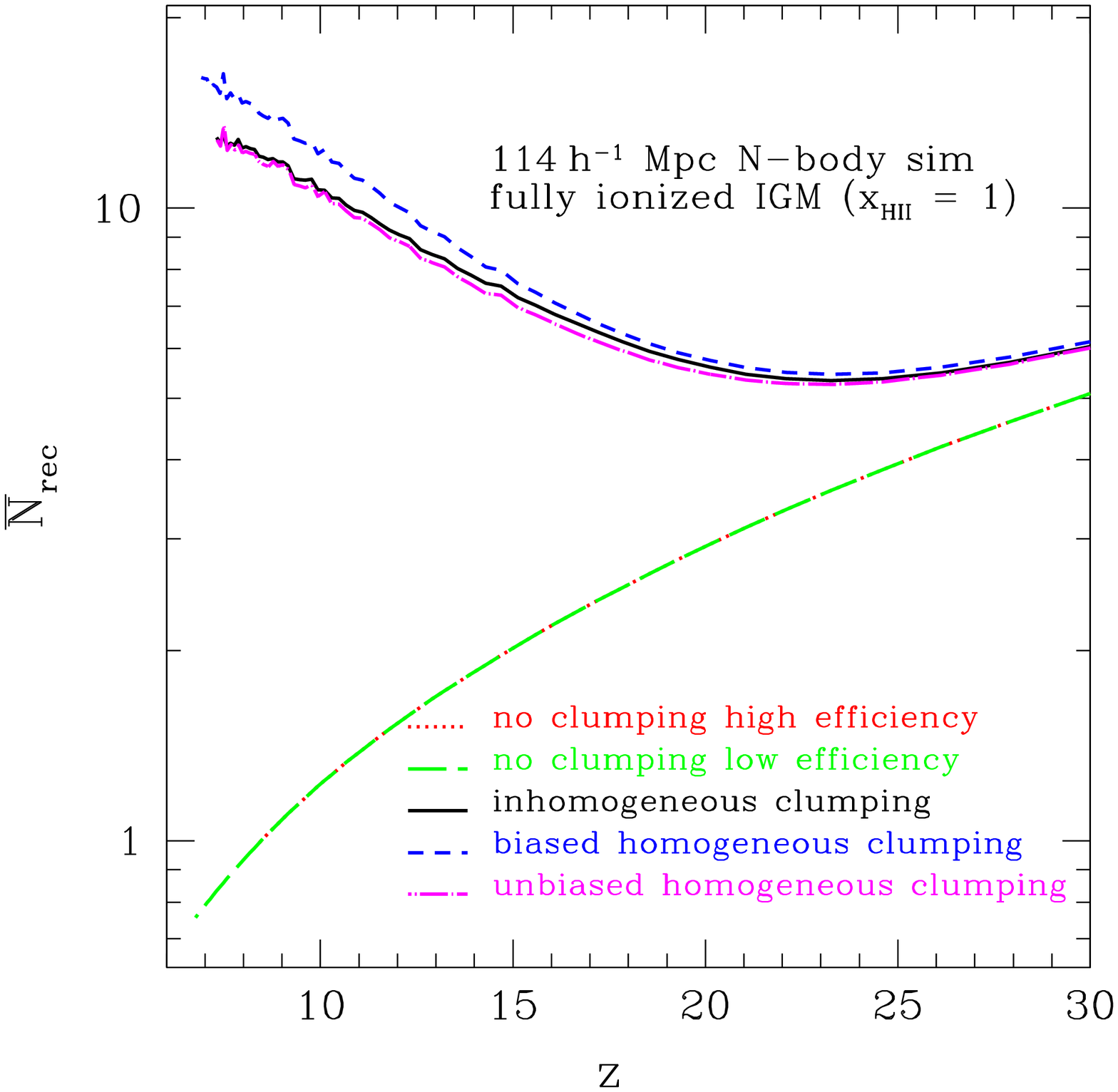} 
  \includegraphics[height=0.3\textheight]{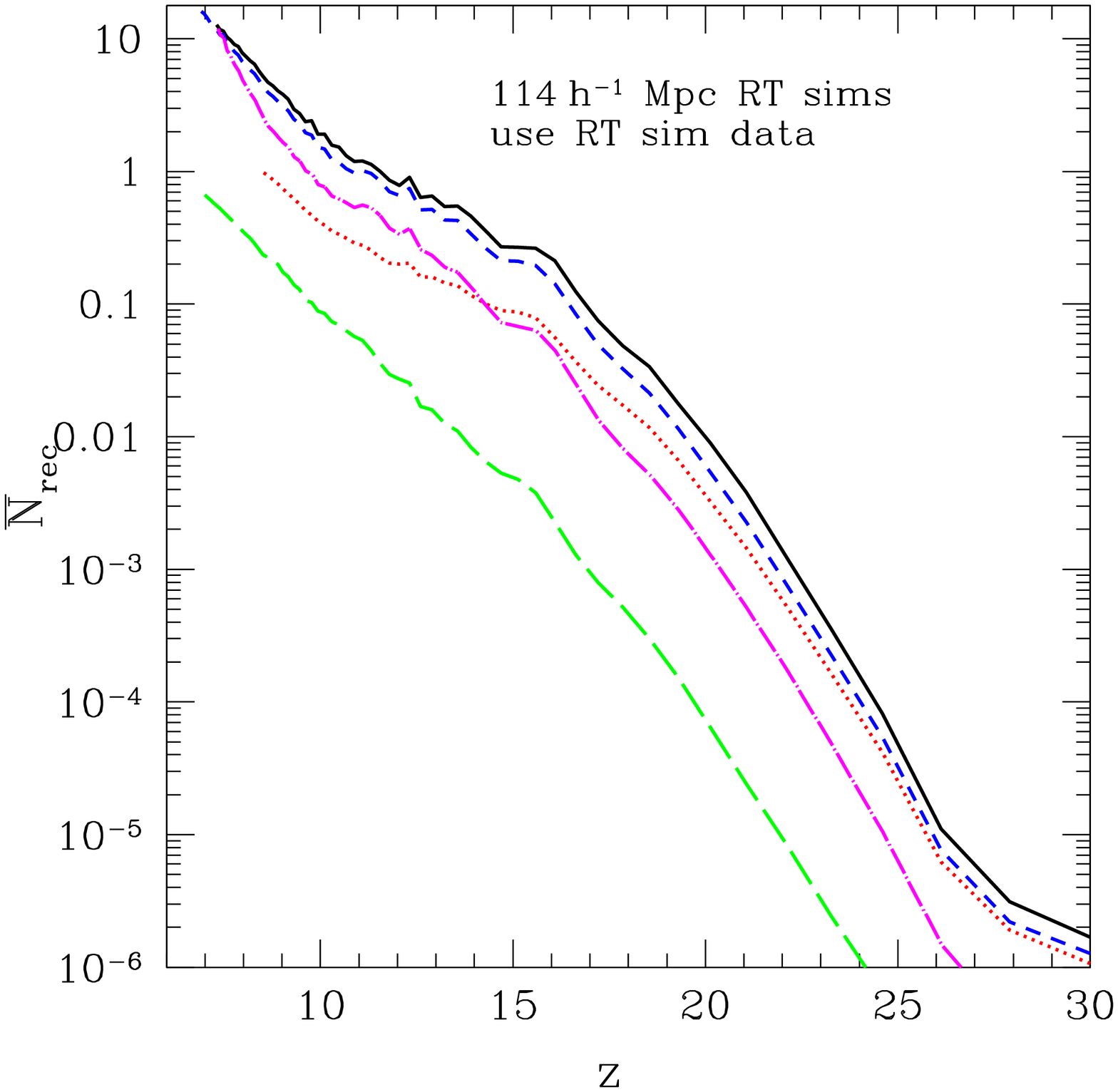}
\end{center}
\caption{The spatially averaged number of recombinations per hydrogen atom per Hubble time at that redshift, $\bar{N}_{\rm rec}$, as a function of redshift, when the IGM is assumed to be fully ionized (left) and the actual ionization field is applied (right).}
\label{fig:recom_sim}
\end{figure*}

\subsubsection{Clumping factor}
\label{subsec:clumping_bhc_vs_ic}

Figure~\ref{fig:CHII_sim} shows the mean clumping factor of the ionized IGM, 
$\bar{C}_{\rm HII}\equiv\overline{n^2}_{\rm HII,IGM}/\bar{n}^2_{\rm HII,IGM}$ as a function of redshift. 
For illustrative purposes, we first consider the case of fully ionized universe 
($x_{\rm HII}=1$ everywhere, $\bar{C}_{\rm HII} = \bar{C}_{\rm H}$, see left panel). 
In this case, the mean clumping factor in the NCHE and NCLE models is almost constant 
(increasing slightly from $\sim 1$ at $z = 30$ to $\sim 1.5$ at $z = 7$), because the subgrid 
clumping factor is assumed to be unity in these two models and the cellwise density fluctuations 
are small on the scale of cell size in the coarse-grained mesh at high redshift. 
The mean IGM clumping factor in the IC model is enhanced significantly, reaching $\sim 20$ at $z\sim 7$, 
which is consistent with $\overline{\hat{C}}(z) $ in Figure~\ref{fig:meanC}. 
(Note that $\overline{\hat{C}} = \bar{C}_{\rm H} \bar{f}_{\rm IGM}^2$, 
where $\bar{f}_{\rm IGM} = \bar{n}_{\rm H,IGM}/\bar{n}_{\rm H,total} = 1 - \bar{f}_{\rm coll} \approx 1 $ 
and the mean collapsed fraction $\bar{f}_{\rm coll}$ averaged over the universe is very small at high redshift.) 

On the other hand, while the BHC and IC models yield the same $\bar{C}_{\rm H}$ at high redshift, the mean clumping factor in the BHC model is up to $40\%$ larger than that in the IC model, for $z\lesssim 20$ when $\bar{x}_{\rm HII,m} \gtrsim 1\%$. The reason is the variation of cellwise IGM densities across the coarse-grained grid cells. In the BHC model, $\bar{C}_{\rm H}\equiv\overline{n^2}_{\rm H,IGM}/\bar{n}^2_{\rm H,IGM}=\overline{\hat{C}}\cdot \left[ \overline{\left<n_{\rm H,IGM}\right>_{\rm cell}^2}/\overline{\left<n_{\rm H,IGM}\right>_{\rm cell}}^2 \right] >\overline{\hat{C}}$. However, in the IC model, $\bar{C}_{\rm H}\equiv\overline{n^2}_{\rm H,IGM}/\bar{n}^2_{\rm H,IGM} \approx \overline{\hat{C}} \times \bar{n}_{\rm H,total}^2/\bar{n}^2_{\rm H,IGM} = \overline{\hat{C}}/\bar{f}_{\rm IGM}^2 \approx \overline{\hat{C}}$. Here we used the fact found in Figure~\ref{fig:meanC} that $\overline{\hat{C}}$ estimated by the IC model using $114\,h^{-1}\,{\rm Mpc}$ simulation and that by direct SPH-like smoothing of $6.3\,h^{-1}\,{\rm Mpc}$ N-body particle data are in agreement. 

Now we consider the actual mean IGM clumping factor {\it in the H~II regions}, $\bar{C}_{\rm HII}$, from the RT simulations. $\bar{C}_{\rm HII}$ is affected by the distribution of ionized regions; even in a universe with uniform density everywhere, $\bar{C}_{\rm HII}$ is boosted from unity by $\bar{C}_{\rm HII}=1/\bar{x}_{\rm HII,m}$. So the quantity $\bar{x}_{\rm HII,m}\,\bar{C}_{\rm HII}$ presents the combined effects of density clumpiness and 
inhomogeneous reionization \citep{2012MNRAS.427.2464F}. We plot this quantity in 
Figure~\ref{fig:CHII_sim} (right panel), 
and find that $\bar{x}_{\rm HII,m}\,\bar{C}_{\rm HII}$ behaves like
$\bar{C}_{\rm H}$ (left panel). 
This is not a coincidence, because it is straightforward to show that 
\begin{equation}
\bar{x}_{\rm HII,m}\,\bar{C}_{\rm HII} = \left( \frac{\bar{x}_{\rm HII,m^2}}{\bar{x}_{\rm HII,m}}\right) \,\bar{C}_{\rm H}\,,
\label{eqn:CHII-CHI-relation}
\end{equation}
where $\bar{x}_{\rm HII,m^2} \equiv \overline{n^2}_{\rm HII,IGM} / \overline{n^2}_{\rm H,IGM} $. If we think of $x_{\rm HII}$ as the probability to find a region as ionized, then $\bar{x}_{\rm HII,m^2}$ is the $n^2_{\rm H,IGM}$ (or, mass squared) weighted mean ionized fraction. It should be of the same order as the mass-weighted mean ionized fraction $\bar{x}_{\rm HII,m}$, so the prefactor $\bar{x}_{\rm HII,m^2}/\bar{x}_{\rm HII,m}$ on the RHS of  equation~(\ref{eqn:CHII-CHI-relation}) is of order unity. 

The comparison between the BHC and IC model, as shown in Figure~\ref{fig:CHII_sim} (right panel), qualitatively confirms the explanations in \S\ref{subsec:reion-history} and \S\ref{subsec:ion-rad}, namely that the BHC model underestimates the mean H~II clumping factor when reionization starts, but overestimates it when $z<8$, corresponding to the intermediate and late stages ($\bar{x}_{\rm HII,m} \gtrsim 0.5$).

\subsubsection{Recombination}

Figure~\ref{fig:recom_sim} shows the spatially averaged number of recombinations per hydrogen atom per Hubble time at that redshift. For illustrative purposes, we first consider the case of fully ionized universe ($x_{\rm HII}=1$ everywhere, see left panel). In the NCHE and NCLE models, the recombination rate decreases with time, because the physical density decreases due to cosmic expansion, as ${(1 + z)^3}$, making the 
recombination time grow as ${(1 + z)^{-3}}$, faster than the Hubble time $H^{-1}$ which grows only as ${(1 + z)^{-3/2}}$.
In the IC model, another effect, which is that the subgrid clumping factor increases with time, is more important. In combination, hence, $\bar{N}_{\rm rec}$ is boosted to $\gtrsim 10 $ at $z\sim 7$ in the IC model. Compared to the IC model, the BHC model overestimates $\bar{N}_{\rm rec}$, for the same reason as the overestimation of $\bar{C}_{\rm H}$.

The actual recombination rate from the RT simulations (see right panel of Figure~\ref{fig:recom_sim}) is affected significantly 
by H~II regions: as H~II regions grow, the mean recombination rate increases monotonically, for all clumping models. 
At late times, larger clumping factor in the IC model further enhances the recombination rate by an order of magnitude, 
over that in the NCHE model. 
Between the BHC and IC models, the recombination rate behaves just like the H~II clumping factor in \S\ref{subsec:clumping_bhc_vs_ic}: 
the former underestimates $\bar{N}_{\rm rec}$ at the early stage of reionization, but overestimates it when $\bar{x}_{\rm HII,m} \gtrsim 0.5$. 

\subsubsection{Does an ``unbiased homogeneous clumping'' model work?}
\label{subsec:UHC}

In \S\ref{subsec:clumping_bhc_vs_ic}, we proved that even in a fully ionized universe, the BHC model does {\it not} 
reproduce the same mean clumping factor as the IC model. It is because the clumping factor is multiplied with the inhomogeneous 
cellwise density in the coarse-grained RT mesh. In other words, on some scales we include the effects of inhomogeneity twice, 
first when smoothing N-body data to compute the mean pseudo-clumping factor, 
and secondly when multiplying this clumping factor with the inhomogeneous density. 
That is why we call this kind of homogeneous clumping model {\it biased}. 
In contrast, one may suggest an {\it unbiased} homogeneous subgrid clumping (``UHC'') model, by assuming 
\begin{equation}
\left<n_{\rm N,IGM}^2\right>_{\rm cell}  =  \overline{\hat{C}} \,\bar{n}_{\rm N,total}^2\,.
\end{equation}
In the UHC model, the density that is multiplied by the clumping factor to compute the recombination rate is the global mean density of total matter, so the inhomogeneous density is only accounted for once (in computing the mean pseudo-clumping factor). 
In the UHC model, it is straightforward to prove that $\bar{C}_{\rm H}=\overline{\hat{C}}/\bar{f}_{\rm IGM}^2$, i.e.\ in principle, it reproduces the same mean clumping factor as the IC model, in the fully ionized case. It is also easy to show that the mean recombination rate, $\bar{N}_{\rm rec}$, is the same in the UHC and IC models, in the fully ionized case. That is why we call this kind of homogeneous clumping model {\it unbiased}. These two identities are tested numerically in Figures~\ref{fig:CHII_sim} (left panel) and \ref{fig:recom_sim} (left panel), and confirmed with small numerical errors. 

\begin{figure*}
\begin{center}
  \includegraphics[height=0.23\textheight]{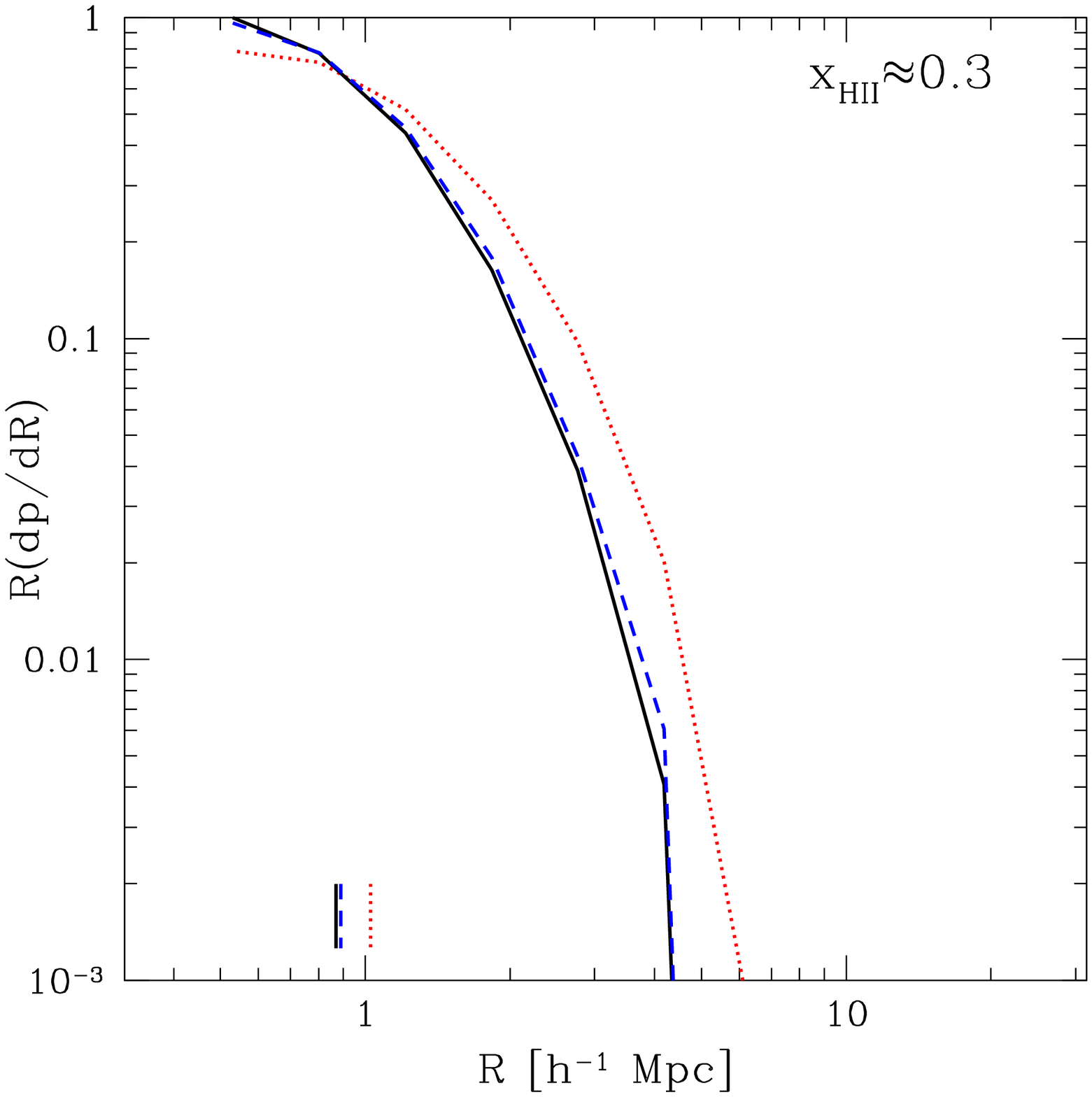} 
  \includegraphics[height=0.23\textheight]{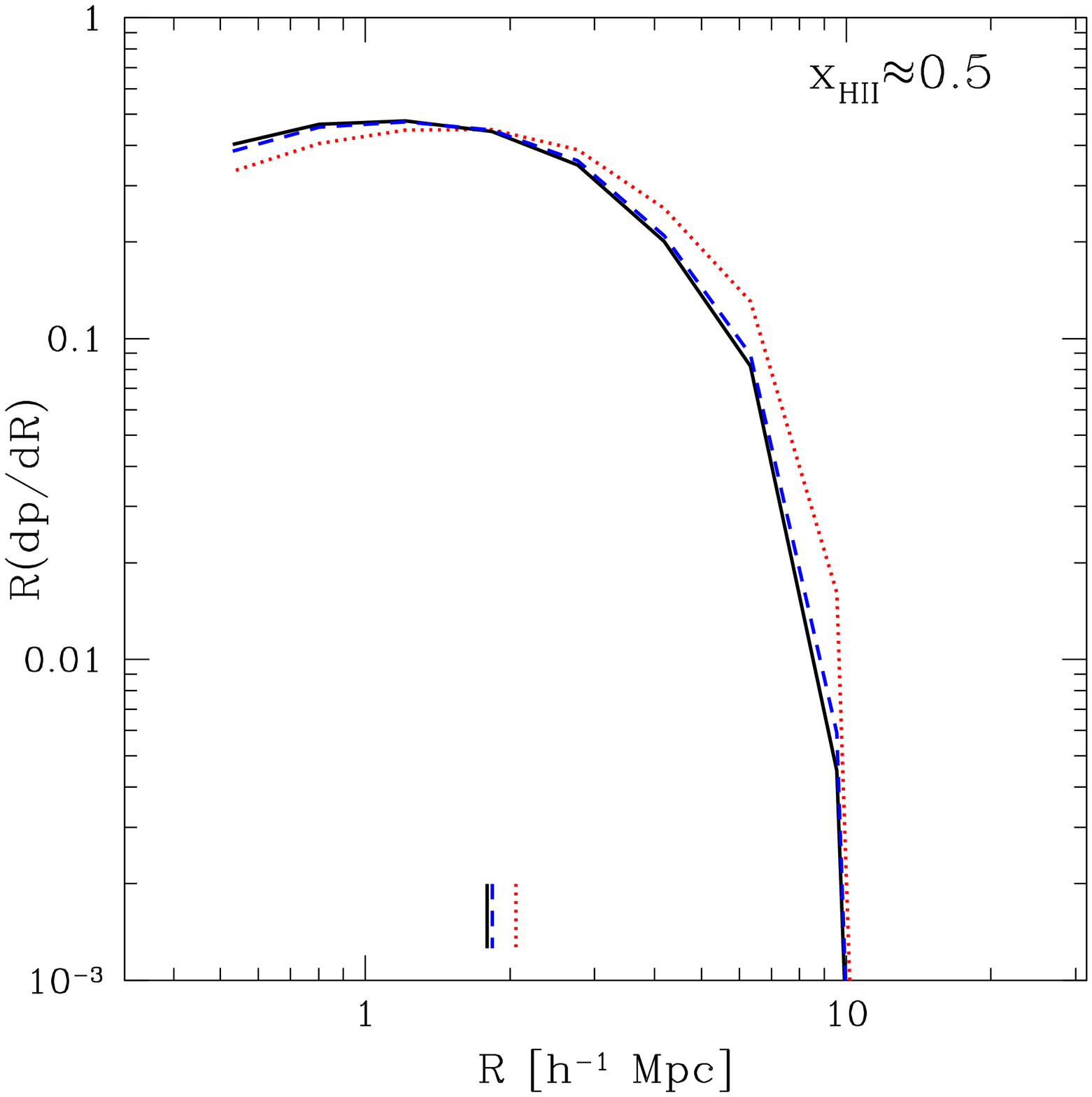}
  \includegraphics[height=0.23\textheight]{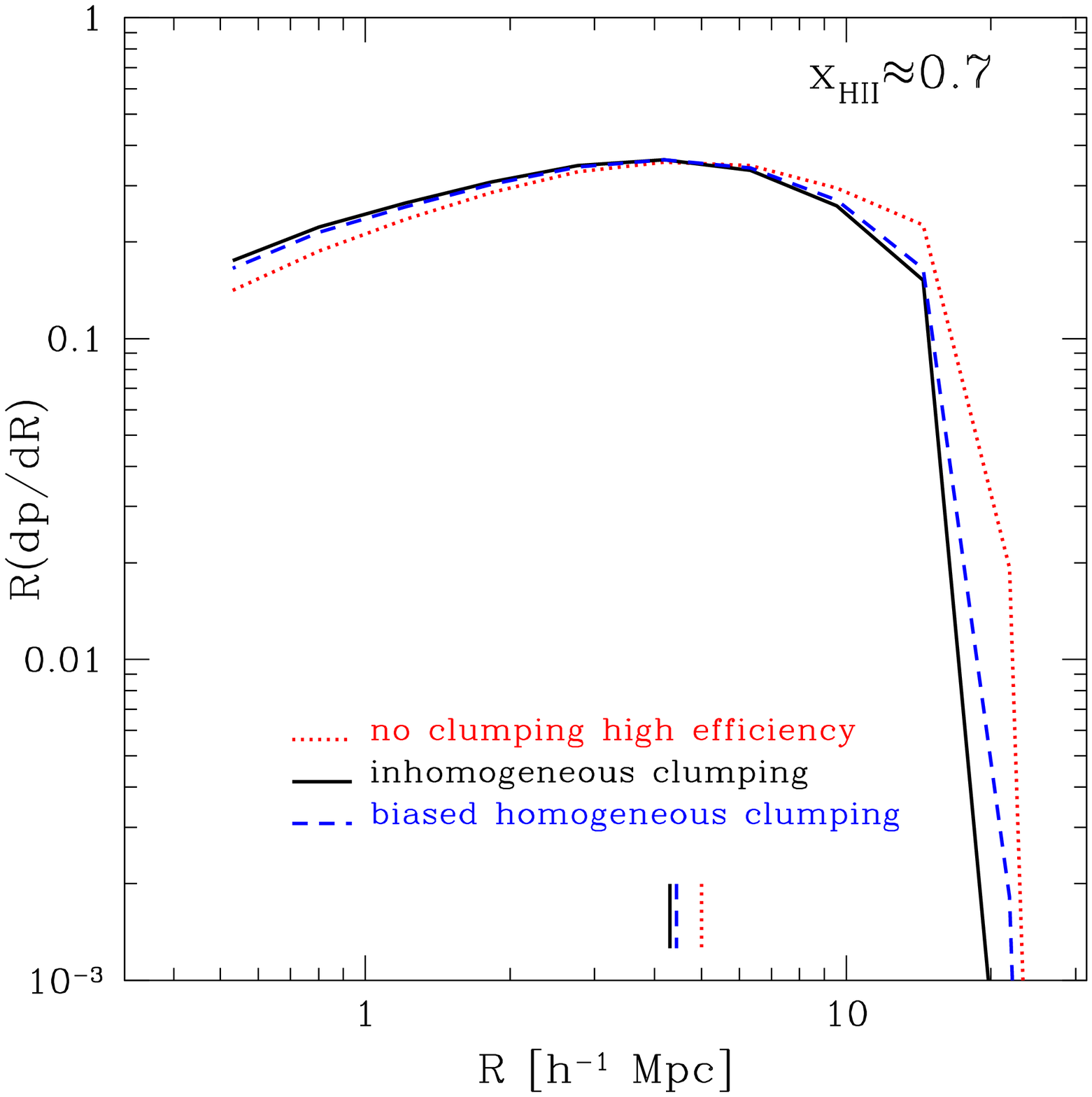}
\end{center}
\caption{Probability distribution function (PDF) per logarithmic radial bin, $R\,dp/dR$, for spherical H~II regions with radius $R$ as given by the spherical average method, based on the ionized distribution given by our simulations. Shown are different stages of the reionization process, for ionized fraction by mass $x_{\rm HII,m} = 0.3$, 0.5 and 0.7, as labelled. We cut off the PDFs at the mesh cell size of reionization simulations, which is the minimum size of H~II regions that could be resolved by our simulations. 
Vertical lines with same colors and line types as the PDF curves mark the mean radius $R_c = \int R(dp/dR)\,dR$ as the characteristic size of H~II regions.}
\label{fig:SPA_size}
\end{figure*}

Given that the UHC model reproduces the mean clumping factor and the mean recombination rate of the IC model {\it in the fully ionized case}, can the UHC model do the same {\it in reionization simulations}? (If so, the UHC model could be an alternative and easier clumping model.) The answer, as told by Figures~\ref{fig:CHII_sim} (right panel) and \ref{fig:recom_sim} (right panel), is {\it no}. Here, instead of running a RT simulation using the UHC model, we assume that the UHC model reproduces the same ionized fraction field as the IC model. We find that the UHC model significantly suppresses both mean clumping factor and mean recombination rate at all redshifts, comparing to the IC model. More importantly, the suppression in the UHC model is much greater than that in the BHC model. The reason is that, since $\left< n_{\rm HII,IGM}^2\right>_{\rm cell} = \left< x_{\rm HII} \right>_{\rm cell}^2\,\left<n_{\rm H,IGM}^2\right>_{\rm cell}$, the UHC model neglected the correlation between the ionized fraction field and the density field in its estimate of the mean clumping factor and mean recombination rate, but the BHC model partially includes this correlation. The upshot is that the BHC model can mimic the IC model better than the UHC model, in terms of their estimates of the clumping factor and recombination rate in reionization simulations.

\subsection{H~II region size distributions}
\label{sec:size_distribution}

Our realistic clumping models, which enhance the clumping factor and recombination rate from the no-clumping models as discussed in the previous section, should affect the large-scale patchiness of reionization. In Figure~\ref{fig:SPA_size}, we illustrate the H~II region size distributions at several stages of reionization, corresponding to the mass-weighted ionized fraction of $\bar{x}_{\rm HII,m} = 0.3$, 0.5 and 0.7. We use the spherical average method \citep{2007ApJ...654...12Z, 2018MNRAS.473.2949G} to find the probability distribution function per logarithmic radial bin, $R\,dp/dR$, for spherical H~II regions with radius $R$ extracted from our simulations. We find that the IC model always yields more numerous small H~II regions at a given $\bar{x}_{\rm HII,m}$ than the NCHE model. Quantitatively, we compute the mean radius $R_c = \int R(dp/dR)\,dR$ as the characteristic size of ionized bubbles, and confirm that, at the same ionized fraction, $R_c$ is always smaller for the IC model than for the NCHE model. This is consistent with the picture we discussed in the previous subsection, namely that larger subgrid clumping increases the recombination rate inside H~II regions. The enhanced recombination can balance the ionizing radiation  that otherwise could be strong enough to completely ionize local hydrogen atoms. The I-fronts thus expand more slowly in the IC model, which results in more numerous small ionized bubbles, as opposed to fewer large bubbles distributed more sparsely, at a given $\bar{x}_{\rm HII,m}$.

For the BHC and IC models, their ionized bubble size distributions are very similar at all times. A further detailed comparison shows that the BHC model always yields slightly more numerous large ionized bubbles than the IC model, at a given $\bar{x}_{\rm HII,m}$. However, Figure~\ref{fig:histories} shows that the BHC model starts to lag slightly behind the IC model when $\bar{x}_{\rm HII,m} \ge 0.2$. Our comparison between the IC and no-clumping model above just suggests a picture in which a model with faster reionization history normally results in more numerous large ionized bubbles, which seems to contradict the comparison here between the BHC and IC model. How do we reconcile this? We first point out that after the BHC model lags behind the IC model after $\bar{x}_{\rm HII,m} \ge 0.2$, the characteristic size of H~II bubbles in the BHC model is indeed smaller than that in the IC model, when they are compared at the same {\it cosmic time}, which is consistent with the reionization history. Since reionization proceeds more slowly at $z\le 10$ in the BHC model, which means that the BHC model reaches the same mean ionized fraction at a later time, source haloes are more massive and luminous and thus H~II bubbles become larger at a given $\bar{x}_{\rm HII,m}$.

\begin{figure}
\begin{center}
  \includegraphics[height=0.27\textheight]{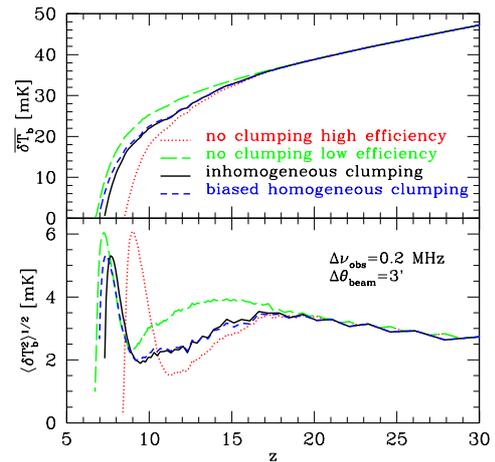} 
\end{center}
\caption{The evolution of (top) the mean 21 cm brightness temperature and (bottom) its RMS fluctuations for Gaussian beamsize $3'$ and bandwidth $0.2\,{\rm MHz}$ with boxcar frequency filter. Here we assume $T_s \gg T_{\rm CMB}$. }
\label{fig:histories3}
\end{figure}

\begin{figure*}
\begin{center}
  \includegraphics[height=0.16\textheight]{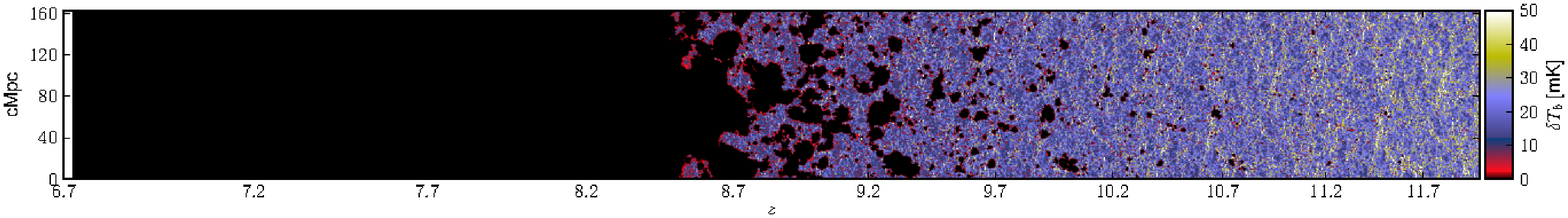} 
  \includegraphics[height=0.16\textheight]{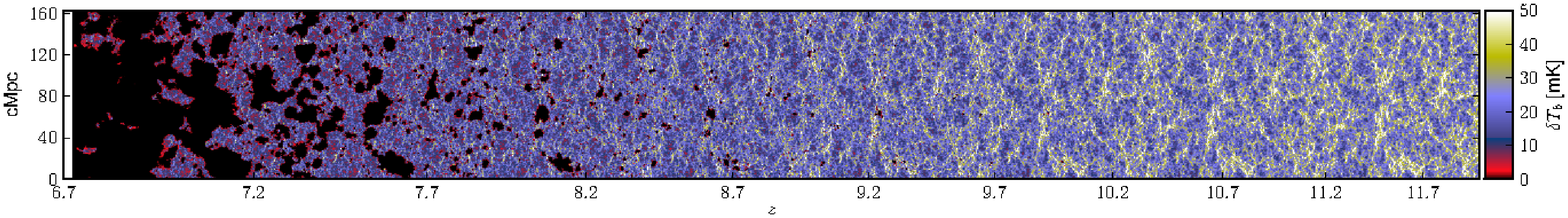}
  \includegraphics[height=0.16\textheight]{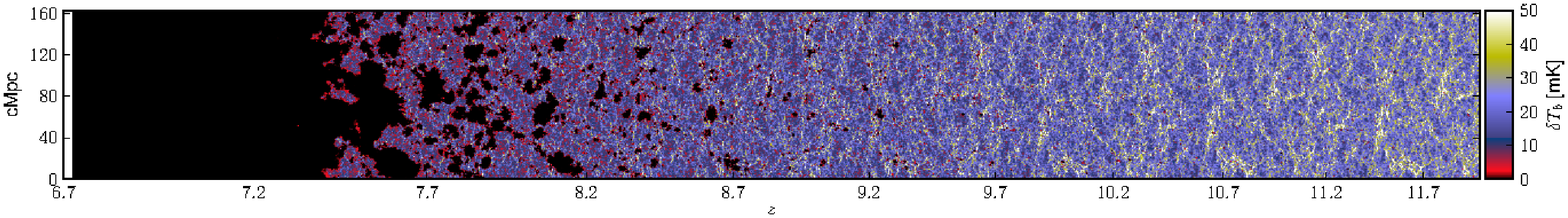}
  \includegraphics[height=0.16\textheight]{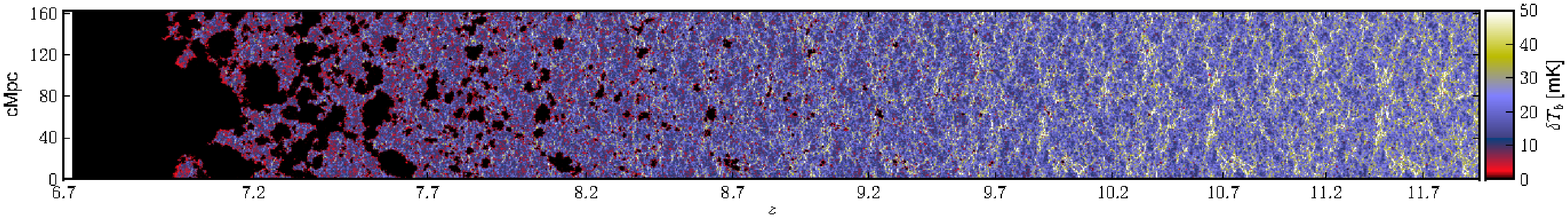}
\end{center}
\caption{Position-redshift slices from our simulations, from top to bottom: no clumping high efficiency, no clumping low efficiency, inhomogeneous clumping, biased homogeneous clumping. These slices illustrate the large-scale geometry of reionization and the significant local variations in reionization history as seen at redshifted 21~cm line. Observationally they correspond to slices through an image-frequency volume of a radio array. 
The images show the differential brightness temperature at the full grid resolution in linear scale. The spatial scale is given in comoving Mpc. We note that for visualization purposes we artificially set $x_{\rm HI}=10^{-5}$ after reionization ($z<z_{\rm ov}$). The redshift-space distortions due to the peculiar velocities are also included.
}
\label{fig:slices}
\end{figure*}


\begin{figure*}
\begin{center}
  \includegraphics[height=0.27\textheight]{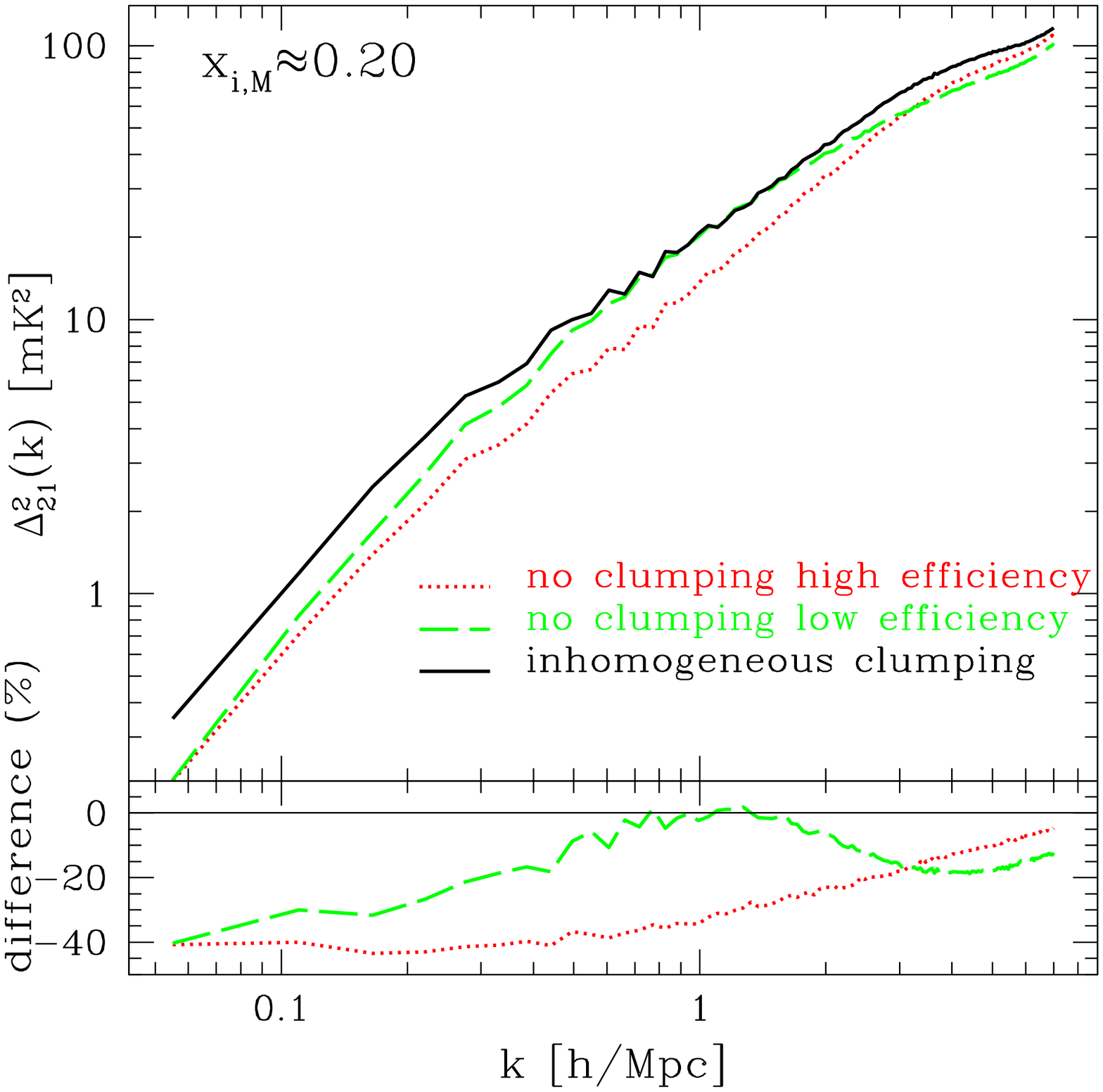} 
  \includegraphics[height=0.27\textheight]{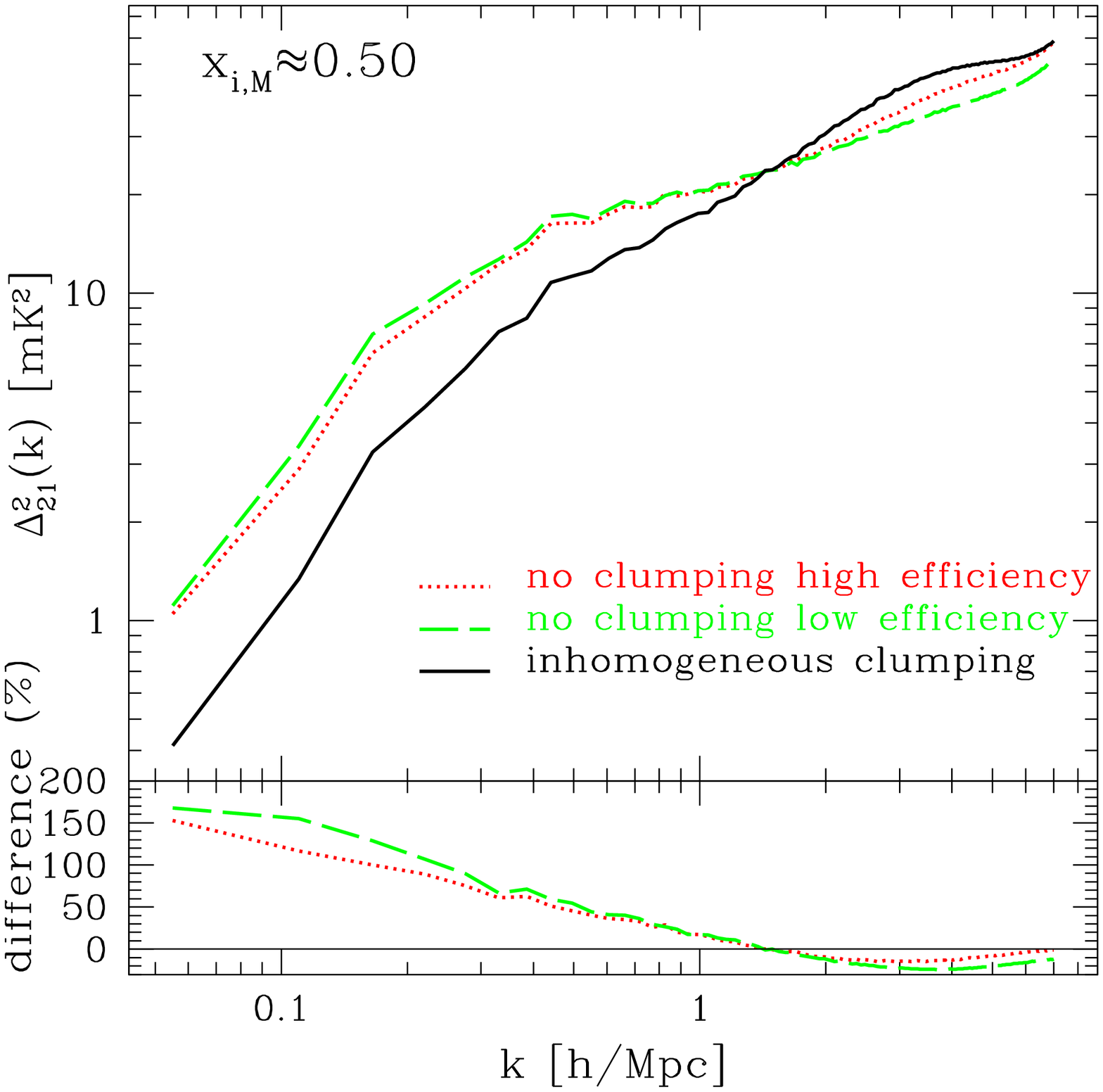}
  \includegraphics[height=0.27\textheight]{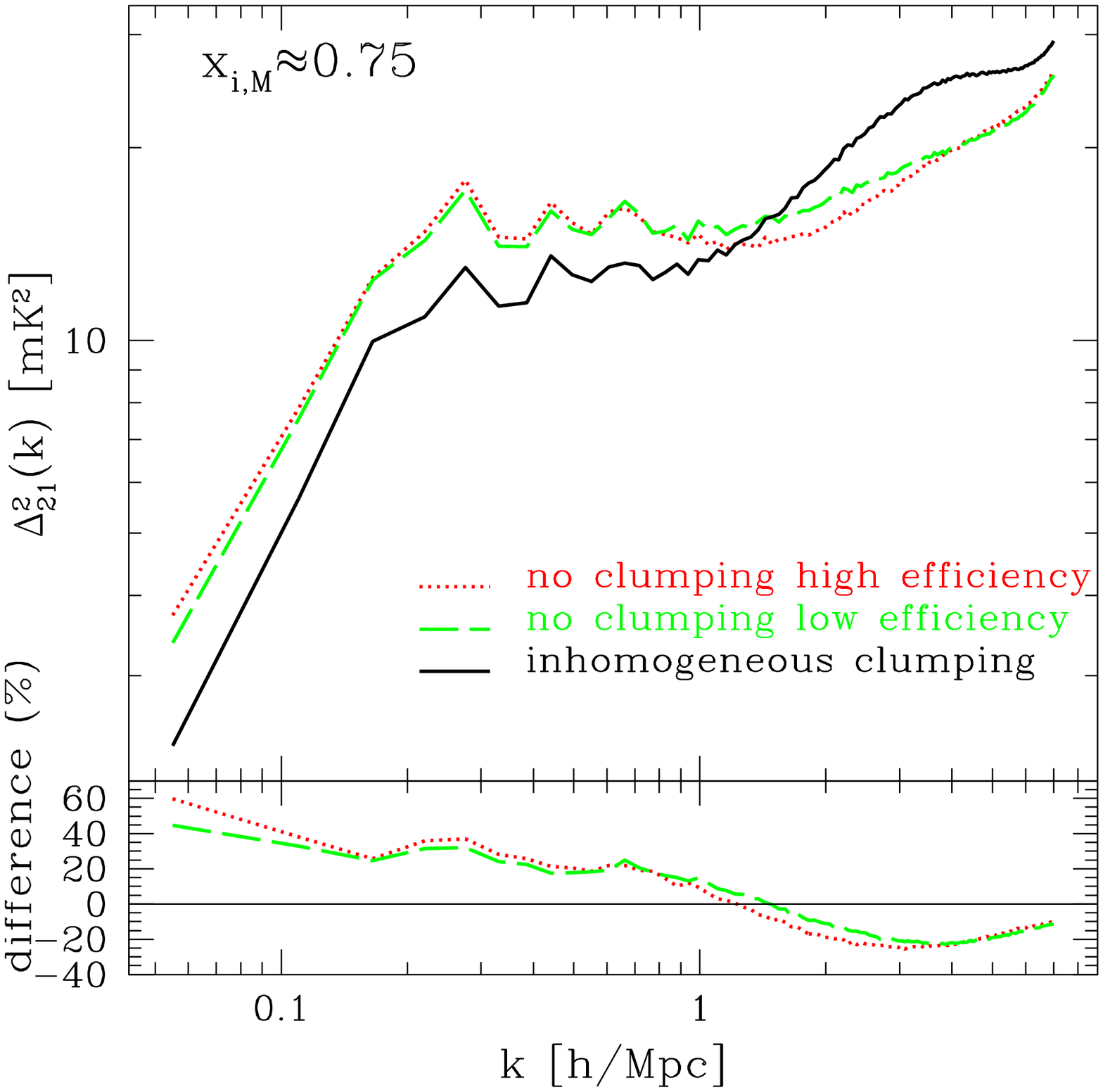}
  \includegraphics[height=0.27\textheight]{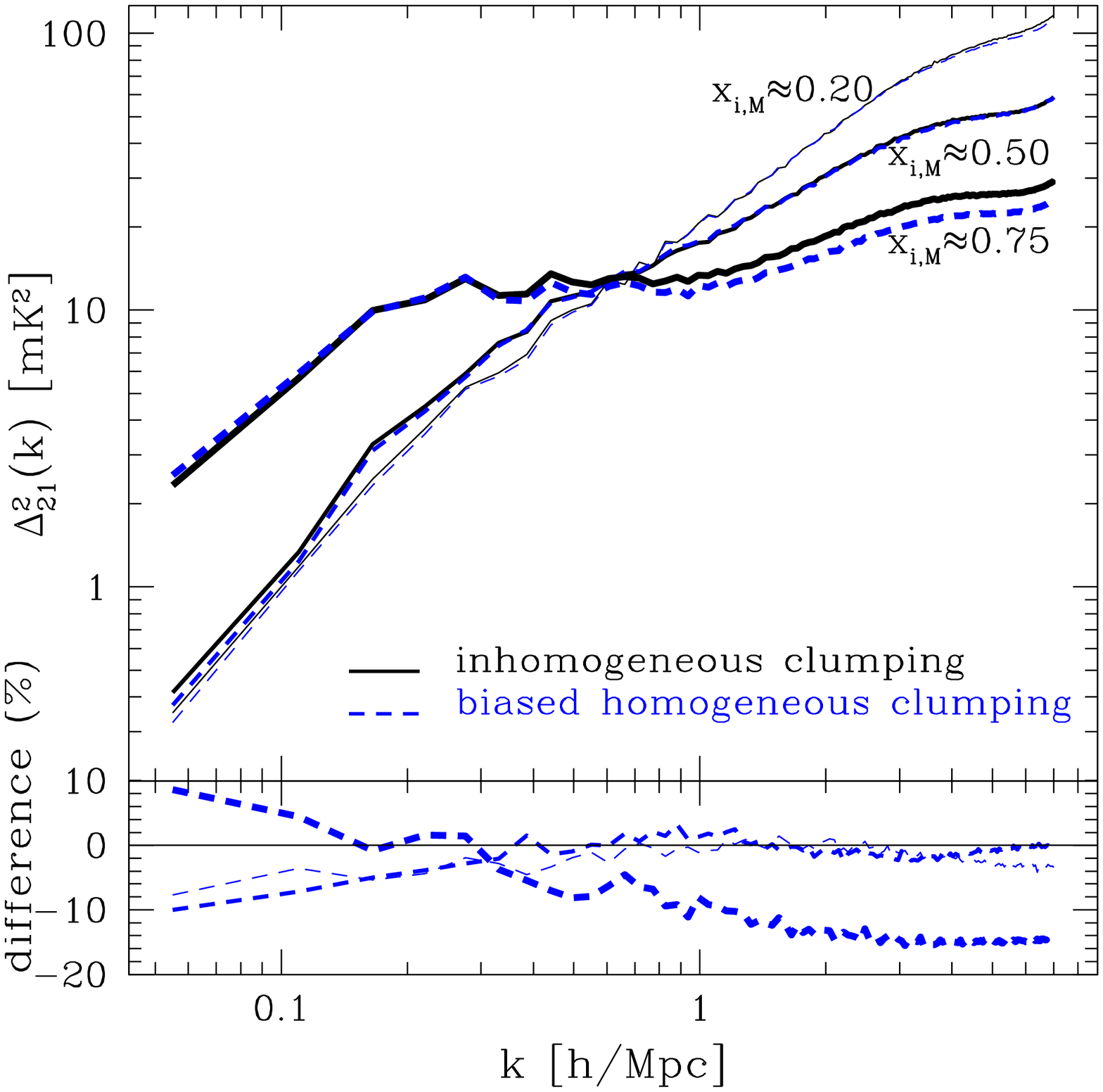} 
\end{center}
\caption{The angle-averaged 21~cm power spectrum $\Delta^2(k) = k^3 P_{\rm 21}(k)/2\pi^2$. We compare the results of the NCHE, NCLE, and IC model at a few key stages of reionization: $\bar{x}_{\rm HII,m} = 0.20$ (top left) ,  $0.50$ (top right), $0.75$ (bottom left), respectively, and show the comparison between the BHC and IC model in the bottom right panel: $\bar{x}_{\rm HII,m} = 0.20$ (thin), $0.50$ (thicker), and $0.75$ (thickest),  respectively. In each panel, the inset shows the fractional difference (in per cent) with respect to the IC model.
}
\label{fig:powers_inh_noclumping}
\end{figure*}

\begin{figure*}
\begin{center}
  \includegraphics[height=0.24\textheight]{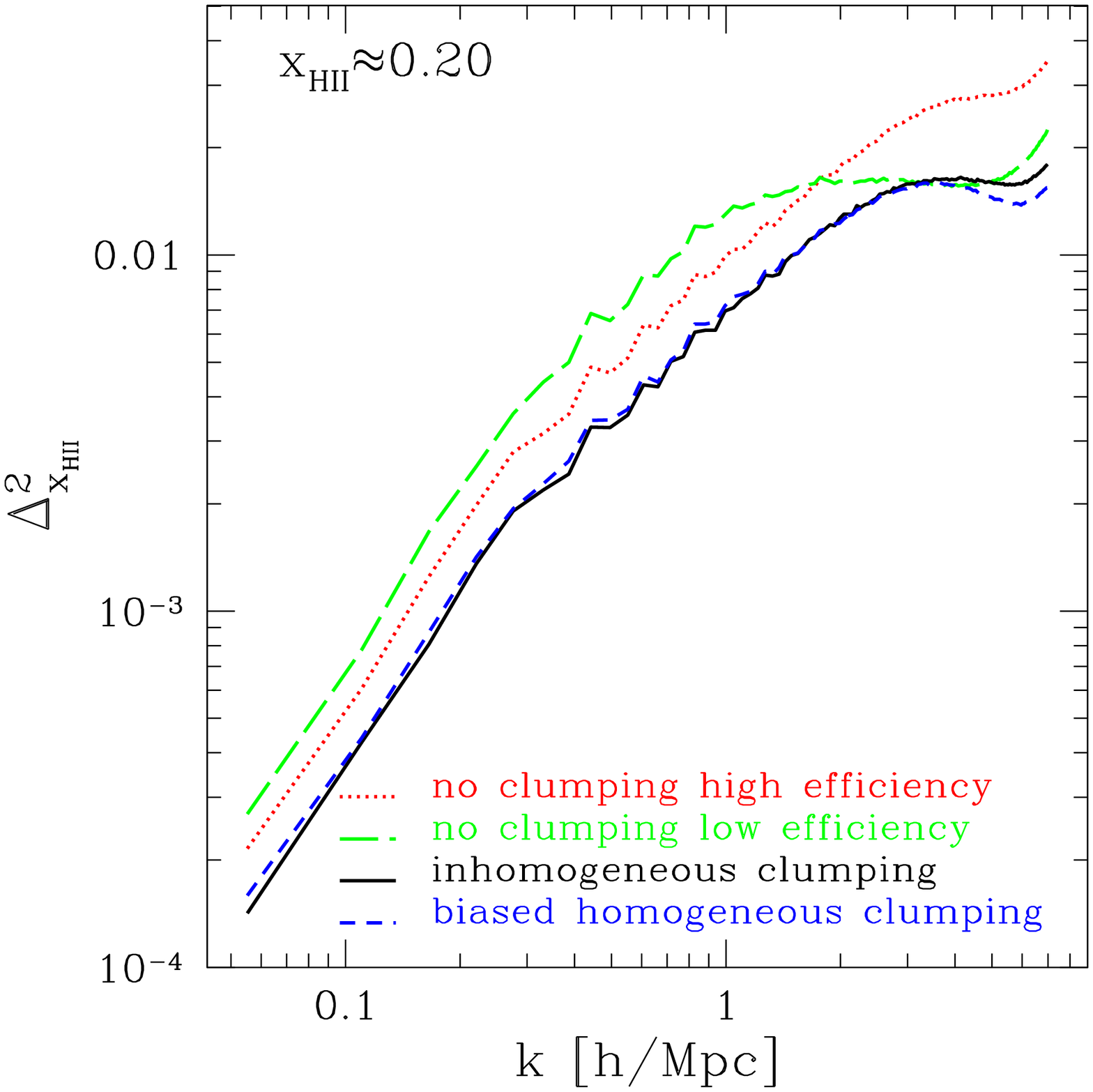} 
  \includegraphics[height=0.24\textheight]{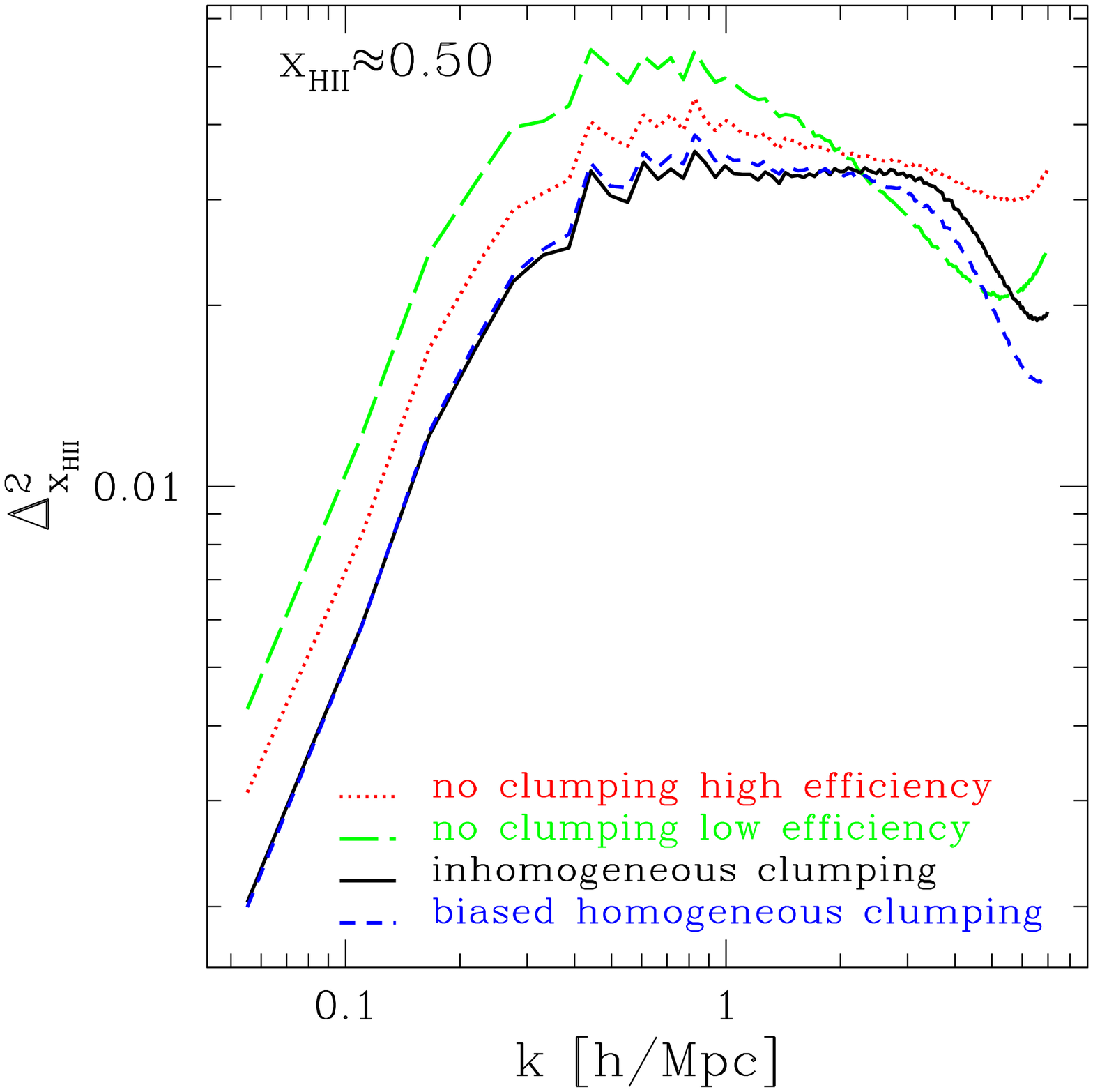}
  \includegraphics[height=0.24\textheight]{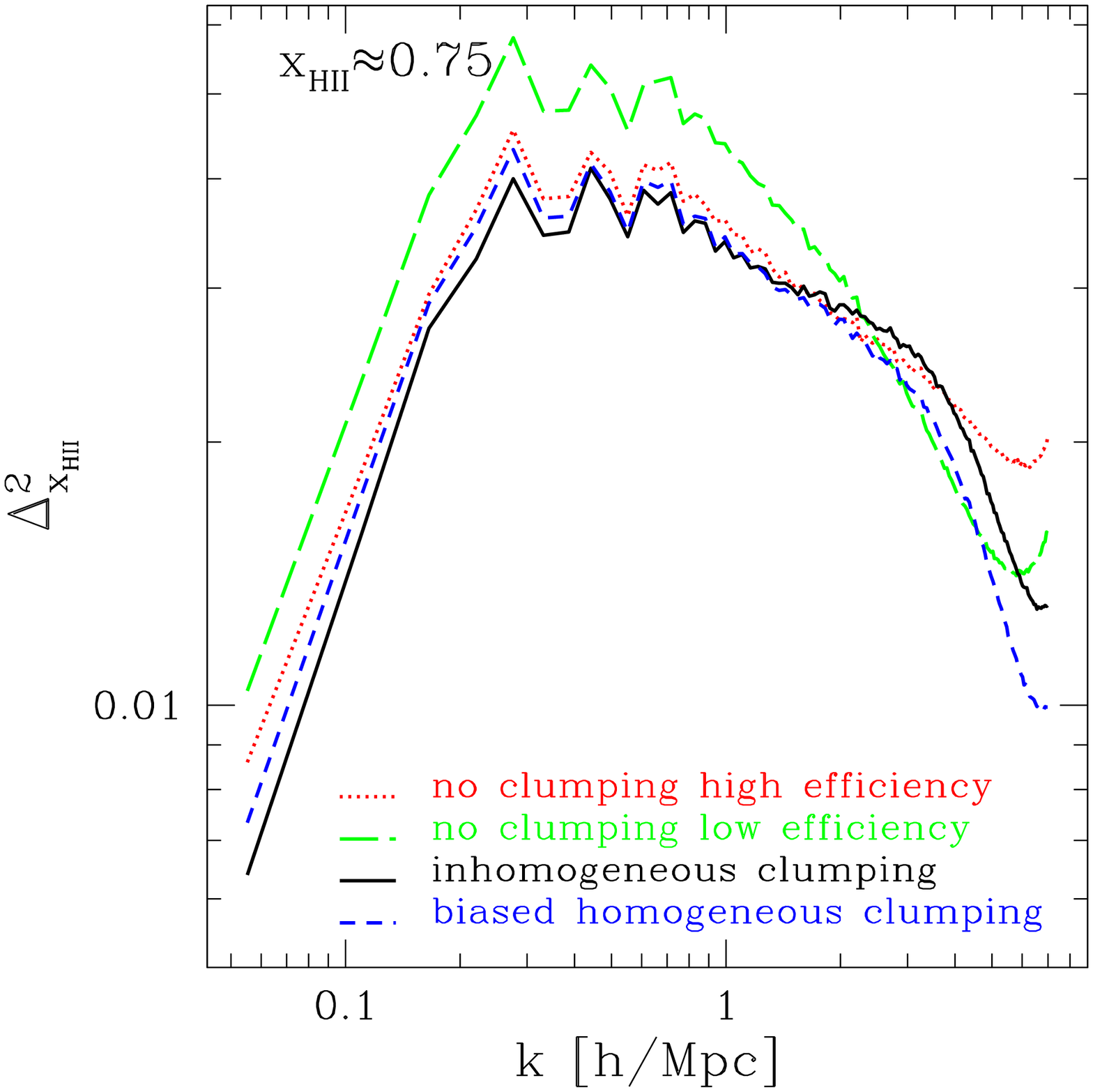}
  \includegraphics[height=0.24\textheight]{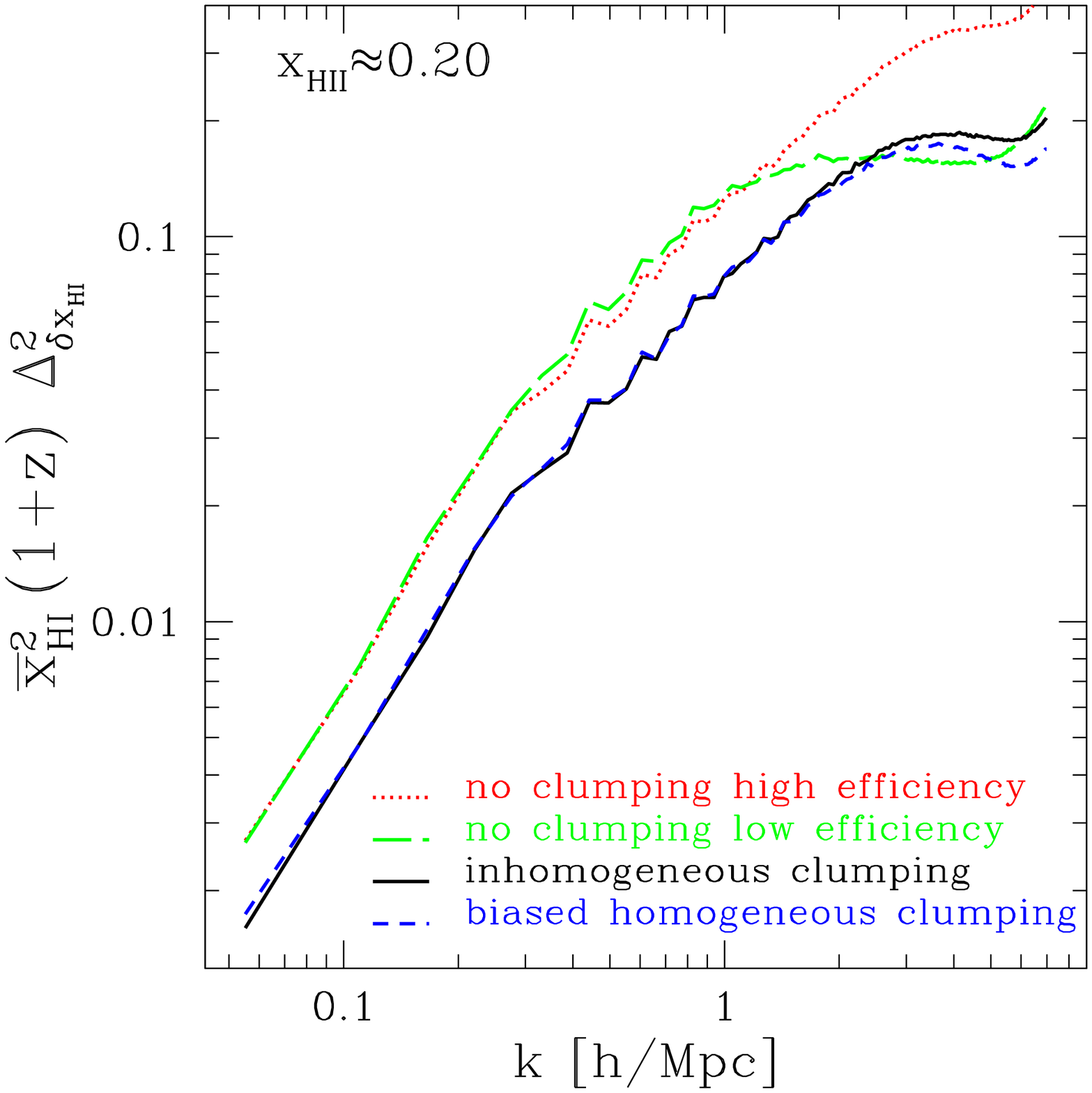} 
  \includegraphics[height=0.24\textheight]{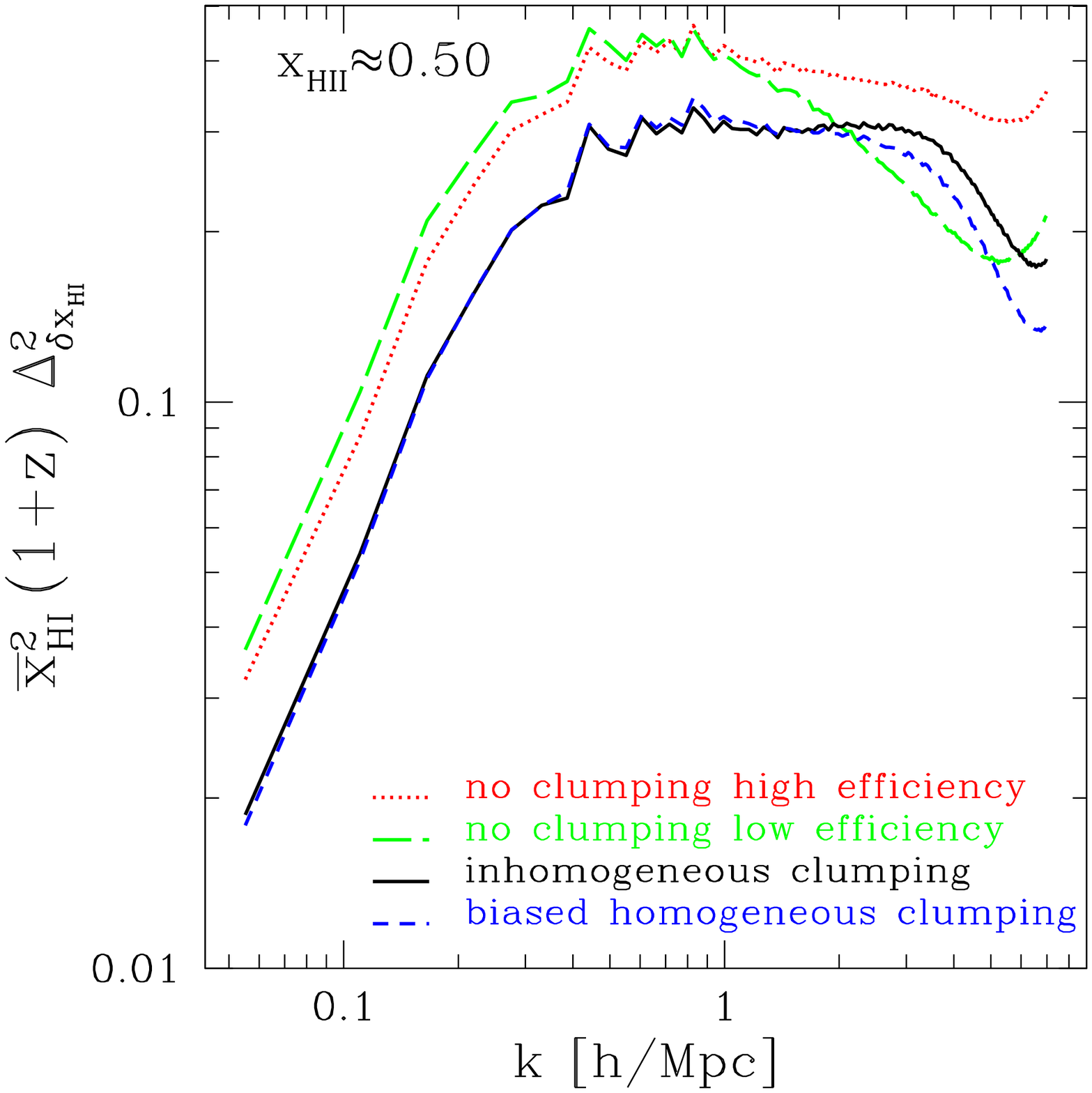}
  \includegraphics[height=0.24\textheight]{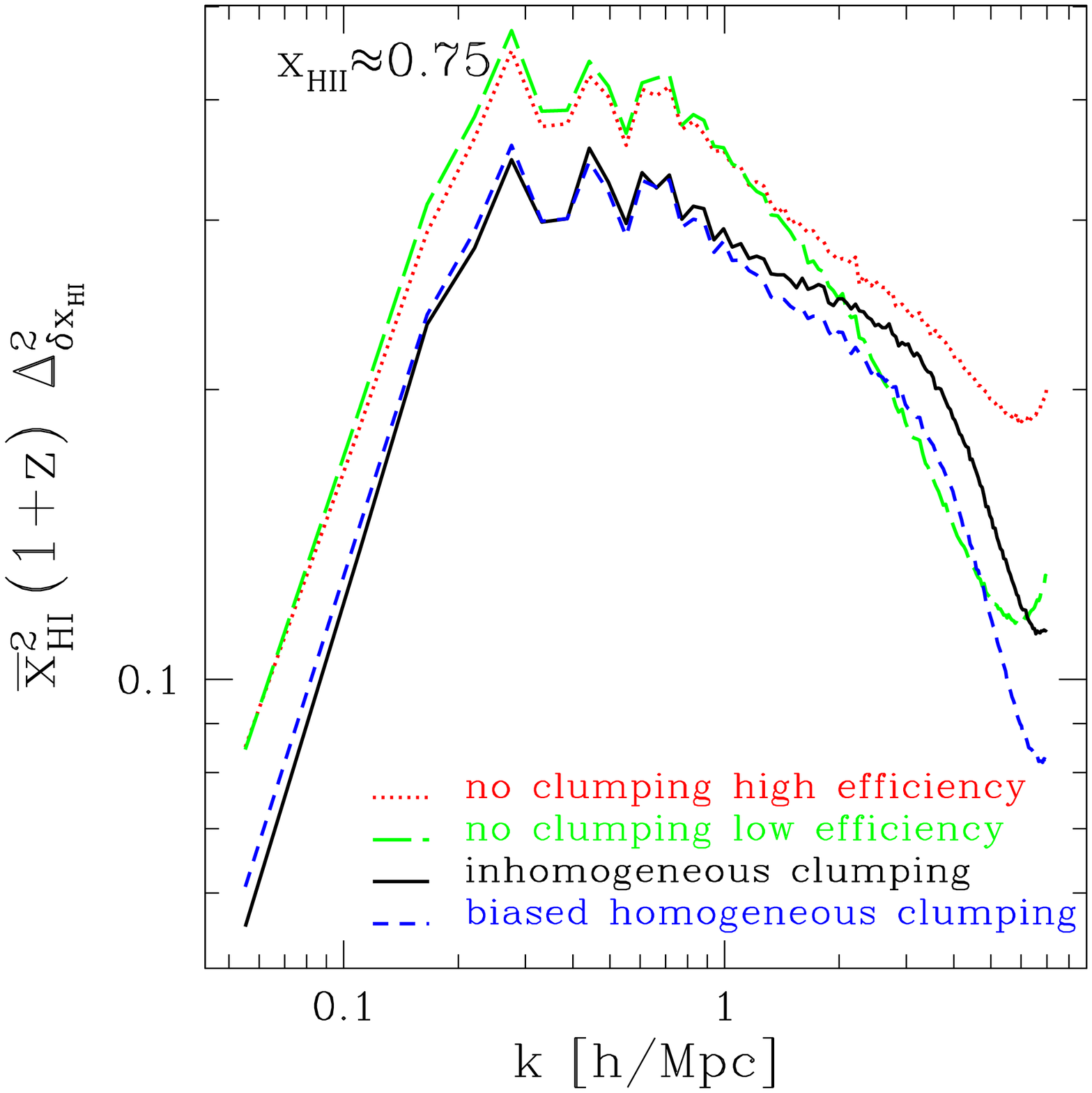}
  \includegraphics[height=0.24\textheight]{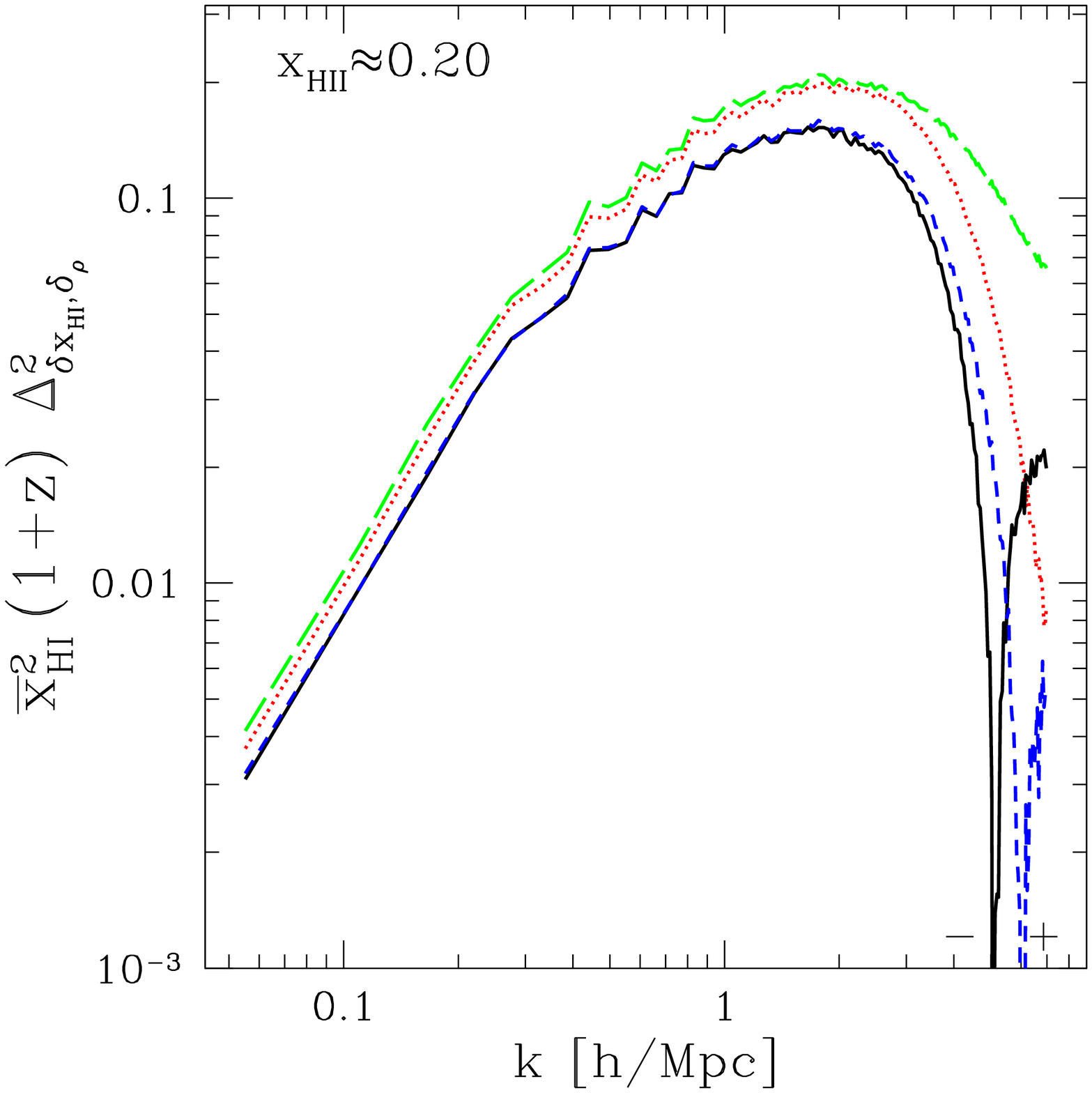} 
  \includegraphics[height=0.24\textheight]{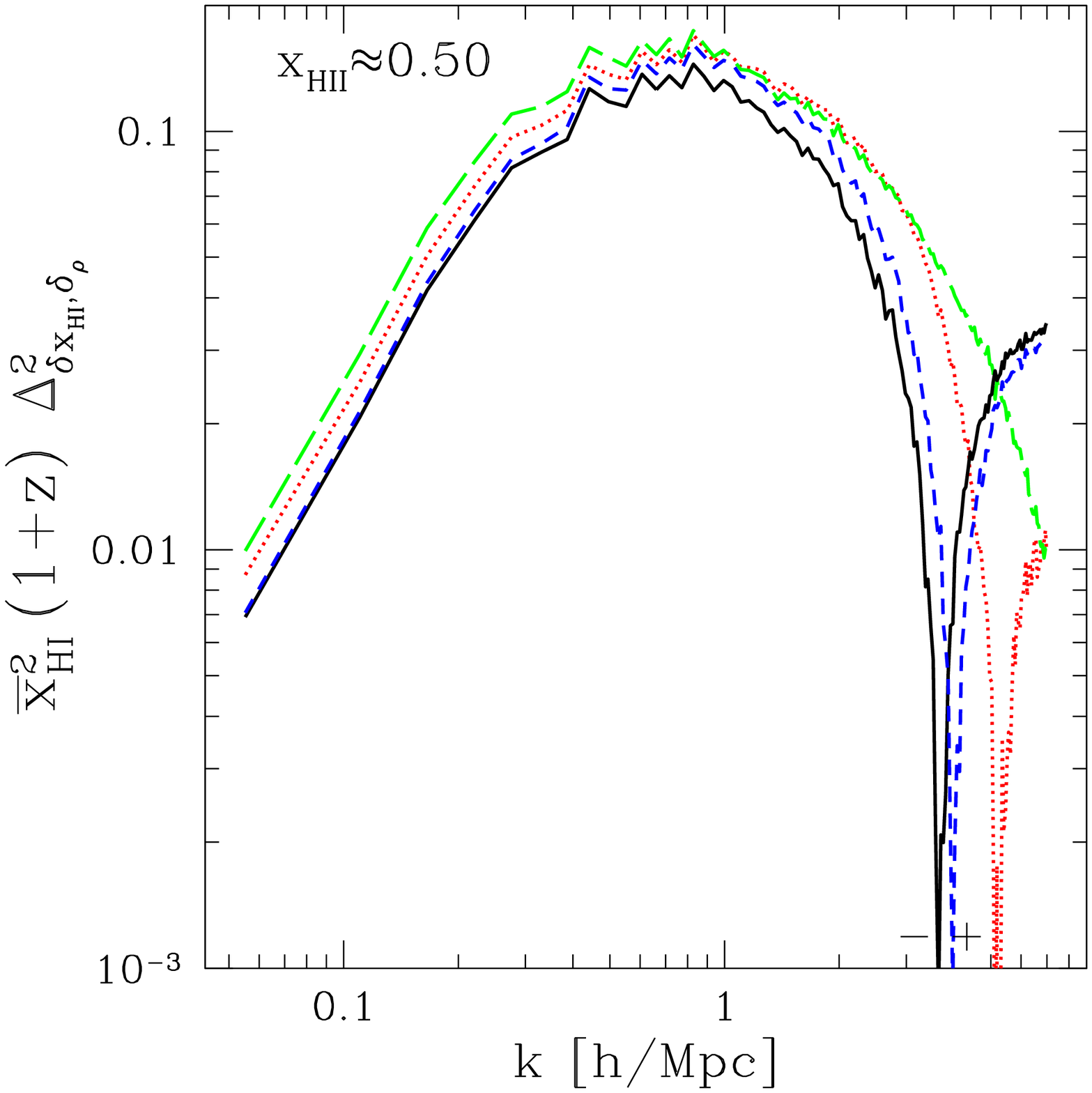}
  \includegraphics[height=0.24\textheight]{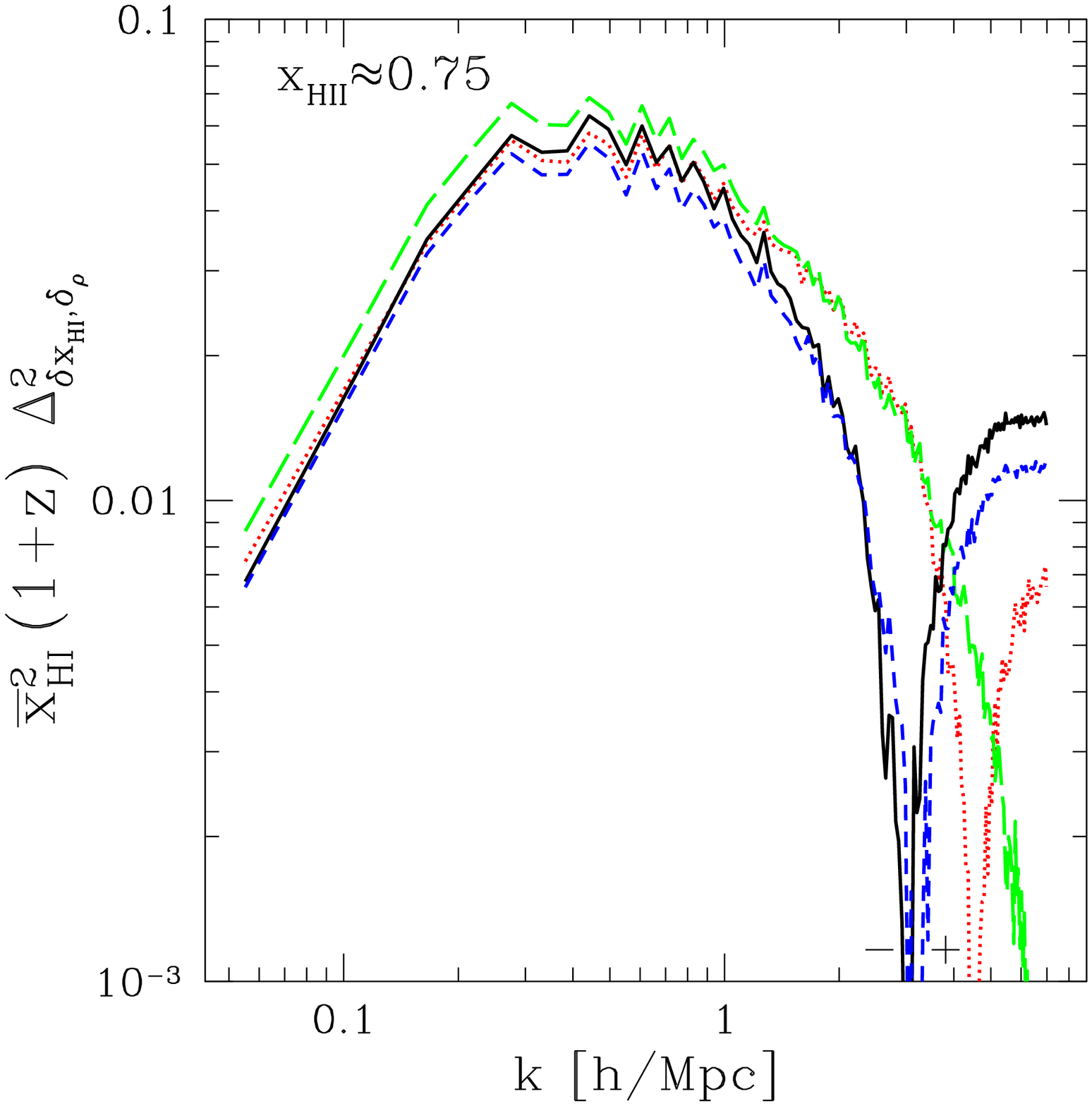}
\end{center}
\caption{Top panels: the auto power spectrum of the neutral fraction (or, equivalently, the ionized fraction) $\Delta^2_{x_{\rm HII}}(k)=\Delta^2_{x_{\rm HI}}(k) = k^3 P_{x_{\rm HI}}(k)/2\pi^2$; middle panels: the auto power spectrum of neutral fraction {\it fluctuations} $\Delta^2_{\delta_{x_{\rm HI}}}(k) = k^3 P_{\delta_{x_{\rm HI}},\delta_{x_{\rm HI}}}(k)/2\pi^2$, scaled by $\bar{x}_{\rm HI}^2(1+z)$; bottom panels: the cross power spectrum between neutral fraction fluctuations and density fluctuations $\Delta^2_{\delta_{x_{\rm HI}},\delta_\rho}(k) = k^3 P_{\delta_{x_{\rm HI}},\delta_\rho}(k)/2\pi^2$, scaled by $\bar{x}_{\rm HI}^2(1+z)$. Shown are results for various reionization models at a few key stages of reionization: $\bar{x}_{\rm HII,m} = 0.20$ (left) ,  $0.50$ (middle), $0.75$ (right). In the bottom panels, all cross power spectra are negative at small $k$ and positive at large $k$. We use ``$-$/$+$'' near the zero crossing to indicate the sign of the cross power spectrum.
}
\label{fig:pxx_pxd}
\end{figure*}

\begin{figure*}
\begin{center}
  \includegraphics[height=0.21\textheight]{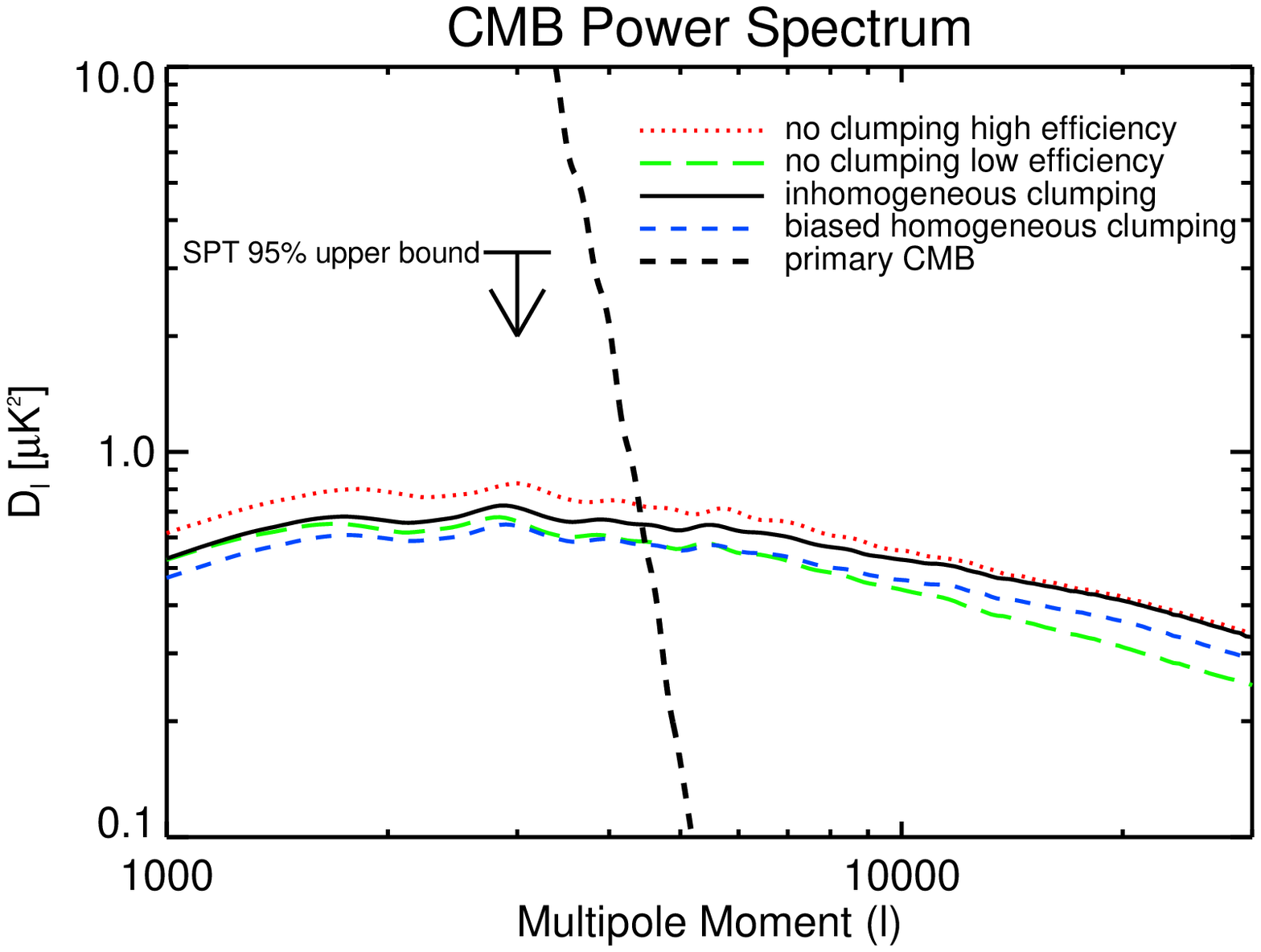} 
  \includegraphics[height=0.21\textheight]{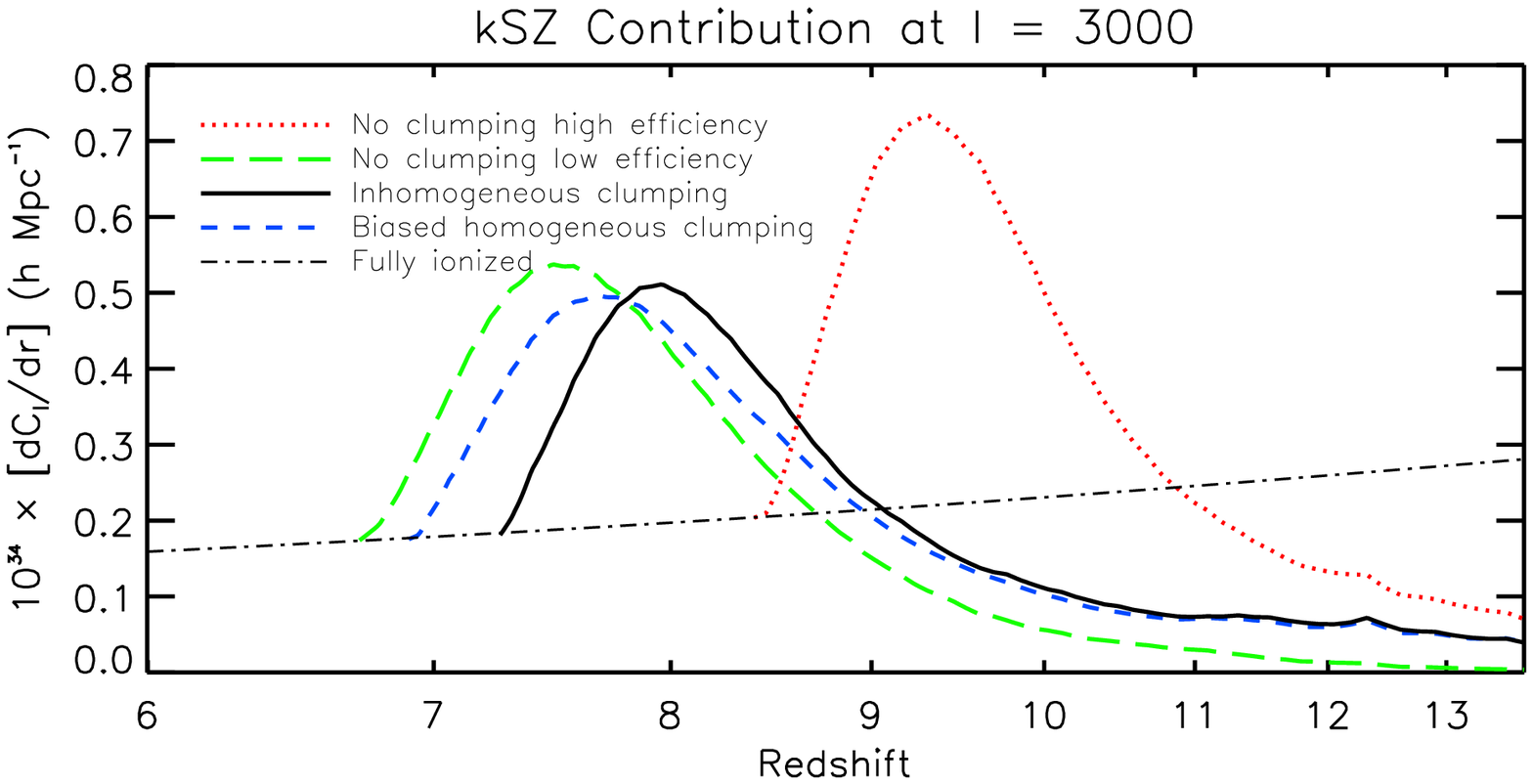} 
\end{center}
\caption{(Left) The kinetic Sunyaev-Zel'dovich effect from $z>z_{\rm ov}$ (see Table~\ref{tab:summary_RT_table} for the value of $z_{\rm ov}$ for each model). The power spectrum of the primary CMB is shown as the thick dashed line for comparison. The 95\% upper bound of $D_{l=3000}$ from the South Pole Telescope measurement \citep{2015ApJ...799..177G} subtracting from it the post-reionization kSZ signal from the cooling and star-formation model of \citet*{2012ApJ...756...15S} re-scaled to our cosmology is shown as a downward arrow. 
(Right) The history of contribution of the kSZ signal at $l=3000$ in terms of the contribution per comoving distance, $dC_l /dr$. The nearly-horizontal dot-dashed line shows the case that assumes all the gas is ionized.}
\label{fig:kSZ}
\end{figure*}

\subsection{Observational signatures}

\subsubsection{The 21~cm background: mean and RMS fluctuations}

The evolution of the mean 21~cm brightness temperature, as shown in Figure~\ref{fig:histories3} (top panel), contains information about the reionization history. Under the assumption $T_s \gg T_{\rm CMB}$, the mean 21~cm brightness temperature decreases as reionization proceeds, since the mean 21~cm signal is proportional to the mean neutral fraction (other than the $\sqrt{1+z}$ dependence). In particular, the zero of brightness temperature corresponds to $z_{\rm ov}$, and its value is consistent with Figure~\ref{fig:histories}. Note, however, that the assumption $T_s \gg T_{\rm CMB}$ breaks down at high redshift, say $z\gtrsim 15$, in which case our prediction of the 21~cm brightness temperature is an overestimate if $T_s >  T_{\rm CMB}$, and $\delta T_b$ can be even negative if $T_s <  T_{\rm CMB}$ (i.e.\ absorption against the CMB).  

Beyond the mean history, the fluctuations in the 21~cm brightness temperature, through the observables like RMS and power spectrum, can reveal geometric information about reionization patchiness, in particular the characteristic sizes of H~II regions during the EOR. Figure~\ref{fig:histories3} (bottom panel) shows the 21~cm RMS fluctuations for a fixed Gaussian beamsize ($3'$) and bandwidth ($0.2\,{\rm MHz}$) with boxcar frequency filter. It is interesting to find that the peak of the RMS fluctuations occurs when $\bar{x}_{\rm HII,m} \simeq 0.75$ in all models. Therefore, this peak appears in the redshift order of the NCHE, IC, BHC, and NCLE models from highest to lowest redshift, respectively, for the reasons explained in \S\ref{subsec:reion-history}. 

Furthermore, we find that the peaks of 21~cm RMS fluctuations depend on whether clumping factor is accounted for, because the peak value for both no-clumping models (NCHE and NCLE models), $\sim 6\,{\rm mK}$, is slightly higher than the peak value for both clumping models (IC and BHC models), $\sim 5.3\,{\rm mK}$. This reflects the fact that the clumping models tend to have smaller H~II regions than no-clumping models at the same mean ionized fraction, as shown in Figure~\ref{fig:SPA_size}. On the other hand, the peak values are about the same for both no-clumping models, and for both clumping models, respectively. This suggests that the peak of 21~cm RMS fluctuations is insensitive to the detail of clumping. 

\subsubsection{The 21~cm background: mock images}

In Figures~\ref{fig:slices}, we illustrate the position-redshift slices cut through the mock image cube, with the spatial dimension on the vertical axis and redshift along the horizontal axis, as a radio telescope would observe it if there were no beam- or bandwidth-smoothing. Images are the 21~cm differential brightness temperature signal extracted from our simulations with different clumping and/or source efficiency models, on a linear scale which reflects neutral structures better, at the full simulation resolution. Note that there is an artificial repetition of structures along the LOS due to the periodic boundary condition, but this does not affect the results as long as the wavenumber is restricted to be larger than the one corresponding to the box size. We do apply the redshift-space distortions due to peculiar velocities, to mimic what an observer would see (after the removal of foregrounds) if there were no beam- or bandwidth-smoothing. 

The 21~cm differential brightness temperature reflects the distribution of neutral hydrogen. The distribution of regions with suppression of the 21~cm signal is a proxy for the distribution of ionized hydrogen. These images are consistent with the histories of reionization found in Figure~\ref{fig:histories} for different reionization models. In addition, the same H~II regions may be visually identified among slices extracted from different simulations, according to their similar shapes, but visual difference in their sizes can be found. We confirm that the IC model yields more numerous small H~II regions to fill the same ionized fraction than the NCHE model. 
On the other hand, the H~II regions appear slightly more fragmented at late times in the IC model than in the BHC model, as shown in Figure~\ref{fig:slices},  
which is consistent with our findings of the H~II bubble size distribution in \S\ref{sec:size_distribution}.

\subsubsection{The 21~cm fluctuation power spectrum} 

The 21~cm power spectrum can provide the geometric information about inhomogeneous reionization in more detail than the 21~cm variance. Figure~\ref{fig:powers_inh_noclumping} shows the 21~cm power spectrum, spherically averaged in the Fourier space, for three key stages of reionization ($\bar{x}_{\rm HII,m} = 0.20$, $0.50$, $0.75$, respectively). On large scales, the 21~cm power spectra for the NCHE and NCLE model show significantly less power than the IC model in the early stages. However, in the middle and late stages the situation is reversed. The fractional error is about tens of per cent at the early and the late stages, but can reach up to $\lesssim 170\%$ in the middle stage of reionization. On the other hand, the 21~cm power spectrum in the BHC model differs from the IC model by $\lesssim 20\%$. 

The ``quasi-linear $\mu_{\bf k}$-decomposition scheme'' \citep{2012MNRAS.422..926M} is a useful tool for providing insight into the trend of 21~cm power spectra. In this scheme, the angle-averaged 21~cm power spectrum is approximately written as 
\begin{equation}
P_{21}^{s,{\rm qlin}} (k) = \widehat{\delta T}_b^2 \left[ P_{\delta_{\rho_{\rm HI}},\delta_{\rho_{\rm HI}}} + \frac{2}{3} P_{\delta_{\rho_{\rm HI}},\delta_{\rho_{\rm H}}} + \frac{1}{5}P_{\delta_{\rho_{\rm H}},\delta_{\rho_{\rm H}}} \right] \,,
\label{eqn:lin-scheme}
\end{equation}
where $P_{a,a}$ denotes the auto power spectrum of the field $a$, and $P_{a,b}$ denotes the cross power spectrum between the fields $a$ and $b$. On large scales, the power spectrum can be further approximated to linear order, $P_{\delta_{\rho_{\rm HI}},\delta_{\rho_{\rm HI}}} \approx P_{\delta_{x_{\rm HI}},\delta_{x_{\rm HI}}} + 
2 P_{\delta_{x_{\rm HI}},\delta_{\rho_{\rm H}}} + P_{\delta_{\rho_{\rm H}},\delta_{\rho_{\rm H}}}$, and 
$P_{\delta_{\rho_{\rm HI}},\delta_{\rho_{\rm H}}}\approx  P_{\delta_{\rho_{\rm H}},\delta_{\rho_{\rm H}}}+ P_{\delta_{x_{\rm HI}},\delta_{\rho_{\rm H}}}$. 
Figure~\ref{fig:pxx_pxd} (top panels) shows the auto power spectrum of the neutral fraction  (or, equivalently, the ionized fraction) field, $P_{x_{\rm HI}}(k) = P_{x_{\rm HII}}(k)$, which reflects the size distribution of the H~II regions. To explicitly compare the components in the 21~cm power spectrum, we show the auto power spectrum of neutral fraction {\it fluctuations}, $P_{\delta_{x_{\rm HI}},\delta_{x_{\rm HI}}} = P_{x_{\rm HI}}(k)/\bar{x}_{\rm HI,m}^2$ (middle panels), and the cross power of neutral fraction fluctuations and total density fluctuations, $P_{\delta_{x_{\rm HI}},\delta_{\rho_{\rm H}}} = P_{x_{\rm HI},\delta_{\rho_{\rm H}}}/\bar{x}_{\rm HI,m}$ (bottom panels), both of which are rescaled by the factor $\bar{x}_{\rm HI}^2(1+z)$, since $\widehat{\delta T}_b \propto \bar{x}_{\rm HI}\sqrt{1+z}$. 

On large scales, the IC model yields less power in neutral fraction, $P_{x_{\rm HI}}$, than the NCHE and NCLE models at all times. This reflects the fact that, given the same source efficiency, the IC model yields more numerous yet smaller H~II regions in the IGM than the NCHE and NCLE model, at the same $\bar{x}_{\rm HII,m}$, which suppresses the power spectrum of the ionized (or, equivalently, neutral) fraction fluctuations. 
On the other hand, comparing the IC and BHC models on large scales, the BHC model always has slightly more power in neutral fraction, which is consistent with our finding in \S\ref{sec:size_distribution} that the characteristic size of H~II regions in the BHC model is always slightly larger than that in the IC model. 

The bottom panel of Figure~\ref{fig:pxx_pxd} shows that the trend of the amplitudes of the cross power spectrum $\left| P_{\delta_{x_{\rm HI}},\delta_{\rho_{\rm H}}} \right| $ is similar to that of the auto power $P_{\delta_{x_{\rm HI}},\delta_{x_{\rm HI}}}$, which confirms the explanation above. Note that the cross power spectrum is negative at small $k$, i.e.\ neutral fraction fluctuations anticorrelate with density fluctuations on large scales, because overdense regions are ionized earlier on average than underdense regions. 

Now we can use equation~(\ref{eqn:lin-scheme}) to explain the trend of the amplitudes of the 21~cm power spectra among different models. At the early phase of reionization, the magnitudes of all three power spectra --- $P_{\delta_{x_{\rm HI}},\delta_{x_{\rm HI}}}$, $\left|P_{\delta_{x_{\rm HI}},\delta_{\rho_{\rm H}}}\right|$, and $P_{\delta_{\rho_{\rm H}},\delta_{\rho_{\rm H}}}$ (density fluctuations power spectrum) --- are comparable. On large scales, although both $P_{\delta_{x_{\rm HI}},\delta_{x_{\rm HI}}}$ and $\left|P_{\delta_{x_{\rm HI}},\delta_{\rho_{\rm H}}}\right|$ are smaller in the IC model than in the NCHE and NCLE model, it is coincidental that the 21~cm power spectrum is larger in the IC model, due to cancellations between these terms. From the intermediate stage, however, the 21~cm power spectrum becomes rapidly dominated by the auto power spectrum $P_{\delta_{x_{\rm HI}},\delta_{x_{\rm HI}}}$. Therefore, the 21~cm power spectrum in the IC model becomes suppressed on large scales with respect to that in the NCHE and NCLE model, following the same trend as $P_{\delta_{x_{\rm HI}},\delta_{x_{\rm HI}}}$.

\subsubsection{The kinetic Sunyaev-Zel'dovich effect}

We plot the kSZ power spectra from our simulations in Figure~\ref{fig:kSZ} (left panel). 
The positive slope of the kSZ power spectrum at $l\lesssim 3000$ reflects the geometry of reionization. If the reionization is dominated by smaller H~II regions, the resulting slope is higher \citep{2013ApJ...769...93P}. We find that the IC model yields a larger slope for the kSZ power spectrum than the NCLE model, which reflects the fact that the IC model yields more numerous small ionized bubbles at a given ionized fraction. Also, the kSZ power spectra of the BHC and IC models have almost identical slopes, which implies that their characteristic sizes of H~II regions are close, though not identical, to each other. 

The right panel of Figure~\ref{fig:kSZ} shows the contribution to the kSZ power spectrum at $l=3000$ from different redshifts. For a given clumping model and its reionization simulation, the kSZ power spectrum is dominated by the contribution from a narrow range of redshifts toward the end of the EOR, as also found in \cite{2013ApJ...769...93P}.  It is interesting that the peak of this distribution appears when $\bar{x}_{\rm HII,m} \gtrsim 0.5$ for all reionization models considered, so it reflects the global history of reionization: the faster reionization proceeds, the earlier the peak contribution of the kSZ power spectrum appears. We find that the redshift of this peak follows exactly the order in $z_{\rm ov}$ (see Figure~\ref{fig:histories}), i.e.\ (from the earliest to the latest arrival) the NCHE, IC, BHC, and NCLE models.  
 
The amplitude of the kSZ power spectrum in Figure~\ref{fig:kSZ} (left panel) depends both on the peak location and on the amplitude of the peak. If the peak amplitude is fixed, then the earlier the peak appears, the larger the total amplitude of the kSZ power spectrum is. If the peak redshift is fixed, then a larger peak amplitude surely enhances the total kSZ amplitude. We find that the NCHE model has the largest total amplitude, the IC model second, and the BHC and NCLE models the smallest\footnote{Careful readers may find that the amplitude of the kSZ power spectrum in the NCLE model is larger than in the BHC model for $l\lesssim 3000$, while reionization proceeds slightly slower in the former. This is because the amplitude of the kSZ peak contribution in the BHC model is smaller.}, which is consistent with the locations of the peak for different models. Note that the amplitude of the kSZ power spectrum in the BHC model is about $10\%$ (relative error) smaller than in the IC model. 

\begin{figure}
\begin{center}
  \includegraphics[height=0.35\textheight]{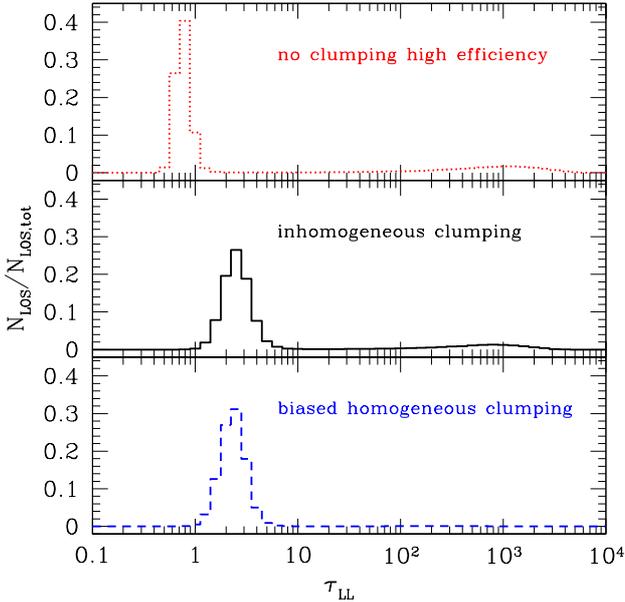} 
\end{center}
\caption{The PDF of the Lyman-limit optical depth at the respective end of reionization for various models: (from top to bottom) no clumping high efficiency ($z_{\rm ov} = 8.4$), inhomogeneous clumping ($z_{\rm ov} = 7.3$), biased homogeneous clumping ($z_{\rm ov} = 6.9$). The optical depth in each case 
is calculated along a LOS thru the
simulation cube, parallel to the $x$, $y$, or $z$-axes. 
The PDF (normalized by the total LOS number) samples all such LOS's along all of those three directions.}
\label{fig:tauLL}
\end{figure}

\subsubsection{End-of-reionization Lyman-limit opacity}
\label{subsubsec:LLO}

Figure~\ref{fig:tauLL} shows the probability distribution function (PDF) of the Lyman limit optical depth, $\tau_{\rm LL}$, for various clumping models at their respective end of reionization. 
We find that the PDF of $\tau_{\rm LL}$ in the NCHE model (at $z_{\rm ov} = 8.4$) is peaked at $\tau_{\rm LL} = 0.8$, so the mean free path of the IGM to ionizing photons at the Lyman limit is $\lambda^{912}_{\rm mfp} = 22$ proper ${\rm Mpc}$. 
This value of mean free path is too large to be favored by the extrapolated value from the low redshift observation of quasar spectra, e.g.\ \cite{2014MNRAS.445.1745W} found that $\lambda^{912}_{\rm mfp} \approx 8.7 - 11.9$ proper ${\rm Mpc}$ at $z=5.2$, and \cite{2010ApJ...721.1448S} found that $\lambda^{912}_{\rm mfp} \approx 4 - 9$ proper ${\rm Mpc}$ at $z=5.7$. 
The Lyman-limit opacity through the IGM during most of the EOR is dominated by the fully neutral patches. However, as we approach the end of reionization, the small residual neutral fraction in the large ionized regions becomes important. In addition there is a contribution from so-called Lyman-limit systems likely in self-shielded regions in galactic haloes and filaments which we do not consider here (but see \citealt{2016MNRAS.458..135S}). Inside these large ionized regions the conditions are close to photoionization equilibrium. Therefore, the small remaining neutral fraction is sensitive to the clumping factor in the IGM and we thus expect larger values for the IGM Lyman-limit opacity when clumping is included.

We indeed find that the PDFs of $\tau_{\rm LL}$ in the IC/BHC model (at $z_{\rm ov} = 7.3/6.9$) both peak at $\tau_{\rm LL} = 2.5$, a value three times larger
than in the NCHE model. The corresponding  mean free paths of ionizing photons in the IC/BHC model are $\lambda^{912}_{\rm mfp} \approx 7.8$ and $8.2$ proper ${\rm Mpc}$, respectively.

We should note that the exact values of $\lambda^{912}_{\rm mfp}$ here are not precise, because we did not consider the Lyman-limit systems in our simulations and these models were not meant to match the end-of-reionization observations. 
In addition, regarding the caveat of our calculation lacking hydrodynamics, it is possible that our IGM-only $\tau_{\rm LL}$ is overestimated. 
Nevertheless, the general trends we find here are reasonable in that the presence of subgrid clumping can decrease the Lyman-limit mean free path substantially, bringing the high-efficiency model, which is otherwise fairly unrealistic with no clumping, more into line with observations. Our results show that the subgrid clumping factor is an important effect to take into account for prediction of the end-of-reionization Lyman-limit opacity. 

\subsubsection{Does the biased homogeneous clumping model work?}

The BHC model has been employed in some previous reionization simulations \citep{2006MNRAS.372..679M,2007MNRAS.376..534I,2007ApJ...657...15K} because its implementation is simple. However, \cite{2011MNRAS.412L..16R} questioned its validity by showing that this simplistic model may lead to significant errors in the estimate of recombination. While we have confirmed in this paper that the BHC model either underestimates or overestimates the mean clumping factor and the recombination rate at different stages of reionization, we find that this model may still be useful in some regimes. Its validity depends on the stage of reionization and on the observables.  
For the mean 21~cm brightness temperature and the 21~cm RMS fluctuations, the BHC model is good to within $20\%$ error 
when $\bar{x}_{\rm HII,m} < 0.5$, but its error increases rapidly at the later stage. 
For the 21~cm power spectrum, the BHC model is a good approximation to within $10\%$ error in the 
range $k< 1\,h\,{\rm Mpc}^{-1}$ at any time or $\bar{x}_{\rm HII,m} \le 0.5$ at all scales considered 
herein $0.06 < k < 7\,h\,{\rm Mpc}^{-1}$. If an error $\lesssim 20\%$ is allowed, then the BHC model 
is good for all stages of reionization throughout this $k$-range considered herein. 
For the end of reionization, the BHC and IC models have an offset of $\Delta z_{\rm ov} = 0.4$, but their CMB optical depths are only different by 3\%, and the PDFs of the Lyman limit optical depth to ionizing photons at the end of reionization are peaked at the same value of $\tau_{\rm LL}$. For the kSZ power spectrum, which integrates over the contributions from all redshifts during the EOR but favors the contribution 
from the epoch $\bar{x}_{\rm HII,m} >  0.5$, the BHC model predicts an amplitude for the kSZ power spectrum $\sim 10\%$ smaller than 
does the IC model, at all scales.

\subsubsection{Are lowered source efficiencies degenerate with enhanced clumping?}

With no clumping, the NCHE model both starts and completes reionization at high redshifts. That results in large CMB optical depth $\tau_{\rm es}$ and high $z_{\rm ov}$, which is inconsistent with current observations. To reconcile the conflicts, both lowered source efficiencies (NCLE model) and enhanced clumping factor (IC and BHC models) can delay the process of reionization. Are they degenerate? Basically, sources in the NCLE model release fewer ionizing photons, so it needs more of the massive haloes to provide just enough ionizing photons, and it takes more time for those additional massive haloes to form, which delays the whole process (i.e.\ both beginning and completion) of reionization. In contrast, in the IC/BHC model, the reionization can start early because of its high source efficiencies, but the completion of reionization is also 
delayed by the consumption of more photons through recombination. The difference between these two delay mechanisms makes the NCLE and IC/BHC model non-degenerate. Specifically, while they complete reionization at similar redshifts, i.e.\ $z_{\rm ov}=6.7/7.3/6.9$ in the NCLE/IC/BHC models, their CMB optical depth predictions ($\tau_{\rm es}=0.058$ and $0.069/0.067$ in the NCLE and IC/BHC models, respectively) are different by about the $1\sigma$ error of the Planck measurement, because reionization in the IC/BHC model is more extended, (i.e. starts at higher redshift). 
For the 21~cm RMS fluctuations, while the peaks for NCLE/IC/BHC appear at similar redshifts, 
the peak amplitudes for the two no-clumping models (NCLE and NCHE) are about the same, and about 20\% larger than that in the IC/BHC model. 
Similarly, the 21~cm power spectra for the two no-clumping models (NCLE and NCHE) are about the same at $\bar{x}_{\rm HII,m} =  0.5$ 
and $0.75$ but clearly different by more than tens of per cent from that in the IC/BHC models.

\section{Summary and conclusions}
\label{sec:conclusion}

\subsection{Summary}

We have investigated the effects of small-scale (typically subgrid) clumping on the progress, 
duration and observational signatures of cosmic reionization. 
Clumping factors were calculated based on a high-resolution N-body simulation of structure formation, 
which resolves haloes down to the Jeans mass scale before reionization. 
We smoothed this N-body particle data onto a grid, using an adaptive kernel, in order to calculate a spatially-varying, 
local subgrid clumping factor to use in boosting the recombination
rates in reionization simulations in which the RT grid is too coarse-grained 
to resolve the full range of this small-scale structure on scales down to the 
prereionization Jeans mass scale. 
We then derived fitting formulae for this local clumping factor as a 
function of the corresponding density of each coarse-grained cell, at each redshift. 
These results were used to run a series of radiative transfer simulations of inhomogeneous cosmic reionization 
with different assumptions about the unresolved gas clumping.  These include cases with no clumping,
in which the recombination rate in each coarse-grained RT cell is based only
on the coarse-grained cellwise density of that cell (which varies from cell to cell but
does not reflect the subgrid density variations within each cell), the case of ``biased
homogeneous clumping'', in which the globally-averaged clumping factor at each
redshift uniformly modifies these coarse-grained recombination rates in each RT cell,
and a fully-inhomogeneous one in which both the coarse-grained density of the RT cells 
varies from cell to cell, just as for the other cases, and the clumping factor is also
spatially-varying according to its dependence on the local overdensity of each RT cell
derived above from the high-resolution N-body simulation. 
 
We find that the simulation for the inhomogeneous clumping model results in a more extended history of reionization than that with no clumping, assuming the same source emissivities, starting reionization at the same time but ending it later. 
Furthermore, the ionized patches are generically smaller and grow slower in the inhomogeneous clumping model -- that is consistent with the picture that higher recombination rates in the former model reduce the characteristic size of H~II regions on average. The 21~cm power spectrum in the inhomogeneous clumping model is suppressed significantly on large scales from the intermediate stage to the end of reionization ($\bar{x}_{\rm HII,m}\ge 0.5$). Also, the IGM Lyman-limit opacity at the end of reionization in the inhomogeneous clumping model is three times larger than that in the no-clumping model. 

We also derived the globally-averaged clumping factor as a function of redshift from the high-resolution N-body simulation, and used it to run the radiative transfer simulation with homogeneous clumping factor. This alternative, simplified, prescription for clumping turns out to be useful for predicting observational signatures of cosmic reionization with modest errors with respect to the inhomogeneous clumping model. For example, for 21~cm power spectrum, its error is within 20\% at all time for all scales of interest to future 21~cm observations.

\subsection{Conclusions}

We have demonstrated that accounting for the local, density-dependent, subgrid clumping is essential for predicting the observational signatures of cosmic reionization correctly. Not only can reionization simulations with inhomogeneous subgrid clumping factor result in an extended history of reionization which can satisfy both high CMB Thomson optical depth $\tau_{\rm es}$ and late end of reionization $z_{\rm ov}$, but also inhomogeneous clumping slows down the expansion of H~II regions and produces more numerous small ionized regions, which, observationally, suppresses the 21~cm power spectrum on large scales when $\bar{x}_{\rm HII,m} \ge 0.5$. Simulations with inhomogeneous subgrid clumping model can also avoid H~II regions with artificially low neutral fraction, which enhances the Lyman-limit opacity at the end of reionization. 
We also provide a simplified prescription with a time-varying, global clumping factor that 
uniformly boosts the recombination rate of the inhomogeneous IGM density field computed with a
lower-resolution simulation (i.e. that misses the subgrid structure) -- the ``biased homogeneous clumping'' model,
which results in fairly modest errors with respect to the inhomogeneous clumping model. 
The first 21~cm measurements by upcoming radio interferometric arrays will allow errors of tens of per cent, 
so the BHC model can be used as a good and easy tool for clumping. However, for precision, percent-level, 
measurements by future 21~cm observations, the inhomogeneous clumping approach is absolutely necessary. 

{\it How do the clumping effects depend on the the reionization model parameters?}  
While our demonstration is based on a limited set of comparisons, the features of clumping we find herein and its impact on the reionization should be generic, because the picture of how and why the inhomogeneous clumping affects the reionization applies generically. Nevertheless, the quantitative impact on cosmological observables may (more or less) depend on the reionization model parameters. When varying the reionization model parameters, the recombination correction inside the H~II regions can be changed in two aspects -- the overall amplitude of clumping averaged over the whole universe, which grows with time as structure formation advances, and the inhomogeneity of that clumping as it corrects the recombination rates inside the H~II regions. If reionization overall is delayed, e.g., by decreasing all source luminosities, then one might expect the overall amplitude of clumping everywhere to be higher at a given stage in the growth of the global ionized fraction. 
Regarding the inhomogeneity of clumping, in that case, the matter inside large H~II regions centered on massive haloes is clustered more strongly at the fixed volume ionized. As such, the effect in which clumping retards the growth of H~II regions might be enhanced relative to our illustrative case. 
However, if we adjust the reionization model parameters in such a way to recover the same global reionization constraints like $\tau_{\rm es}$ or the lower limit on $z_{\rm ov}$ so that reionization is not delayed overall, then the globally-averaged clumping factor would be similar if the H~II region volume filling factor is similar at the same redshift. 
On the other hand, by boosting the relative importance of rarer, higher-mass haloes over less-rare lower-mass haloes, the reionization duration tends to be smaller, reflecting the later, rapid rise of the higher-mass haloes, when we do not account for the inhomogeneous clumping, and with that comes a smaller $\tau_{\rm es}$, if $z_{\rm ov}$ is held fixed. With clumping correction added, since clumping grows with time, the retardation of the growth of the H~II regions relative to the case without clumping is enhanced, so the source luminosities must be adjusted to be higher than it would be, so that $z_{\rm ov}$ is higher if $\tau_{\rm es}$ is held fixed, instead. 
In that case, the clumping correction is smaller than it would have been at lower redshift. This kind of self-consistent adjustment suggests that the relative differences made by taking the inhomogeneous clumping into account and comparing to the results for the reionization model parameters which are adjusted to recover the same global reionization constraints might actually be, not only qualitatively generic, but even quantitatively similar.

{\it How to account for subgrid clumping factor of total density?} While we only demonstrate the case for the cell size of $\sim 0.45\,h^{-1}$ comoving ${\rm Mpc}$ in the coarse-grained mesh, we suggest a fitting formula that relates the locally-averaged subgrid clumping factor to the locally-averaged density. This correlation can be used to calculate the subgrid clumping factor from the local overdensity, since the latter can be easily computed from theory or simulations by smoothing density fluctuations over the coarse-grained resolution. On the other hand, if a modest systematic error, e.g.\ at the level $\sim 20\%$ for 21~cm power spectrum prediction, is allowed, the biased homogeneous clumping model is a convenient, alternative, method for clumping. It is worth noting that the values of global homogeneous clumping factor in this paper are independent of the coarse-grained mesh resolution we adopted for post-processing the high-resolution N-body data, and therefore can be applied elsewhere to a mesh with different grid resolution. 

{\it What are the caveats?} While we have explored the dependence of subgrid clumping factor on local overdensity and on redshift, it can depend, in principle, on three other things: the coarse-grained resolution (or the smoothing scale), the local ionization fraction and local gas temperature, and the stochasticity of clumping. (1) We leave it to future work to investigate the dependence on the mesh cell size, so that the fitting formula can be generalized to a wider range of smoothing scales, which facilitates its application to reionization simulations. 
(2) Our paper assumed that the dependence on ionization fraction and gas temperature is negligible, 
so our simulations contain only dark matter particles and no gas particles (i.e.\ only $N$-body simulations and no hydrodynamics). 
However, when other works have looked into this effect (e.g.\ \citealt{2009MNRAS.394.1812P,2012MNRAS.427.2464F}) using hydrodynamical 
simulations, they do not generally have sufficient resolution to
capture the small-scale structure all the way down to 
the prereionization Jeans scale, and, in addition, are restricted to 
small volumes, in which case the variations of local H~II 
clumping factor are overlooked. 
\cite{2016ApJ...831...86P} performed the first fully-coupled radiation-hydrodynamics simulation of the hydrodynamical back-reaction
of reionization on this small-scale structure, in extremely small volumes that more than resolve the prereionization Jeans scale,
volumes comparable to the size of a single coarse-grained RT cell in the reionization simulations presented here, 
including the time-dependent impact of hydrodynamics on the subgrid clumping factor and its dependence on the mean overdensity of the simulated volume.
We leave it to future work to investigate the impact of this back-reaction and its inhomogeneity 
on the large-scale simulation of reionization discussed here. 
(3) While we assumed in our simulation for the inhomogeneous clumping model that the subgrid clumping factor can be interpolated if the overdensity and redshift are given, we found in Figure~\ref{fig:scatterplot} that there exists a stochastic scatter of subgrid clumping factor at any given overdensity bin which gets stronger at the lower redshift. We leave it to future work to investigate the effect of this stochasticity on cosmic reionization.


\section*{Acknowledgments} 
YM is supported by the National Key R\&D Program of China (Grant No.2017YFB0203302, No.2018YFA0404502), 
the National Natural Science Foundation of China (NSFC Grant No.11761141012, 11673014, 11821303, 11543006), 
the Chinese National Thousand Youth Talents Program, 
and by the Opening Project of Key Laboratory of Computational Astrophysics, National Astronomical Observatories, Chinese Academy of Sciences. 
JK is supported by MUIR PRIN 2015 `Cosmology and Fundamental Physics: illuminating the Dark Universe with Euclid' and Agenzia Spaziale Italiana agreement ASI/INAF/I/023/12/0. 
PRS was supported in part by U.S.~NSF grant AST-1009799, NASA grant NNX11AE09G, NASA/JPL grant RSA Nos.~1492788 and 1515294, and supercomputer resources from
NSF XSEDE grant TG-AST090005 and the Texas Advanced Computing Center (TACC) at the University of Texas at Austin. 
ITI was supported by the Science and 
Technology Facilities Council [grant numbers ST/I000976/1, ST/F002858/1 and
ST/P000525/1]; 
and The Southeast Physics Network (SEPNet). GM was supported 
in part by Swedish Research Council grant 60336701. 
KA was supported by NRF (Grant No.~NRF-2016R1D1A1B04935414). 
This work was supported by World Premier International Research Center Initiative (WPI), MEXT, Japan. 
The authors acknowledge the Texas Advanced 
Computing Center (TACC) at the University of Texas at Austin for providing 
HPC resources that have contributed to the research results reported within 
this paper. This research was supported in part by an allocation of advanced 
computing resources provided by the National Science Foundation through TACC. 
The authors gratefully acknowledge the Gauss Centre for Supercomputing e.V.~(www.gauss-centre.eu) for funding this project by providing computing time through the John von Neumann Institute for Computing (NIC) on the GCS Supercomputers JURECA and JUWELS at J\"ulich Supercomputing Centre (JSC).

\bibliography{my_central_ref}
\bibliographystyle{mnras} 

\begin{table*}
\begin{center}
\begin{minipage}{0.8\linewidth}
\caption{The redshift dependent fitting of the cellwise pseudo-clumping factor $\hat{C}_{\rm cell}$ as a function of local overdensity $\left<\delta\right>_{\rm cell}=\left<n_{\rm N,total}\right>_{\rm cell}/\bar{n}_{\rm N,total}-1$, $y = a_0 + a_1\,x + a_2\,x^2$, where $x = \log_{10}(1+\left<\delta\right>_{\rm cell})^2$ and $y=\log_{10} \hat{C}_{\rm cell}$. The data is based on a coarse-grained mesh in which each cell is $0.45\,h^{-1} {\rm cMpc}$ on each side, using the $6.3\,h^{-1}$\,Mpc N-body simulation which can resolve haloes down to the Jean mass before reionization ($10^5\,M_\odot$).}
\label{tab:fitting}
\begin{tabular}{@{}cccccccccccccc}
 \hline
$z$ & $a_0$ & $a_1$ & $a_2$        & & $z$ & $a_0$ & $a_1$ & $a_2$       & & $z$ & $a_0$ & $a_1$ & $a_2$ \\
\hline
60.000 & 0.00124 & 0.0463 & 0.0594 & & 13.914 & 0.533 & 0.586 & -0.159	 & & 8.515 & 0.944 & 0.412 & -0.215    \\   
41.106 & 0.0257 & 0.0802 & 0.0899  & & 13.557 & 0.557 & 0.581 & -0.173	 & & 8.397 & 0.952 & 0.399 & -0.208    \\  
38.919 & 0.0307 & 0.0885 & 0.100   & & 13.221 & 0.580 & 0.581 & -0.172	 & & 8.283 & 0.962 & 0.401 & -0.202    \\  
36.996 & 0.0358 & 0.0977 & 0.112   & & 12.903 & 0.602 & 0.567 & -0.192	 & & 8.172 & 0.973 & 0.394 & -0.206    \\  
35.289 & 0.0411 & 0.108 & 0.128    & & 12.603 & 0.623 & 0.558 & -0.205	 & & 8.064 & 0.981 & 0.387 & -0.202    \\  
33.761 & 0.0466 & 0.120 & 0.145    & & 12.318 & 0.642 & 0.558 & -0.183	 & & 7.960 & 0.989 & 0.378 & -0.205    \\  
32.385 & 0.0524 & 0.132 & 0.167    & & 12.048 & 0.662 & 0.558 & -0.198	 & & 7.859 & 0.999 & 0.380 & -0.195    \\  
31.137 & 0.0584 & 0.147 & 0.190    & & 11.791 & 0.680 & 0.539 & -0.205	 & & 7.760 & 1.005 & 0.364 & -0.200    \\  
30.000 & 0.0647 & 0.163 & 0.216    & & 11.546 & 0.698 & 0.533 & -0.202	 & & 7.664 & 1.014 & 0.363 & -0.203    \\  
27.900 & 0.0795 & 0.202 & 0.269    & & 11.313 & 0.715 & 0.524 & -0.195	 & & 7.570 & 1.024 & 0.356 & -0.211    \\  
26.124 & 0.0964 & 0.247 & 0.319    & & 11.090 & 0.733 & 0.519 & -0.200	 & & 7.480 & 1.035 & 0.362 & -0.204    \\  
24.597 & 0.116 & 0.297 & 0.349     & & 10.877 & 0.748 & 0.507 & -0.207	 & & 7.391 & 1.042 & 0.350 & -0.217    \\  
23.268 & 0.138 & 0.347 & 0.359     & & 10.673 & 0.763 & 0.498 & -0.203	 & & 7.305 & 1.047 & 0.348 & -0.204    \\  
22.100 & 0.163 & 0.395 & 0.350     & & 10.478 & 0.780 & 0.497 & -0.203	 & & 7.221 & 1.057 & 0.344 & -0.206    \\  
21.062 & 0.190 & 0.440 & 0.320     & & 10.290 & 0.795 & 0.485 & -0.217	 & & 7.139 & 1.065 & 0.334 & -0.212    \\
20.134 & 0.219 & 0.481 & 0.283     & & 10.110 & 0.807 & 0.480 & -0.198	 & & 7.059 & 1.070 & 0.334 & -0.197    \\  
19.298 & 0.249 & 0.513 & 0.224     & &  9.938 & 0.823 & 0.469 & -0.213	 & & 6.981 & 1.078 & 0.328 & -0.198    \\  
18.540 & 0.279 & 0.539 & 0.162     & &  9.771 & 0.834 & 0.472 & -0.191	 & & 6.905 & 1.086 & 0.328 & -0.201    \\  
17.848 & 0.310 & 0.563 & 0.114     & &  9.611 & 0.851 & 0.460 & -0.212	 & & 6.830 & 1.094 & 0.317 & -0.203    \\  
17.215 & 0.341 & 0.579 & 0.0623    & &  9.457 & 0.863 & 0.452 & -0.220	 & & 6.757 & 1.103 & 0.316 & -0.202    \\  
16.633 & 0.371 & 0.588 & 0.0147    & &  9.308 & 0.875 & 0.441 & -0.225	 & & 6.686 & 1.109 & 0.310 & -0.208    \\  
16.095 & 0.400 & 0.594 & -0.0191   & &  9.164 & 0.887 & 0.446 & -0.199	 & & 6.617 & 1.112 & 0.311 & -0.200    \\  
15.596 & 0.428 & 0.598 & -0.0472   & &  9.026 & 0.897 & 0.440 & -0.193	 & & 6.549 & 1.121 & 0.302 & -0.210    \\  
15.132 & 0.456 & 0.596 & -0.0890   & &  8.892 & 0.908 & 0.429 & -0.197	 & & 6.483 & 1.125 & 0.289 & -0.214    \\  
14.699 & 0.486 & 0.601 & -0.107    & &  8.762 & 0.918 & 0.423 & -0.193	 & &	     &       &       &         \\
14.294 & 0.509 & 0.588 & -0.151    & &  8.636 & 0.931 & 0.418 & -0.211   & & 	     &       &       &         \\
\hline
\end{tabular}
\end{minipage}
\end{center}
\end{table*}

\begin{table*}
\begin{center}
\begin{minipage}{0.85\linewidth}
\caption{The global mean pseudo-clumping factor $\overline{\hat{C}}(z)\equiv \overline{n^2}_{\rm N,IGM}/\bar{n}_{\rm N,total}^2$ as a function of redshift $z$, calculated using the $6.3\,h^{-1}$\,Mpc simulation data.}
\label{tab:meanC}
\begin{tabular}{@{}ccccccccccccccccc}
 \hline
$z$ & $\overline{\hat{C}}$  & & $z$ & $\overline{\hat{C}}$  & & $z$ & $\overline{\hat{C}}$ & & $z$ & $\overline{\hat{C}}$ & & $z$ & $\overline{\hat{C}}$ & & $z$ & $\overline{\hat{C}}$ \\
\hline
60.000 & 1.009 & & 22.100 & 1.648 & & 13.914 & 4.566 & &  10.478 & 9.053 & & 8.515  & 13.88 & &  7.221  & 19.07   \\
41.106 & 1.077 & & 21.062 & 1.789 & & 13.557 & 4.866 & &  10.290 & 9.302 & & 8.397  & 14.27 & &  7.139  & 19.07   \\
38.919 & 1.092 & & 20.134 & 1.949 & & 13.221 & 5.220 & &  10.110 & 9.938 & & 8.283  & 14.93 & &  7.059  & 20.04   \\
36.996 & 1.108 & & 19.298 & 2.122 & & 12.903 & 5.488 & &  9.938  & 9.996 & & 8.172  & 15.27 & &  6.981  & 20.35   \\
35.289 & 1.125 & & 18.540 & 2.310 & & 12.603 & 5.761 & &  9.771  & 10.67 & & 8.064  & 15.64 & &  6.905  & 20.71   \\
33.761 & 1.143 & & 17.848 & 2.521 & & 12.318 & 6.211 & &  9.611  & 10.81 & & 7.960  & 15.82 & &  6.830  & 21.02   \\
32.385 & 1.163 & & 17.215 & 2.740 & & 12.048 & 6.511 & &  9.457  & 11.12 & & 7.859  & 16.58 & &  6.757  & 21.58   \\
31.137 & 1.184 & & 16.633 & 2.971 & & 11.791 & 6.805 & &  9.308  & 11.43 & & 7.760  & 16.58 & &  6.686  & 21.99   \\
30.000 & 1.207 & & 16.095 & 3.217 & & 11.546 & 7.222 & &  9.164  & 12.33 & & 7.664  & 17.20 & &  6.617  & 22.48   \\ 
27.900 & 1.264 & & 15.596 & 3.476 & & 11.313 & 7.580 & &  9.026  & 12.80 & & 7.570  & 17.11 & &  6.549  & 22.02   \\
26.124 & 1.335 & & 15.132 & 3.725 & & 11.090 & 7.946 & &  8.892  & 12.98 & & 7.480  & 19.02 & &  6.483  & 21.96   \\
24.597 & 1.422 & & 14.699 & 4.060 & & 10.877 & 8.187 & &  8.762  & 13.50 & & 7.391  & 17.95 & &  	&         \\
23.268 & 1.526 & & 14.294 & 4.251 & & 10.673 & 8.604 & &  8.636  & 13.50 & & 7.305  & 18.65 & &  	&         \\
\hline
\end{tabular}
\end{minipage}
\end{center}
\end{table*}

\begin{appendix}

\section{Subgrid clumping data}

We list the best-fit coefficients of the clumping-overdensity correlation in Table~\ref{tab:fitting}, and the numerical result of the global mean pseudo-clumping factor $\overline{\hat{C}}(z)$ in Table~\ref{tab:meanC}. These results are obtained from high resolution N-body simulations (with minimum mass halo resolved at the Jeans mass before reionization). When applying the data in Table~\ref{tab:fitting} to other simulations, note that the cell size in the coarse-grained mesh must be $\sim 0.45\,h^{-1}$ comoving ${\rm Mpc}$. The data in Table~\ref{tab:meanC} is independent of the coarse-grained resolution. For the convenience of readers, the evolution of this global mean pseudo-clumping factor with redshift is well fit by
\begin{equation}
\overline{\hat{C}}(z) = 158.\,\exp{(-0.334\, z + 0.00576 \, z^2)}\,.
\label{eqn:meanCfit}
\end{equation}
(Note that the BHC model in our paper applied the data in Table~\ref{tab:meanC} directly, not this best-fit formula in Eq.~\ref{eqn:meanCfit}.) 


\end{appendix}


\label{lastpage}

\end{document}